\documentclass[]{article}

\newcommand{\chat}[1]{ }
\def\lesssim{\stackrel{<}{_\sim}}
\def\gtrsim{\stackrel{>}{_\sim}}

\def\bbox#1{\bm{#1}}
\def\be{\begin{equation}}
\def\ee{\end{equation}}
\def\bes{\begin{equation*}}
\def\ees{\end{equation*}}
\def\bea{\begin{eqnarray}}
\def\eea{\end{eqnarray}}
\def\beas{\begin{eqnarray*}}
\def\eeas{\end{eqnarray*}}
\usepackage[]{graphicx}
\usepackage{bm}
\begin{document}



\title{{\bf Rapidly Rotating Atomic Gases}}

\author{N. R. Cooper\\
\hskip0cm 
{\small {\it T.C.M. Group, Cavendish Laboratory, J.J. Thomson
Avenue,  
Cambridge, CB3 0HE, United Kingdom}}\\
{\small {\it and 
L.P.T.M.S., Universit{\' e} Paris-Sud, 91405 Orsay, France.}}}

\date{17 October 2008}


\maketitle

\begin{abstract}

  This article reviews developments in the theory of rapidly rotating
  degenerate atomic gases. The main focus is on the equilibrium properties of
  a single component atomic Bose gas, which (at least at rest) forms a
  Bose-Einstein condensate. Rotation leads to the formation of quantized
  vortices which order into a vortex array, in close analogy with the
  behaviour of superfluid helium. Under conditions of rapid rotation, when the
  vortex density becomes large, atomic Bose gases offer the possibility to
  explore the physics of quantized vortices in novel parameter regimes. First,
  there is an interesting regime in which the vortices become sufficiently
  dense that their cores -- as set by the healing length -- start to overlap.
  In this regime, the theoretical description simplifies, allowing a reduction
  to single particle states in the lowest Landau level. Second, one can
  envisage entering a regime of very high vortex density, when the number of
  vortices becomes comparable to the number of particles in the gas. In this
  regime, theory predicts the appearance of a series of strongly correlated
  phases, which can be viewed as {\it bosonic} versions of fractional quantum
  Hall states. This article describes the equilibrium properties of rapidly rotating atomic
  Bose gases in both the mean-field and the strongly correlated
  regimes, and related theoretical developments for Bose gases in  lattices, 
for multi-component Bose gases, and for atomic Fermi gases. The current
  experimental situation and outlook for the future are discussed in the light
  of these theoretical developments.

\bigskip


\end{abstract}

\newpage

\tableofcontents 


\section{Introduction}

\label{ch:introduction}

One of the most remarkable characteristics of a Bose-Einstein condensate
(BEC) is its response to rotation. As was first understood in the context of
superfluid helium-4\cite{donnelly}, a Bose-Einstein condensate does not rotate
in the manner of a conventional fluid, which undergoes rigid body rotation.
Rather, the rotation of superfluid helium leads to the formation of an array of quantized vortex
lines. Quantized vortices appear also in the superfluid states of helium-3 and
in type-II superconductors in applied magnetic field,\footnote{The connection
  between a rotating neutral fluid and a charged fluid in a magnetic field
  will be clarified below.} which may be viewed as condensates of pairs of
fermionic atoms or electrons.

The experimental achievement of Bose-Einstein condensation in
ultra-cold atomic gases --
formed either from the condensation of atomic bosons\cite{cornellweimannobel,ketterlenobel}, or of pairs of
atomic fermions\cite{blochdz} -- opens up a wide range of new features
in the physics of quantized vortices and vortex arrays, allowing
access to parameter regimes that are inaccessible in the
helium superfluids or type-II superconductors.  This can lead to novel
properties of the rotating groundstates, including the possibility of
exotic strongly correlated phases.  
In recent years, there have been advances in experimental capabilities
and in the theoretical understanding of rotating ultra-cold atomic
gases in these unconventional regimes. The aim of this article is to
review the theoretical developments. It  focuses mainly on
situations in which there are large vortex arrays in a regime of
rapid rotation, and on the theoretical predictions of the novel phases that can
appear in this regime.

 We start with a brief introduction to the
properties of Bose-Einstein condensates and to the physics of quantized
vortices.

\subsection{Atomic Bose-Einstein Condensates}

Consider an ideal (non-interacting) gas in three dimensions, of identical particles of mass $M$, at number density $\bar{n}$, and in equilibrium at a
temperature $T$.  The temperature sets the thermal de Broglie wavelength,
$\lambda_T$, via $\hbar^2/M\lambda_T^2 \sim k_B T$, and the density the mean
inter-particle spacing $\bar{a}\sim \bar{n}^{-1/3}$. At low temperatures, when
$\lambda_T \gtrsim \bar{a}$ the gas must be described by quantum theory, and
its properties depend strongly on the statistics of the particles. For
bosons there is a phase transition when $\lambda_T\sim \bar{a}$, at
a critical temperature
\begin{equation}
T_c  =  \frac{2\pi \hbar^2}{M k_B}\left(\frac{\bar{n}}{\zeta\left(\frac{3}{2}\right)}\right)^{2/3} \,.
\end{equation}
For $T<T_{\rm c}$ the gas is a {\it Bose-Einstein
  condensate} (BEC), characterized by a non-zero fraction of the
particles occupying the same quantum state.

Until recently there was only one experimental realization of an atomic
BEC. The transition of helium-4 into a superfluid state below $T_{\rm c} =
2.17\mbox{K}$ is known to be associated with Bose-Einstein
condensation,\footnote{Bose-Einstein condensation has been measured by neutron
  scattering, yielding a condensate fraction at low temperatures of about
  $9$\%\cite{sokol}.}  albeit in a system in which the inter-particle interactions are
relatively large. However, in recent years,
Bose-Einstein condensation has been achieved in a wide variety of atomic species.
These are prepared
as metastable low density gases, $\bar{n}\sim 10^{12}-10^{15}\mbox{cm}^{-3}$,
confined in magnetic or optical traps. At such low densities, the BEC
transition temperature is extremely small, $T_{\rm c} \simeq
100\mbox{nK}$.  Nevertheless, by a combination of laser and evaporative
cooling these low temperatures can be routinely achieved.

At the low temperatures involved, the thermal de Broglie wavelength,
$\lambda_T \gtrsim \bar{a}\simeq 0.1-1\mu\mbox{m}$, is much larger than the
typical range of the inter-atomic potential. The two-particle scattering is
therefore dominated by $s$-wave scattering, with a scattering length $a_{\rm s}$
that is typically of order a few nanometres (for $^{87}$Rb, $a_{\rm s}\simeq 5
\mbox{nm}$). A typical atomic gas of bosons is weakly interacting, in the
sense that $\bar{n}a_{\rm s}^3 \ll 1$. Consequently, it can be well described as an
ideal Bose gas with very small condensate depletion. (This is in contrast to
superfluid helium, for which the strong interactions cause significant
condensate depletion.) That said, inter-atomic interactions are still present
and are important for many physical properties. 
Non-zero repulsive interactions will cause the gas to behave as a superfluid,
with non-zero critical velocity\cite{Leggett01}. This, in turn, leads to the appearance of
quantized vortices when the superfluid is forced to rotate.

\subsection{Quantized Vortices}

\subsubsection{Quantized Vortex Line}

\begin{figure}
\center\includegraphics[width=6cm]{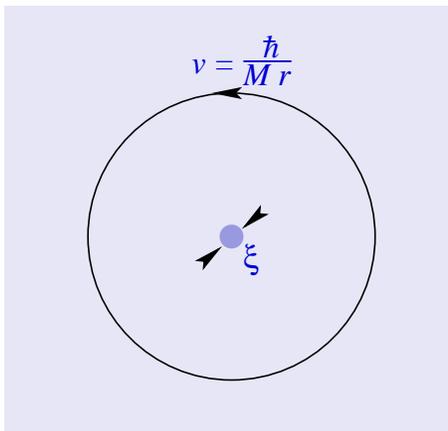}
\caption{Schematic diagram of the velocity field around a quantized vortex
line with one quantum of circulation, $\kappa = \frac{h}{M}$.
The superfluid density falls to zero within the vortex core, which has 
a scale set by the healing length $\xi$.}
\label{fig:vortex}
\end{figure}

It was noted by Onsager and Feynman that superfluid helium cannot rotate as
a conventional fluid. A conventional fluid rotating at angular frequency
$\bm{\Omega}$ has the velocity field of rigid body rotation
\begin{eqnarray}
\label{eq:rigid1}
\bbox{v} & = & \bm{\Omega}\times\bm{r}
\end{eqnarray}
for which the ``vorticity'' of the flow, $\bbox{\nabla}\times \bbox{v}$,
is uniform
\begin{equation}
 \bbox{\nabla}\times \bbox{v}   =  2 \bm{\Omega}\,.
\label{eq:rigid2}
\end{equation}
If, as is believed to be the case, the
superfluid is described by a superfluid wavefunction
$\psi_{s} = \sqrt{n_s} e^{i\phi(\bbox{r})}$
then the  superfluid velocity is
\begin{equation}
\bbox{v_s}  =   \frac{\hbar}{M} \bbox{\nabla}{\phi} \,.
\end{equation}
Hence, the fluid vorticity apparently vanishes:
\begin{equation}
  \bbox{\nabla}\times \bbox{v_s} = 
\frac{\hbar}{M} \bbox{\nabla}\times \bbox{\nabla}\phi = 0 \,.
\label{eq:zero}
\end{equation}
The last equality relies on $\phi$ being a smooth function of position and 
 overlooks the
possibility that the phase $\phi$ might have line-like singularities
(point-like in 2D) around which $\phi$ changes by an integer multiple of
$2\pi$. These are the {\it quantized vortex lines}\cite{donnelly}.  Integrating the vorticity
over a 2D surface containing such a singularity gives
\begin{equation}
\int \bbox{\nabla}\times \bbox{v}_s \cdot d \bbox{S} = \oint \bbox{v_s}\cdot
d\bbox{l} = \frac{h}{M} \times \mbox{integer}
\end{equation}
indicating a delta-function contribution to the fluid vorticity on
the vortex line.  The ``circulation" of the vortex, defined as
\begin{equation}
\kappa \equiv  \oint \bbox{v_s}\cdot d\bbox{l}
 =  \frac{h}{M} \times \mbox{integer}
\end{equation}
is therefore quantized in units of $\frac{h}{M}$.  This leads to a
characteristic velocity profile, with an azimuthal velocity $|\bm{v}_s|$
which diverges as $r\to 0$, see Fig.~\ref{fig:vortex}.  In order to avoid the infinite
kinetic energy associated with this divergent velocity, in the core of the
vortex the superfluid density $n_s$ tends to zero, over a
lengthscale of the {\it healing length}, $\xi$.

The healing length is an important characteristic of a superfluid.
Within the Gross-Pitaevskii mean-field theory\cite{Leggett01} for a homogeneous
interacting Bose condensate the energy is
\begin{equation}
E - \mu N = \int \left[ \frac{\hbar^2}{2M} \left|\bbox{\nabla}\psi\right|^2
  +\frac{1}{2}g\left|\psi\right|^4
-\mu \left|\psi\right|^2 \right] \;d^3\bbox{r}
\label{eq:gp}
\end{equation}
where $\psi$ is the condensate wavefunction. This is written for an
atomic Bose gas with $s$-wave interactions, with the contact
interaction strength, $g=4\pi\hbar^2a_{\rm s}/M$, chosen to reproduce the
$s$-wave scattering length $a_{\rm s}$ [see Eqn.(\ref{eq:gas})]. In many situations, Eqn.(\ref{eq:gp}) is an accurate microscopic description of an atomic BEC,
owing to the small condensate depletion $\bar{n}a_{\rm s}^3\ll 1$.
For a uniform fluid,
minimization with respect to $|\psi|^2$ gives $\mu =
g|\psi|^2$. There is therefore one characteristic lengthscale of
the equilibrium fluid, the healing length,
set by
\begin{equation}
 \frac{\hbar^2}{2M\xi^2}  =
g |\psi|^2   =
\mu \quad \Rightarrow \quad \xi   \equiv   \sqrt{\frac{\hbar^2}{2M\mu}}
= \frac{1}{\sqrt{8\pi \bar n a_{\rm s}}} \,.
\label{eq:healinglength}
\end{equation}
For an atomic gas, the healing length is typically of order $\xi
\simeq 0.5\mu\mbox{m}$ and is large compared to the typical particle
separation, $\xi/\bar{a} \sim \sqrt{\bar{a}/a_{\rm s}} \gtrsim 1$.  In
the case of superfluid helium, for which interactions are strong, the
healing length is very short, $\xi \sim 0.8$\AA; the vortex cores
therefore have a size of order the inter-particle spacing.

\subsubsection{Vortex Lattice}

If superfluid helium is forced to rotate, for example by cooling
liquid helium in a rotating vessel from above the superfluid
transition (where the conventional fluid rotates as a rigid body) into
the superfluid phase\cite{hallvinen}, the superfluid establishes an
array of singly quantized vortex lines oriented along the rotation axis.

For dilute vortices, spaced by a distance $a_{\rm v}\gg \xi$, the
arrangement of vortices is dominated by the influence of the kinetic
energy of the superfluid flow.  Minimizing the kinetic energy of the
superfluid flow in the rotating frame leads to the
conclusions\cite{cooperleshouches} that:

(i) The mean vortex density in the plane perpendicular to the
rotation axis is set by the condition that the mean
superfluid flow (averaged over a lengthscale large compared to the
vortex spacing) mimics rigid body rotation.  The integral of the
vorticity of a uniformly rotating body (\ref{eq:rigid1}) over an area $A$ is $2A\Omega$;
setting this equal to the circulation of a superfluid containing
$N_{\rm v}$ vortices in this area, $N_{\rm v} h/m$, leads to Feynman's
result for the mean vortex density\cite{donnelly}
\begin{equation}
 n_{\rm v} \equiv \frac{N_{\rm v}}{A} = 
  \frac{2M\Omega}{h} \;.
\label{eq:feynman}
\end{equation}

(ii) The vortices experience pairwise logarithmic repulsive
interactions, with an interaction energy per unit length of 
\begin{equation}
\label{eq:log}
-n_s \frac{h^2}{2\pi M}\sum_{i<j} \ln \left(\left|\bbox{R}_i -
  \bbox{R}_j\right|/\xi\right)
\end{equation}
where $n_s$ is the superfluid density and $\bm{R}_i=(X_i,Y_i)$ is the
position of the $i^{\rm th}$ vortex in the plane perpendicular
to the rotation axis.  The groundstate
configuration of a set of classical particles interacting with
logarithmic repulsion is a triangular lattice. One therefore expects
the vortices to arrange in a triangular lattice at low
temperature\cite{campbellziff}.

The direct observation of vortex arrays in rotating superfluid helium
poses a very significant challenge: it requires the measurement
of a set of very small vortex cores, of size $\xi \sim 0.8\mbox{\AA}$, spaced
by the much larger lengthscale $a_{\rm v} \simeq 1/\sqrt{n_{\rm v}}\sim 1\mbox{mm}$ (at the
achievable rotation rates of superfluid helium).  Images of the
positions of quantized vortices emerging from the surface of rotating helium
have been obtained in very remarkable experiments\cite{yarm,yarm1982},
showing arrays of small numbers of vortices.

\subsection{Rotating Atomic Gases}

Vortex lattices may be generated in atomic BECs by confining the gas
in a magnetic trap of cylindrical symmetry, and stirring the gas with
a rotating deformation\cite{MadisonCWD00,abos01,hodby:010405}.  The
formation of vortices is driven by dynamical instability of the
surface
modes\cite{PhysRevLett.86.564,PhysRevA.63.011601,madinstab,PhysRevLett.92.020403}
followed by a period of turbulent flow before steady state is reached.

We shall focus on the equilibrium states that can be achieved,
assuming that the system reaches a steady state at a rotation rate
$\Omega$. Under typical conditions, the rotation rates are $\Omega
\simeq 2\pi \times 100\mbox{Hz}$, setting a mean inter-vortex spacing
of $a_{\rm v}\sim 2\mu\mbox{m}$. This is small compared to the condensate
size, allowing large numbers of vortices
to be created in the condensate (observations of up to 160 vortices
were reported in Ref.\cite{abos01}).  The gases are
typically in the regime $\xi\lesssim a_{\rm v}$ for which the vortex
lattice structure is determined by the kinetic energy of superfluid
flow (\ref{eq:log}) and one expects a triangular lattice ordering at low
temperatures.  Indeed, images of the particle density, taken following
expansion of the cloud, show large numbers of vortices ordered in
a triangular lattice arrangement\cite{MadisonCWD00,abos01,hodby:010405,coddington:100402,schweikhard:040404}.

There are many reasons why cold atomic gases are very interesting
systems in which to study the properties of superfluid vortices and vortex arrays.
The aspects that we shall focus on in this review are:

\begin{itemize}

\item
{\bf High vortex density.}  In atomic BECs the inter-vortex spacing
$a_{\rm v} \simeq 2 \mu\mbox{m}$ is comparable to the vortex core
size, as set by the healing length which is typically $\xi \simeq
0.5\mu\mbox{m}$.  One can therefore envisage\cite{wgs} entering a
novel regime in which the rotation frequency is sufficiently high that
the separation between the vortices $a_{\rm v}$ becomes less than the
healing length $\xi$. Indeed, this new regime has been achieved for a
Rb condensate, by using an ingenious technique to achieve very high
rotation rate and low density\cite{schweikhard:040404}.  In superfluid
helium, where the zero temperature healing length is $\xi \simeq
0.8\mbox{\AA}$, achieving this regime would require $\Omega \gtrsim
10^{13}\mbox{rads}^{-1}$!  It is of interest to understand the
properties of vortex lattices in the regime, $a_{\rm v} \ll \xi$, which
differ markedly from those of superfluid helium for which $\xi \ll a_{\rm v}$.

\item
{\bf strongly correlated phases.}  One can envisage entering a regime
of even higher vortex density, when the number of vortices is
comparable to the number of particles\cite{cwg}.  In this case, theory predicts
the formation of strongly correlated phases.  These can be
understood as {\it bosonic} analogues of the strongly correlated
phases that are responsible for the fractional quantum Hall effect
(FQHE) of electrons in semiconductors\cite{prangeandgirvin,dassarmapinczuk}. These states have many
interesting and novel properties.

\item {\bf Tunable interactions.}  There is the possibility to tune
the inter-particle interactions: both the strength and the qualitative
form of the interaction, by use of a Feshbach resonance and/or dipolar
interactions. The groundstate of a rapidly
rotating Bose gas is sensitive to the nature of the interaction.

\item
{\bf Lattice potentials.}  The imposition of a rotating optical lattice\cite{tung:240402,hafezi-2007}, or the use of ``artificial
gauge fields''\cite{JakschZoller,mueller,sorensen:086803}, allows
studies of the interplay of vortex lattices with external lattice
potentials. This raises interesting issues involving the effects of
commensurability of the two lattice periods. Similar systems may also help
to stabilize strongly correlated phases in
experiment\cite{sorensen:086803,palmer:180407,palmer:013609,hafezi-2007}.

\item {\bf Multi-component (spinor) gases.} The trapping and rotation of Bose
  condensates with multiple components is possible. This can give rise to more
  complex forms of vortex lattice order, including topological spin-textures,  and raises the possibility of novel  strongly correlated phases.

\item {\bf Fermi gases.}  Atomic Fermi gases can have interesting and
unusual response to rotation. In particular, strongly interacting
Fermi gases in the BEC-BCS regime behave as superfluids. Rotation
leads to the formation of quantized vortices. Rapid rotation raises the
possibility of interesting strongly correlated phases involving the
interplay of the physics of superconductivity with the physics of the
fractional quantum Hall effect (FQHE).

\end{itemize}

As is clear from the above list of topics, this review will focus  mainly on
one aspect of the physics of rapidly rotating gases: the possibility of the
appearance of strongly correlated equilibrium phases in novel parameter
regimes accessible in experiments. Recent progress in this field has involved
many other interesting developments. These include the non-equilibrium
properties of rotating gases, dynamical excitations, the effects of
confinement and trap geometry, 
 and the many experimental developments. In the
present article, these subjects will either be omitted or only touched upon
briefly. The reader is referred to Refs.\cite{blochdz,fetter} for excellent
recent reviews that cover these topics in more detail.

\chat{
Under typical conditions, the rotation rates are $\Omega \simeq
2\pi \times 100\mbox{Hz}$, setting a mean inter-vortex spacing of
$\n_{\rm v}\sim 2\mu\mbox{m}$ which is somewhat larger than the
typical core size $\xi\sim 0.5\mu\mbox{m}$.  Thus, the gases are in
the regime $\xi\gtrsim a_{\rm v}$ for which the vortex lattice
structure is determined by the kinetic energy of superfluid
flow\ref{eq:log} and one expects a triangular lattice ordering at low
temperatures.  
At faster rotation rates allow entry into a novel regime in which
$\xi\gtrsim a_{\rm v}$.  This has been achieved by using a very
ingenious technique\cite{schweikhard:040404}. After some initial
stirring, the rotating deformation is removed to leave the gas in a
state of high angular momentum, which is conserved due to the
rotational symmetry of the trap. Atoms at the centre of the trap,
which carry little angular momentum are then selectively removed and
the system allowed to equilibrate.  The gas is still at
(approximately) the same angular momentum but, due to the loss of
atoms, has a smaller moment of inertia, thus leading to an increase in
the angular velocity $\Omega$.

 }

\section{Rapidly Rotating Atomic Bose Gases}

\label{sec:rapid}

The possibility of achieving high vortex density, and entering a regime where
the mean vortex spacing $a_{\rm v}$ is small compared to the healing length
$\xi$\cite{wgs}, is one of the most striking differences between the physics
of atomic BECs and that of rotating helium. We now describe the theoretical
description of atomic gases in this rapid rotation regime. This discussion
will form the basis for many of the theoretical results described in this
review.

\subsection{Rapid Rotation Limit}

\label{sec:2dlll}

The Hamiltonian describing an interacting atomic gas in an
axisymmetric harmonic trap is
\begin{equation}
H  =  \sum_{i=1}^N \left[\frac{|\bbox{p}_i|^2}{2M} + \frac{1}{2}M
  \omega_\perp^2(x_i^2+y_i^2) + \frac{1}{2}M \omega_\parallel^2z_i^2  \right]+
\sum_{i<j} V\left(\bm{r}_i-\bm{r}_j\right) \,.
\label{eq:ham}
\end{equation}
Most commonly, the interaction between ultra-cold atoms can be represented
by a contact interaction\cite{Leggett01}
\begin{equation}
V(\bm{r}) = 
g\delta^{(3)}(\bm{r})
\label{eq:contact}
\end{equation}
with the interaction strength chosen as
\begin{equation}
g = \frac{4\pi \hbar^2a_{\rm s}}{M}
\label{eq:gas}
\end{equation}
to reproduce the $s$-wave scattering length. However, in later
sections we shall also consider situations involving dipolar
interactions or a Feshbach resonance in the interaction.

The Hamiltonian (\ref{eq:ham}) describes $N$ identical particles which
are confined in a harmonic trap, with natural frequencies
$\omega_\perp$ and $\omega_\parallel$. Note that the potential is
axisymmetric about the $\hat{\bm{z}}$-axis. We therefore (for now)
neglect the influence of any deformation that is required  to
stir the gas.  One can imagine that the system has been stirred, such
that it has picked up a large angular momentum about the
$\hat{\bm{z}}$-axis, but this stirring potential has been removed. Due
to the axial symmetry, the angular momentum will be conserved. The
question we shall address is: What is the groundstate of a Bose gas
that has been prepared in this way?  Equivalently: what is the lowest
energy state of (\ref{eq:ham}) as a function of the angular momentum
about the $\hat{\bm{z}}$-axis?

For much of this review, we shall consider situations in which the
interactions are relatively weak. In this limit, the relevant single particle
states simplify. To motivate the form of the single particle states it is
convenient to work not in terms of fixed angular momentum $L$, but in terms of
its conjugate thermodynamic variable which is the rotation rate $\Omega$.

\subsubsection{Rotating Frame of Reference}

In a frame of reference rotating about the ${z}$-axis with angular momentum
$\bm{\Omega} = \Omega\bm{\hat{z}}$, the Hamiltonian is\cite{LLStatMech}
\begin{equation}
H_\Omega = H-\bbox{\Omega}\cdot\bbox{L} \,.
\end{equation}
In this frame, the one-body terms can be written in a suggestive way
\bea
H^{(1)}_\Omega & = &  \frac{|\bbox{p}|^2}{2M}+\frac{1}{2}M\omega_\perp^2(x^2+y^2) +\frac{1}{2}M\omega_\parallel^2 z^2   - \bbox{\Omega}\cdot\bbox{r}\times\bbox{p}\\
 &  = &
 \frac{|\bbox{p}-M\bbox{\Omega}\times\bbox{r}|^2}{2M}+\frac{1}{2}M(\omega_\perp^2-\Omega^2)(x^2+y^2) + \frac{1}{2}M\omega_\parallel^2z^2 \,.
\label{eq:hamrot}
\eea
The kinetic term in this Hamiltonian is equivalent to that of a particle of
charge $q^*$ experiencing a magnetic field $\bm{B}^*$ with
\begin{equation}
q^*\!\bbox{B}^* = 2M\bbox{\Omega} \,.
\label{eq:qB}
\end{equation}
This connection shows that the Coriolis force in the rotating frame plays the
same role as the Lorentz force on a charged particle in a uniform magnetic
field\cite{frohlich1994}. Much use shall be made of the equivalence of these
two effects in this review. Indeed, I shall frequently refer to ``rotation''
or ``magnetic field'' interchangeably, assuming that the connection is clear.

One difference arises from the centrifugal force of the rotation, which has no
analogue for a charged particle in a magnetic field. The centrifugal force
acts to reduce the harmonic confinement potential. Stability of a harmonically
trapped gas requires that the confinement is retained in (\ref{eq:hamrot}),
which requires that the rotation rate is below the ``centrifugal limit'' 
\be
\Omega \leq \omega_\perp \,.
\label{eq:centrifugal}
\ee 

Analogies with the properties of a charged particle in a uniform magnetic
field are most evident in the limit of a large number of vortices, $N_{\rm v}
\gg 1$. In this case, very general considerations lead to the conclusion that
$\Omega \simeq \omega_\perp$ [see \S\ref{sec:lda}, and Eqn.(\ref{eq:close})].
Essentially, one can then neglect any residual confinement and set $\Omega =
\omega_\perp$, viewing the system as being uniform in the plane perpendicular
to the rotation axis. (The analogy holds also for finite $N_{\rm v}$ with
$\Omega < \omega_\perp$, but involves the Fock-Darwin spectrum for a charged
particle in a magnetic field with harmonic confinement\cite{fock,darwin},
which makes the present discussion somewhat less clear. This case is covered
in the following section.) For $\Omega = \omega_\perp$, the Hamiltonian
(\ref{eq:hamrot}) describes a quasi-2D system of particles in a uniform
magnetic field. The energy spectrum takes the form familiar from studies of
the QHE\cite{prangeandgirvin} \be E_{n,m,n_\parallel} =
2\hbar\omega_\perp\left(n +\frac{1}{2}\right) +
\hbar\omega_\parallel\left(n_\parallel +\frac{1}{2}\right)
\label{eq:fullllspectrum}
\ee where
$n=0,1,2,\ldots $ is the
Landau level index, $m=-n,-n+1,-n+2,\ldots$ the angular momentum quantum number about the rotation axis, and
$n_\parallel=0,1,2,\ldots$ is the subband index for motion along the rotation axis. The
spectrum is highly degenerate, with the single-particle states of different
angular momentum $m$ having the same energy. The effective ``cyclotron
frequency'' is $q^*B^*/M = 2\omega_\perp$.

The weak interaction limit occurs when the  chemical potential $\mu \sim
g\bar{n}$ is small compared to the single-particle level spacings\footnote{For the strong inequalities in
  (\protect\ref{eq:weak}) to hold, the factor of $2$ in this expression is
  clearly unimportant. We leave this factor in this formula merely as a
  reminder that  the lowest energy single particle excitation out of the
  lowest Landau level
  has energy $2\hbar\omega_\perp$. In practice, one expects the
physics derived by theoretical studies in the regime (\protect\ref{eq:weak})
to be at least qualitatively accurate even in experiments in which 
these inequalities are not well satisfied.}
\begin{equation}
\mu \ll \hbar\omega_\parallel, 2\hbar\omega_\perp \,.
\label{eq:weak}
\end{equation}
These conditions are equivalent to the conditions that the healing length is
large compared to the inter-vortex spacing, $\xi\gg a_{\rm v}$,  and to the subband thickness in the
$z$-direction, $\xi\gg a_\parallel$.
 Although (\ref{eq:weak}) is typically not satisfied for a
non-rotating gas, for a rapidly rotating gas the centrifugal forces spread the
cloud out, the density falls and the system tends towards this weakly
interacting regime\cite{schweikhard:040404}.

Under these conditions (\ref{eq:weak}),\footnote{The validity of the
  restriction of single particle states to the 2D LLL has been explored
  in Ref.\protect\cite{morris:033605}. The numerical results are
  interpreted to indicate disagreement with the condition
  (\protect\ref{eq:weak}) for restriction to the LLL, for the case of
  incompressible states at fixed filling factor $\nu=N/N_{\rm v}$, Eqn.(\ref{eq:fillingfactorN}). However, the criterion applied -- the shift of the
  rotation frequency $\Omega$ -- appears to overlook the dependence of
  $\Omega$ on $N_{\rm v}$, e.g. Eqn.(\ref{eq:close}) 
for $N_{\rm v}\gg 1$. An analysis
based on Eqn.(\ref{eq:close}) together with the assumption that  
$\mu \sim g\bar{n}$ with a correction due to LL mixing $\delta\mu \sim
(g\bar{n})^2/\hbar\omega_\perp$ that is small ($\delta\mu\ll \mu$)  leads
  to the conclusion that the quantity $g_{\rm max}$ defined in
  Ref.\cite{morris:033605} should vary as $g_{\rm max}\sim N_{\rm v}^2/N$.
  Thus $g_{\rm max} \propto 1/N$ at fixed $N_{\rm v}$ and $g_{\rm max} \propto
  N$ at fixed $\nu$, in rough agreement with the numerical
  results\cite{morris:033605}.}
to a good approximation
 the single particle states
are restricted to quasi-2D ($n_\parallel=0$) and to  the lowest Landau
level (LLL, $n=0$). The single particle wavefunctions are\cite{wgs}
\begin{equation}
\psi_m(\bm{r})  \propto  (x+iy)^m
\;  e^{-(x^2+y^2)/2a_\perp^2}\;e^{-z^2/2a_\parallel^2}
\label{eq:2dlll}
\end{equation}
where
\begin{equation}
a_{\perp,\parallel}\equiv
  \sqrt{\frac{\hbar}{M\omega_{\perp,\parallel}}}
\end{equation}
are the trap lengths in the radial and axial directions. 
We refer to this regime (\ref{eq:weak}) as the {\bf ``2D LLL regime''}.

To make connections to the quantum Hall effect as
clear as possible, we shall introduce the complex representation 
\begin{equation}
\zeta  \equiv  \frac{x+iy}{\ell} 
\end{equation}
where $\ell$ is the conventional magnetic length associated with $q^*B^* = 2M\omega_\perp$
\be
\ell  \equiv  \sqrt{\frac{\hbar}{2M\omega_\perp}}  = \frac{a_\perp}{\sqrt{2}} \,.
\label{eq:ell}
\ee
The (normalized) 2D LLL basis states are then
\begin{equation}
\psi_m(\zeta,z)  = \frac{1}{\sqrt{2\pi 2^{m} m!}} \;\zeta^m
\;\times
\left[ \frac{e^{-|\zeta|^2/4}}{\ell}
\frac{1}{(\pi a_\parallel^2)^{1/4}}
e^{-z^2/2a_\parallel^2}
\right] \,.
\label{eq:2dlllzeta}
\end{equation}
For simplicity, in the following when writing wavefunctions in the 2D LLL, we
shall  omit the ubiquitous [bracketed] exponential terms, focusing only
on the prefactors that are polynomial in $\zeta$.

\subsubsection{Laboratory Frame of Reference}

An alternative way to understand the restriction to the 2D LLL 
is to work in the laboratory frame, and consider the total angular
momentum as the control parameter\cite{wgs}. 
This approach makes the appearance of Landau level wavefunctions somewhat less
evident, but it has the advantage of clarifying that $N_{\rm v}\gg 1$ is not
required for the restriction to the states (\ref{eq:2dlll}) in the 2D LLL
regime (\ref{eq:weak}).

For the harmonically trapped gas (\ref{eq:ham}) the single particle energy
spectrum is
\begin{equation}
E = \hbar\omega_\perp(2n_\perp  +|m| + 1) + \hbar\omega_\parallel(n_\parallel +
1/2)
\label{eq:spec}
\end{equation}
where $m=0,\pm 1, \pm 2, \ldots$ is the angular momentum quantum number (about the $z$-axis)
and $n_\perp, n_\parallel = 0,1,2,\ldots$ are the radial and axial quantum
numbers. For a set of particles $i=1,N$ occupying these states, 
the total
angular momentum (in units of $\hbar$)  
is
\begin{equation}
L = \sum_{i=1}^N m_i
\label{eq:l}
\end{equation}
where $m_i$ is the  angular momentum of particle $i$.

For weak interactions (\ref{eq:weak}), to determine the groundstate at fixed
total angular momentum $L$, we must first minimize the total kinetic and
potential energies. Clearly, from (\ref{eq:spec}), the particles then must
only occupy single-particle states with $n_\perp=n_\parallel=0$. In addition, one can show
that at fixed $L$ (and assuming $L\geq 0$ without loss of generality)
particles must only occupy states with $m_i\geq 0$\cite{wgs}. The resulting
set of single particle states -- those states with $n_\perp=n_\parallel=0$ and
$m_i\geq 0$ -- is precisely the set of 2D LLL states (\ref{eq:2dlll},\ref{eq:2dlllzeta}).

To see why one requires $m_i\geq 0$, let us first write down the energy for a
collection of $N$ particles with total angular momentum (\ref{eq:l})
restricted to the 2D LLL ($n_\perp=n_\parallel=0$, $m_i\geq 0$). The total single-particle energy is
\begin{equation}
E  =   \sum_{i=1}^N \left[\hbar\omega_\perp(m_i + 1) +
  \hbar\omega_\parallel/2\right] = 
 \left(\hbar\omega_\perp + \frac{1}{2}\hbar\omega_\parallel\right) N +
\hbar \omega_\perp L\,.
\label{eq:spe}
\end{equation}
Now, consider moving one particle, $i=1$, from its state
$m_1\geq 0$ to a new state $m_1'$, which may have $m_1'<0$ and then be outside of the
2D LLL. To conserve the total angular momentum $L$,
we shall, in addition, need to add $m_1-m_1'$ units of angular momentum to the
other particles. Keeping these particles in states with $m\geq 0$ leads to an
overall change in energy
\be
\Delta E = 
\hbar\omega_\perp \left[(-|m_1| + |m_1'|) + (m_1-m_1')\right] = \hbar\omega_\perp
\left[|m_1'| - m_1'\right]
\ee
which is an energy {\it increase} if $m_1'<0$. Hence, at fixed $L\geq 0$ the lowest
energy states must have $m_i\geq 0$.

\subsubsection{Effects of Weak  Interactions}

The inclusion of weak interactions poses a theoretical problem that is very
closely related to that for electrons in the
FQHE\cite{prangeandgirvin,dassarmapinczuk}. For a system of $N$ particles with
a given total angular momentum, $L$, one must distribute the particles within
the 2D LLL orbitals such that the angular momentum (\ref{eq:l}) is fixed. The
single particle (kinetic+potential) energy of all these states (\ref{eq:spe})
is the same, leading to a very high degeneracy at the single-particle level.
The groundstate is determined by the action of the interactions within this
degenerate space. This is a fundamentally non-perturbative problem, as in the
FQHE. The differences are firstly in the nature of the inter-particle forces,
and secondly, and most importantly, that here we are studying bosons, not fermions. We are
interested in the nature of the phases that can emerge for rotating bosons
for the specific forms of interactions that are physically
relevant in ultra-cold atomic gases.

For the most part, we shall focus on the case of contact interactions
(\ref{eq:contact}) between bosons. The interaction between
particles is fully specified by the Haldane pseudo-potentials (see
\S\ref{sec:haldane}), which for contact interactions reduce to the
single parameter \be V_0 = \sqrt{\frac{2}{\pi}}\frac{\hbar^2 a_{\rm
    s}}{Ma_\perp^2a_\parallel} \,.
\label{eq:v0}
\ee  
This parameter sets the energy scale of the rotating bosons in the 2D LLL.

For a set of $N$ particles at total angular momentum $L$, we write the
interaction energy of the groundstate of the contact interactions as
$E_I(L,N)\equiv V_0
\epsilon_I(L,N)$ where $\epsilon_I(L,N)$ is a dimensionless function. 
In the laboratory frame of reference, the total energy (single particle and interaction), is then, using
Eqn.(\ref{eq:spe})
\be
E(L,N) = \left(\hbar\omega_\perp + \frac{1}{2}\hbar\omega_\parallel\right) N
+ \hbar \omega_\perp L + V_0\epsilon_I(L,N) \,.
\ee
In the rotating frame of reference, the energy is
\be
E_\Omega(L,N) = \left(\hbar\omega_\perp + \frac{1}{2}\hbar\omega_\parallel\right) N
+ \hbar (\omega_\perp -\Omega) L + V_0\epsilon_I(L,N)\,.
\label{eq:tote}
\ee At given $\Omega$, the groundstate angular momentum $L(\Omega)$ is
obtained by minimizing (\ref{eq:tote}) over the allowed integer values of $L$.
Thus, in the rotating frame of reference, the groundstate angular momentum is
a function of the dimensionless control parameter $\hbar(\omega_\perp -
\Omega)/V_0$.
From the viewpoint of the laboratory frame, in which the control parameter is
angular momentum, the rotation frequency is a derived
quantity, set by
 \be \Omega = \frac{V_0}{\hbar}\frac{\partial
  \epsilon_I}{\partial L} + \omega_0 \,. \ee This is a standard thermodynamic relation
stating that  $\Omega$ is the conjugate thermodynamic variable to $L$
(expressed at zero temperature and in terms of the parameters we have defined).

For a highly anisotropic trap, with $\omega_\parallel \ll \omega_\perp$, there
is the possibility to explore a related regime\cite{Ho01} 
\be
\hbar\omega_\parallel \ll \mu \ll 2 \hbar\omega_\perp \,.
\label{eq:q3dregime}
\ee The interactions are then small compared to the cyclotron splitting
$2\hbar\omega_\perp$, so the motion perpendicular to the rotation
axis is still described by the LLL. However,
since the interactions are large compared to the subband spacing, many
subbands of the axial confinement are relevant. We shall refer to this regime
(\ref{eq:q3dregime}) as the
``{\bf 3D LLL limit}''. Following the approach proposed in Ref.\cite{Ho01},
within mean-field theory in the 3D LLL regime, 
one can expect the density distribution along the
rotation axis to adopt a Thomas-Fermi profile, with overall length $2W_z$.
As described in \S\ref{sec:haldane} the properties are then the same as in the 2D LLL, but with an overall
change in interaction constant (\ref{eq:v0tf}), to
$V_0^{\rm TF} =    \frac{6}{5}\frac{\hbar^2 a_{\rm
s}}{Ma_\perp^2 W_z}$. (See \S\ref{sec:3dmft} for a discussion of the validity
of mean-field theory in the 3D LLL regime.)

In the above we have discussed the rapid rotation limit of atomic bosons. The
same considerations apply equally well to rotating fermions, leading to 2D LLL
and 3D LLL limits at sufficiently low density and weak interactions that
(\ref{eq:weak}) or (\ref{eq:q3dregime}) apply. In \S\ref{sec:fermions} we
shall describe the properties of rapidly rotating fermions in the LLL.

\subsection{Gross-Pitaevskii Mean-Field Theory}

The Gross-Pitaevskii mean-field theory for an interacting Bose gas amounts to
the assumption that the many-body state is a pure condensate, in which all
particles occupy the same single particle state $\psi(\bm{r})$
\begin{equation}
  \Psi({\{\bbox{r}_i\}}) = \prod_{i=1}^N
  \psi(\bbox{r}_i)\,.
\end{equation}
As an approximation to the groundstate, the (normalized) condensate
wavefunction $\psi$ is chosen to minimize the expectation value of the Hamiltonian.
(We shall discuss the limits of applicability of this ansatz in \S\ref{sec:beyond}.)

For a rapidly rotating Bose gas  the single particle
orbitals consist only of the states (\ref{eq:2dlll}).  Thus, the condensate
wavefunction can be expanded in terms of these states
\begin{equation}
\psi(\zeta,z) = \sum_{m\geq 0} c_m \psi_m
\label{eq:expand}
\end{equation}
with $\sum_m |c_m|^2 =1$.  The mean-field groundstate is obtained by choosing
the coefficients $c_m$ to minimize the expectation value of the interaction
energy, which for contact interactions
(\ref{eq:contact}), is 
\be
E_I^{\rm GP}(L,N) = \frac{1}{2}g N^2 \int 
|\psi({\bm r})|^4\; d^3{\bm r}
\label{eq:egp}
\ee
with a constraint on the (average)  angular momentum,
\begin{equation}
L = N  \sum_m m |c_m|^2\,.
\end{equation}
The mean-field groundstate therefore depends only on $L/N$. The determination of
the condensed LLL state 
that minimizes the interaction energy at
fixed $L/N$ constitutes the LLL mean-field theory. {The application of GP mean-field theory within the 2D LLL regime
(\ref{eq:weak}), or the 3D LLL regime with TF profile
(\ref{eq:q3dregime}), is sometimes referred to as the ``mean-field quantum
Hall'' regime in the literature. We avoid use of this terminology,
preferring the term ``mean-field LLL'' regime. 
As discussed below, the physics of the mean-field
LLL regime is in  close analogy with that of the Abrikosov lattice in
type-II superconductors (predating quantum Hall effects significantly!). 
We 
reserve the term ``quantum Hall regime'' to
the regime of strong
correlations, described in \S\ref{sec:stronglycorrelated}, where there
are close analogies with
 the fractional quantum Hall effect.}

Note that, in general, Eqn.(\ref{eq:expand}) is not an eigenstate of angular
  momentum, while the Hamiltonian conserves total angular momentum. The fact
  that the condensate wavefunction does not preserve the rotational symmetry
  of the Hamiltonian should not necessarily be viewed as a deficiency. Rather this
  wavefunction correctly captures the fact that the system  spontaneously
  breaks rotational symmetry, in the limit $N\to\infty$ with $L/ N$
  fixed. This will be discussed further in \S\ref{sec:symmetry}.

Noting that the condensate wavefunction (\ref{eq:expand}) is a polynomial in $\zeta$, it may be expressed in terms of its zeros\cite{Tesanovic91} as
\begin{equation}
\psi(\zeta,z) = A \prod_{\alpha=1}^{m_{\rm max}} \left(\zeta-\zeta_\alpha\right) 
\label{eq:condensate}
\end{equation}
where $m_{\rm max}$ is introduced as a cut-off in the degree of
the polynomial (\ref{eq:expand}), and $A$ is a normalization factor. These zeros are the complex positions of
the quantized vortices.  Thus the LLL Gross-Pitaevskii wavefunction is fully
described by the positions of the vortices.  
The process of
choosing the $m_{\rm max}+1$ complex coefficients $c_m$ (with normalization) is equivalent to the
choice of the locations of $m_{\rm max}$ vortices.

\begin{figure}
\center\includegraphics[width=7cm]{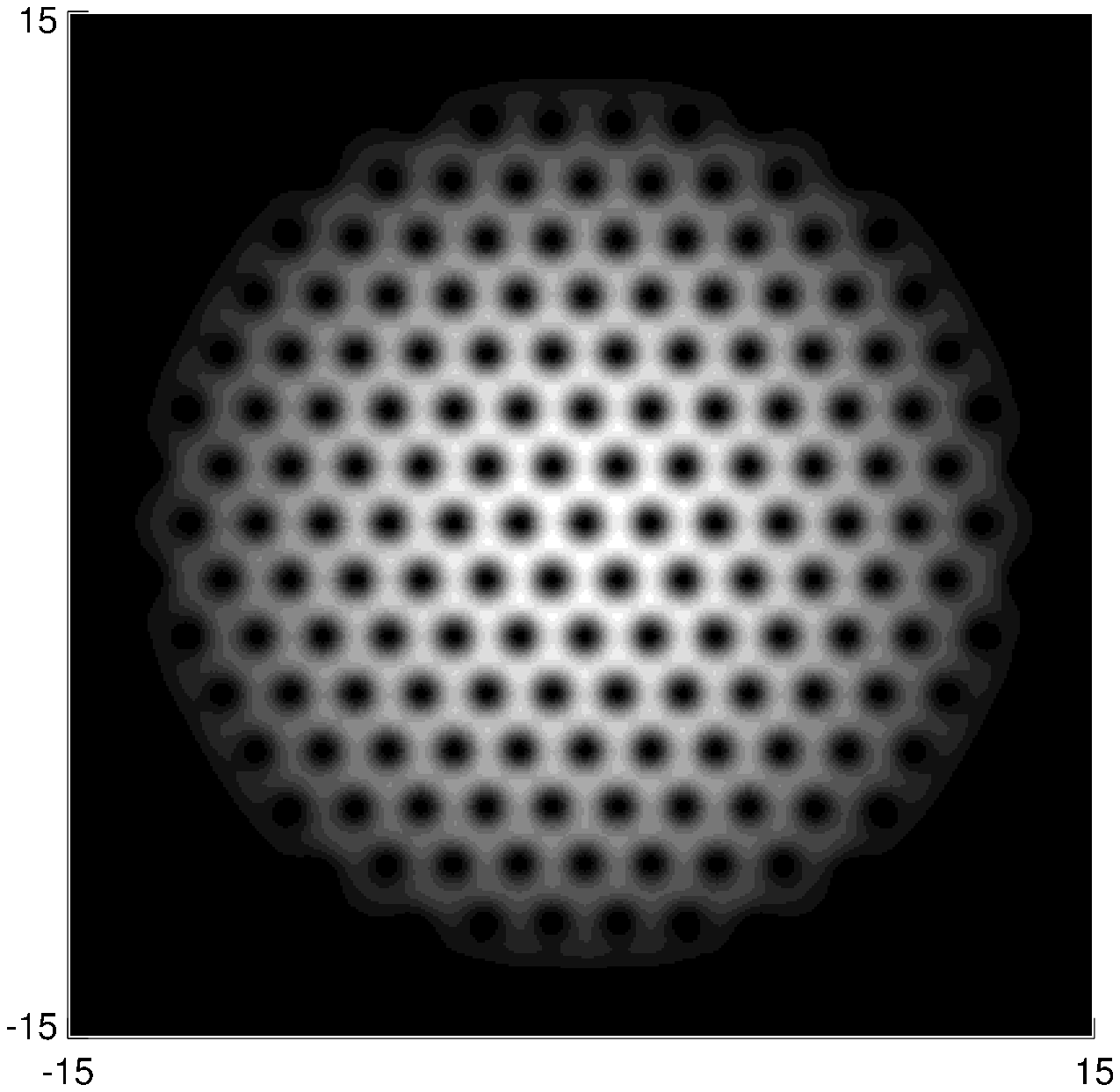}\includegraphics[width=7cm]{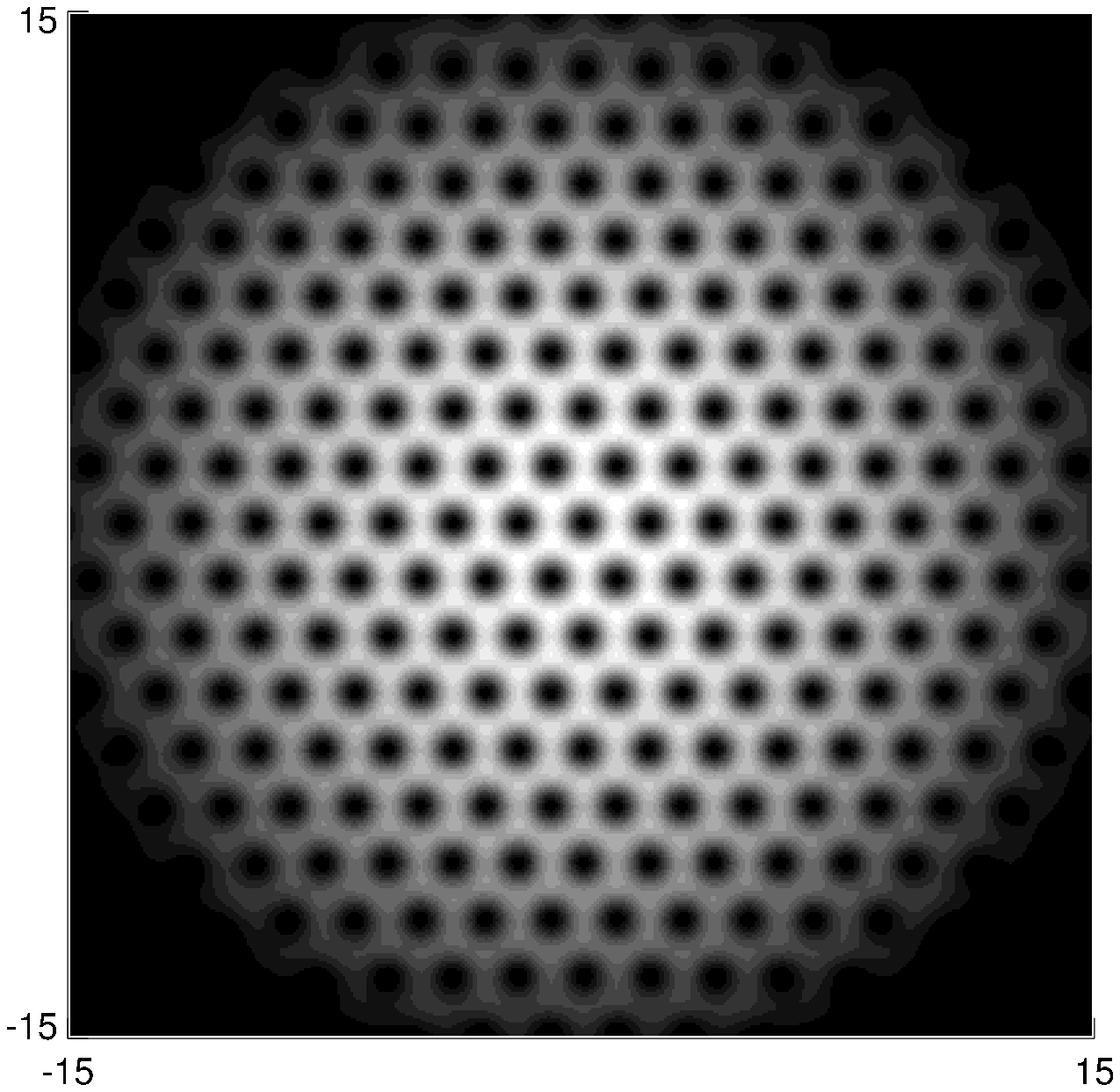}
\caption{Calculations of the particle density of an atomic BEC in the mean-field LLL regime, for two values of the angular momentum per
particle $L/N=60$ and 
$90$, from Ref.\protect\cite{CooperKR}.  The lattice structure is determined only by the  interactions,
which are here chosen to be contact interactions (\ref{eq:contact}).
Close to the centre of the trap, the  vortices form a regular triangular
lattice. 
[Reprinted figure from: N.R. Cooper, S. Komineas, and N. Read, 
 Phys. Rev. A {\bf 70}, 033604 (2004). Copyright (2004) by the American
 Physical Society. 
]
}
\label{fig:vl}
\end{figure}

The minimization for contact interactions has been implemented
numerically by several
authors\cite{ButtsR99,CooperKR,aftalion:023611}.  A simple result is
for the state at $L/N=1$, for which the lowest energy condensate wavefunction is
found to be
\begin{equation}
\psi(\zeta,z) = A \; \zeta 
\label{eq:leq1}
\end{equation}
in which all particles are condensed in the $m=1$ orbital. For other values of $L/N$ the
system spontaneously breaks rotational symmetry. For large $L/N$ the number of
vortices grows, as $N_{\rm v} \simeq 3 L/N$, and forms a triangular vortex
lattice  that is weakly distorted by the confinement,
see Fig.~\ref{fig:vl}.
Although small, the distortions of the triangular lattice are crucial
to obtain the correct density distribution in the trap. Fixing the
positions of the vortices on a triangular lattice leads to a density
distribution that, when averaged over a lengthscale larger than the
vortex spacing, is Gaussian\cite{Ho01}. The true averaged density
distribution is Thomas-Fermi like (i.e. an inverted
parabola)\cite{watanabe} as is found from numerical
studies\cite{CooperKR}. The density distribution in this regime is 
discussed further in \S\ref{sec:lda}.

The appearance of this triangular vortex lattice is very closely related to the
appearance of a triangular lattice of flux lines in type-II superconductors
close to the upper critical field. In that case, the order parameter is
determined by minimizing the Ginzburg-Landau energy functional (per unit length $L$ along the
field direction)
\begin{equation}
\frac{F - F_n}{L} \sim \int \left[
    \frac{1}{2M}|(-i\hbar\bbox{\nabla} + 2e\bbox{A})\psi_s|^2 + \alpha|\psi_s|^2 +
    \frac{\beta}{2}|\psi_s|^4\right] d^2\bbox{r}\,.
\end{equation}
Close to $H_{c2}$ the order parameter is small, so to a first approximation
one can neglect the quartic term and minimize the quadratic terms. These
describe electron pairs (hence charge $-2e$) in a uniform magnetic field: the
lowest-energy states are the (degenerate) lowest Landau level orbitals. As
explained by Abrikosov\cite{Abrikosov57} the groundstate is selected as the
linear combination of the lowest Landau level orbitals that minimizes the
quartic term. The solution is that the groundstate is a {\it triangular}
vortex lattice\cite{KleinerRA64}, and is characterized by the parameter 
\be \beta_A
\equiv \frac{\langle |\psi_s|^4\rangle_{\rm av}}{\langle |\psi_s|\rangle_{\rm
    av}^2} \simeq 1.1596 \,,
\label{eq:abb}
\ee
where $\langle\ldots \rangle_{\rm av}$ denotes the (spatial) average over the
unit cell of the lattice.
This is mathematically equivalent to the LLL mean-field theory 
for rotating bosons, at least in
the uniform case (i.e. close to the centre of a trap containing many
vortices).
The interaction energy  (\ref{eq:egp}) is
\be
E_I^{\rm GP} = \frac{1}{2} g_{\rm 2d} N^2 \int |\psi(x,y)|^4 d^2\bm{r} \,,
\ee
where $g_{\rm 2d} \equiv g/(\sqrt{2\pi} a_\parallel)$ is the 2D coupling
constant, and $\psi(x,y)$ is 2D component of the condensate wavefunction, which is restricted to
states in the LLL.
Using the above result (\ref{eq:abb}), one concludes that the groundstate is a
triangular lattice, and its energy (for a system of $N$ particles in a total
area $A$) is
\be
E_I^{\rm GP} = \frac{1}{2} g_{\rm 2d} \beta_A \frac{N^2}{A} = \beta_A V_0
\frac{N^2}{N_{\rm v}} \,.
\label{eq:egplattice}
\ee
In the last equality, we have replaced the area $A$ by the number of vortices $N_{\rm v} = A/(\pi a_\perp^2)$.

\subsubsection{Landau Level Mixing}

The limit of weak interactions (\ref{eq:weak}) where the LLL regime is
applicable, corresponds to the case of dense vortices, $\xi \gg a_\perp$. For
dilute vortices, $\xi \ll a_\perp$, interactions lead to contributions from
higher Landau levels to the condensate wavefunction.
We have explained above that for both $\xi \ll a_\perp$ (dilute vortices) and $\xi \gg a_\perp$
(dense vortices) the mean-field groundstate is a triangular vortex
lattice.  As $\xi/a_\perp$ varies between these two limits, there is a
smooth crossover\cite{fischerbaym,cozzini:023615}. This crossover has
been observed in experiments of Eric Cornell's group, which have
achieved a chemical potential $\mu \sim g\bar{n}$ that is less than the
effective cyclotron splitting $2\hbar\omega$.  In these experiments,
$\mu$ remains of order $1.5\hbar\omega_\parallel$, so it is approaching
the quasi 2D limit.
Experimental evidence for the crossover into the LLL regime has been
found in the apparent core size of the
vortices\cite{coddington:063607}, which shows the variation expected
within calculations in the LLL\cite{fischerbaym,cozzini:023615}.  The
frequencies of the collective modes were predicted to show a reduction
on entry into the LLL regime\cite{shm1,baym}.  The
experimental results appear to confirm the expected
variation\cite{schweikhard:210403}, but this conclusion has been queried in a subsequent
theoretical study that reassesses the expected shear
modulus\cite{sonin:021606} and indicates that in the experiments
of Ref.\cite{schweikhard:210403} the collective mode
frequencies have not reached the limiting values expected in the LLL regime.

Very different physics controls the energetics in the two limits: for strong
interactions $\xi\lesssim a_\perp$ the vortex lattice is determined by the
kinetic energy, for weak interactions in the LLL regime $\xi\gtrsim a_\perp$
the vortex lattice is determined entirely by the interactions. One should
therefore view the smooth crossover in the properties as somewhat of a
coincidence. As discussed in \S\ref{sec:dipolar}, changing the form of the
interaction potential can lead to large changes in the vortex lattice
structure in the LLL regime, $\xi\gtrsim a_\perp$.

\subsubsection{Anisotropic Traps}

In order to model the process of stirring the trapped atomic BEC, one
should include a potential that breaks the rotational symmetry of the
Hamiltonian
(\ref{eq:ham}). A simple form is to introduce an elliptical
perturbation of the transverse confinement, which is at rest in the
rotating frame of reference. This gives two (slightly) different
oscillator frequencies $\omega_x, \omega_y$ in place of the single
$\omega_\perp$.

For this model, there exists a generalization of the LLL description
for a weakly interacting gas\cite{fetter:013620}. The centrifugal
limit (\ref{eq:centrifugal}) is now set by the condition that $\Omega \leq
\mbox{min}(\omega_x,\omega_y)$. Close to this limit, the gas elongates
in one of the directions transverse to the rotation axis leading to a
quasi-1D
geometry\cite{linn,PhysRevA.69.023618,sinha:150401,sanchez-lotero:043613}
for rapid rotation.

\subsubsection{Dipolar Interactions}

\label{sec:dipolar}

\begin{figure}
\center\includegraphics[width=5cm]{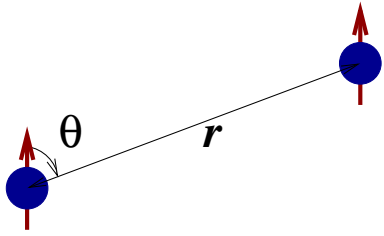}
\caption{Geometry of two atoms with aligned dipole moments, as discussed in
  the text.}
\label{fig:dipolar}
\end{figure}

Under certain circumstances, it has been proposed that the atoms (or
molecules) in a cold trapped gas may carry an electric or magnetic
dipole moment\cite{BaranovPS}.  One should then add to the usual contact interaction
the long-range dipole-dipole interaction.  For two dipolar atoms with
aligned dipole moments the effective interaction is
\begin{equation}
{V({\bm r}) = \frac{4\pi\hbar^2 a_{\rm s}}{M} \,\delta^3(\bm{r}) +
C_{\rm d}\frac{1-3\cos^2\theta}{r^3} }
\label{eq:dipolar}
\end{equation}
where $\theta$ is the angle between the dipole moment and the line separating
the atoms,
see Fig.~\ref{fig:dipolar}.

A BEC of atoms with significant dipolar interactions has been
realized by condensing chromium-${52}$\cite{GriesmaierWHSP05}, which is an atom with a
very large magnetic dipole moment, $\mu = 6 \mu_{\rm B}$, such that $C_{\rm d}
= \frac{\mu_0 \mu^2}{4\pi}$. The relative size of dipolar and contact
interactions is parameterized by the dimensionless ratio
\begin{equation}
\epsilon_{\rm dd}\equiv \frac{C_{\rm d} \,M}{3\hbar^2 a_{\rm s}} 
\end{equation}
which is $\epsilon_{\rm dd}\simeq 0.16$ for native conditions in $^{52}$Cr, and has
been further increased to $\epsilon_{\rm dd} \simeq 1$ by using a
Feshbach resonance to reduce $a_{\rm s}$\cite{lahaye2007}. Very recently, there
has been the exciting achievement of the condensation of hetero-nuclear
molecules in their ground ro-vibrational state\cite{ni-2008}. Such
systems can have large {\it electric} dipole moments, leading to
very strong electric dipole interactions\cite{BaranovPS}.

Consider a rapidly rotating atomic gas (in the 2D LLL limit), and for
simplicity let us choose the dipole moments to be directed parallel to the
rotation axis\cite{crs}. The mean-field theory requires one to minimize the
expectation value of the interaction energy
\begin{equation}
\frac{1}{2}\int\int |\psi({\bm r})|^2
  V({\bm r}-{\bm r}') |\psi({\bm r}')|^2\; d^3{\bm r}\; d^3{\bm r}'
\end{equation} 
for $\psi(\bm{r})$ in the lowest Landau level and for the interaction potential  (\ref{eq:dipolar}). Although this is a simple
generalization of the Abrikosov problem, it is one that does not naturally
appear in that context, where the microscopic physics determining the
superconductivity acts on scales much less than the
vortex lattice period. 

\begin{figure}
\center\includegraphics[width=14cm]{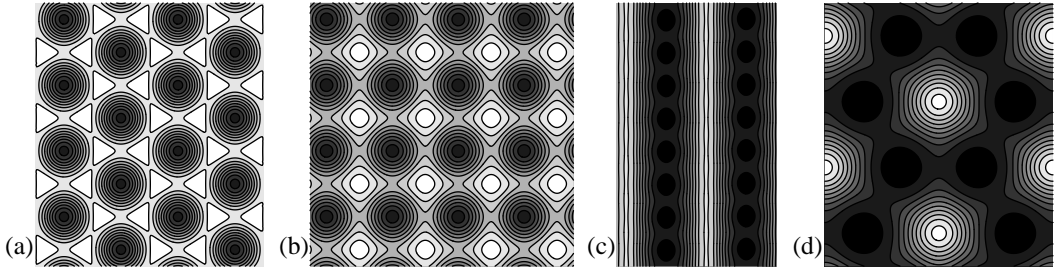}
\caption{Mean-field groundstates of a rotating BEC in the 2D LLL, for 
particles interacting by both contact and dipolar interactions. The relative
size of these is controlled by $\alpha = V_2/V_0$. The structure
of the vortex lattice varies from (a) triangular ($0\leq \alpha\leq 0.20$)
to (b) square ($0.20\leq \alpha\leq 0.24$) to (c)
``stripe crystal'', with  simple rectangular unit cell, ($0.24\leq \alpha\leq
0.60$) to ``bubble crystal'' phases ($\alpha \geq 0.60$) the simplest of which
is shown in (d). 
[Reprinted figure from: N.R. Cooper,  E.H. Rezayi, and S.H. Simon,
 Phys. Rev. Lett {\bf 95}, 200402 (2005). Copyright (2005) by the American
 Physical Society. 
]
}
\label{fig:dipolargs}
\end{figure}

The mean-field groundstates have been found  for the case of a
translationally invariant vortex lattice\cite{crs,zhang:200403}. The results\cite{crs} are shown in Fig.~\ref{fig:dipolargs}, as a
function of
\begin{equation}
\alpha \equiv \frac{V_2}{V_0}
\label{eq:gamma}
\end{equation}
where\cite{crs} 
\bea V_0 & = &
\sqrt{\frac{2}{\pi}}\frac{\hbar^2 a_{\rm s}}{Ma_\perp^2a_\parallel} +
\sqrt{\frac{2}{\pi}}\frac{C_{\rm d}}{a_\perp^2a_\parallel} -
\sqrt{\frac{\pi}{2}}\frac{C_{\rm d}}{a_\perp^3}\\ V_{m> 0} & = &
\sqrt{\frac{\pi}{2}} \frac{(2m-3)!!}{m!\,2^m}\frac{C_{\rm d}}{a_\perp^3}
\label{eq:haldanedip}
\eea 
are the Haldane pseudo-potentials for the dipolar interaction
(\ref{eq:dipolar}) in the limit $a_\parallel/a_\perp\ll 1$. (See
\S\ref{sec:haldane} for definitions of the Haldane
pseudo-potentials.)

The results show that the mean-field groundstate is very sensitive to
long-range repulsion, passing through a series of vortex lattice phases as
$\alpha$ increases.
The contact interaction aims to make $|\psi(\bm{r})|^2$ as uniform as possible,
while the long-range repulsion causes $|\psi(\bm{r})|^2$ to
cluster, leading to crystalline phases of clusters of particles at large
$\alpha$. These are referred to as ``bubble crystal'' phases\cite{crs}, in analogy with the
terminology used for structures of similar symmetry in 2D electron gases in
high Landau levels\cite{KoulakovFS96,MoessnerC96}.
\begin{figure}
\center\includegraphics[width=8cm]{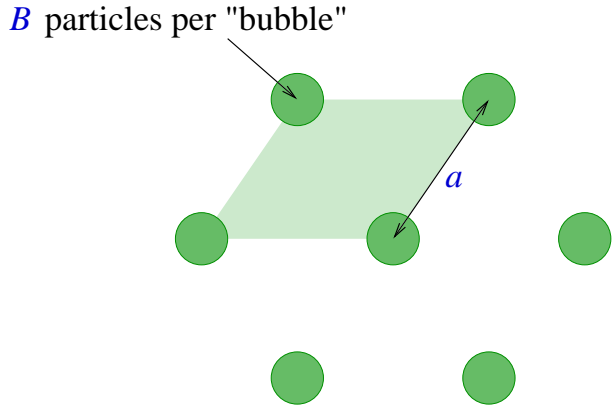}
\caption{Schematic diagram of the bubble crystal phase.}
\label{fig:bubble}
\end{figure}
There is a sequence of
bubble crystal phases
as $\alpha$ increases,
which are parameterized by the number of vortices per bubble, $q$.
For 2D particle density $n_{\rm 2d}$ the number of particles per bubble is
$B = n_{\rm 2d} \frac{\sqrt{3}}{2} a^2$,
where $a$ is the lattice constant of the bubble crystal,  assumed
triangular (see Fig.~\ref{fig:bubble}). For $B\gg 1$, the energy per particle is 
\begin{equation}
\frac{E}{N}   = 
  \frac{1}{2} V_0 B + \frac{1}{2} K \frac{C_{\rm d}}{a^3} B  \; = \;   
\frac{\sqrt{3}}{4} n_{\rm 2d}\left[V_0 {a^2} +  
  \frac{K C_{\rm d}}{ a}\right]
\end{equation}
where $K=11.034$ is the Madelung constant for dipolar interactions on
the triangular lattice\cite{madelunghere}. The energy is minimized for
$a_{\rm min} = \left(\frac{K C_{\rm d}}{2 V_0}\right)^{1/3}$, such
that the number of vortices per bubble is
\begin{equation}
q \equiv  n_{\rm v} \frac{\sqrt{3}}{2} a_{\rm min}^2 
= \frac{1}{\pi a_\perp^2} \frac{\sqrt{3}}{2} \left(\frac{K C_{\rm d}}{2 V_0}\right)^{2/3}
= \sqrt{3}\left(\frac{2K}{\pi^2}\right)^{2/3} \left(\frac{V_2}{V_0}\right)^{2/3} \,.
\end{equation}

Vortex lattices have not, as yet, been created in experiments on the dipolar
condensates. If these condensates can be made to enter the regime of rapid
rotation that has been achieved in rubidium\cite{schweikhard:210403} one
expects not just to see a crossover in the physical properties, but a true
phase transition into a vortex lattice of different symmetry that depends on
the ratio of short- to long-range interactions (\ref{eq:gamma}). The
dependence of the mean-field phases on the degree of Landau level mixing,
as quantified by $\mu/\hbar\omega_\perp$, has been studied in
Ref.\cite{komineas:023623}.

\subsection{Beyond Mean-Field Theory}

\label{sec:beyond}

I now turn to describe the properties of rapidly rotating atomic BECs
beyond the Gross-Pitaevskii mean-field approach. Much of our
understanding has been derived from numerical exact diagonalization
studies. These numerical techniques are described
in Appendix \ref{sec:numerical}. 

In the remainder of this section I describe some of the results for rotating
bosons in a harmonic trap (\ref{eq:ham}) in the weak interaction limit
(\ref{eq:weak}), and concentrate on the case of contact repulsive interactions
(\ref{eq:contact}).

\subsubsection{Low Angular Momentum $L\leq N$}

\label{sec:exactlowl}
For small values of the total angular momentum $L$ the exact
groundstates of a rapidly rotating Bose gas are known analytically.
The following states (\ref{eq:l0},\ref{eq:zcstate},\ref{eq:sympolys})
 have been shown to be exact eigenstates of the
contact repulsion by both analytical\cite{SmithW00} and
numerical\cite{BertschP99} studies, and to be the groundstates for a
class of repulsive interactions\cite{HusseinV02,VorovHI03} that
includes contact repulsion (\ref{eq:contact}).

\begin{itemize}

\item
For $L=0$ there is only one state in the 2D LLL: the pure condensate in the
$m=0$ orbital
\begin{eqnarray}
\Psi_{L=0}(\{\zeta_i\}) & \propto  & 1 \,.
\label{eq:l0}
\end{eqnarray}

\item
For $L=1$ the groundstate is the centre-of-mass excitation of the $L=0$ state 
\begin{eqnarray}
\label{eq:zcstate}
\Psi_{L=1}(\{\zeta_i\}) & \propto  & \zeta_c \\
\zeta_c & \equiv & \frac{1}{N} \sum_{i=1}^N \zeta_i 
\label{eq:zc}
\end{eqnarray}
for which the interaction energy is the same as for the $L=0$ groundstate,
$\epsilon_I(1,N) = \epsilon_I(0,N)$.

\item
For  $2\leq L\leq N$ the groundstates are the elementary symmetric
polynomials in the variables $(\zeta_i-\zeta_c)$
\begin{eqnarray}
\Psi_L(\{\zeta_i\}) 
& \propto & \sum_{p_1<p_2<p_L} (\zeta_{p_1} - \zeta_c)(\zeta_{p_2} - \zeta_c)\ldots
(\zeta_{p_L} - \zeta_c) \,.
\label{eq:sympolys}
\end{eqnarray}

\end{itemize}

For contact interactions, the energies of the states with $L\leq N$
($L\neq 1$) are\cite{SmithW00,BertschP99,kavoulenergy} \be E_I(L,N) = V_0 \epsilon_I(L,N) = \frac{1}{2} V_0 N(N-1-L/2) \,,
\label{eq:exacte}
\ee

A close connection of these exact states for finite $N$ to the
condensed wavefunctions of the Gross-Pitaevskii theory can be made
by considering the limit $N\to \infty$.
Consider first the case $L=N$, for which
\begin{equation}
\Psi_{L=N}(\{\zeta_i\})   \propto  \prod_i (\zeta_i - \zeta_c) \,.
\label{eq:leqn}
\end{equation}
If the co-ordinate $\zeta_c$ were simply a number this would be the
wavefunction of a pure condensate with a vortex at complex position $\zeta_c$,
Eqn.(\ref{eq:condensate}). The fact that $\zeta_c$ is the centre-of-mass
co-ordinate (\ref{eq:zc}), and therefore a function of the particle
co-ordinates, means that this state is not fully condensed.  Nevertheless,
the fluctuations of the centre-of-mass from its
average value $\langle \zeta_c \rangle =0$, computed for the state $\Psi_{L=N}$,  are
\begin{equation}
\langle \zeta_c^2 \rangle \sim 1/N \,.
\label{eq:zetafluct}
\end{equation}
Thus, in the limit $N\to\infty$ 
\begin{equation}
\Psi_{L=N}(\{\zeta_i\})  \stackrel{N\to\infty}{\longrightarrow} \prod_i \zeta_i 
\label{eq:lnlimit}
\end{equation}
which is the fully condensed state with a single vortex at the origin, (\ref{eq:leq1}).
This observation -- that one recovers a fully condensed state in the
limit $N \to \infty$, keeping $L/N = 1$ fixed and staying within the 2D LLL
($g \bar{n} \ll \hbar\omega_\perp, \hbar\omega_\parallel$) -- is a
specific example of the result that Gross-Pitaevskii theory is exact \cite{LiebS06} in the limit $N\to \infty$ with $g\bar{n}$
and $L/N$ finite.  One may generalize this to cases with $L\leq
N$. To this end, we note that in the large $N$ limit of
(\ref{eq:l0},\ref{eq:zcstate},\ref{eq:sympolys}), obtained by setting $\zeta_c\to
0$, the wavefunctions (\ref{eq:l0},\ref{eq:zcstate},\ref{eq:sympolys}) are the set of all $N+1$ symmetric polynomials of
$\{\zeta_i\}$. These polynomials can be generated by the condensate wavefunction
\begin{equation}
\Psi_{Z}(\{\zeta_i\})  =  \prod_i (\zeta_i - Z)
\label{eq:onevortex}
\end{equation}
where $Z$ is a complex co-ordinate which may be viewed as the location of a
single vortex. Thus, we conclude that, in the
2D LLL, the lowest energy condensate wavefunction for $L/N\leq 1$ is precisely 
described as a {\it single} vortex state. For $|Z|$ finite and non-zero,
the condensed state contains a single vortex located away from
the origin, so it breaks rotational symmetry.

The relation between the exact spectra and the condensed states has been
studied in detail for $L=2N$ in Ref.\cite{JacksonKMR01}.

\subsubsection{Spontaneous Symmetry Breaking}

\label{sec:symmetry}

It is a general feature of the Gross-Pitaevskii theory for a rotating gas in
an axisymmetric trap that the lowest-energy condensed state spontaneously
breaks the rotation symmetry of the Hamiltonian. For example, for (\ref{eq:ham}) the general
mean-field state is not rotationally invariant about the $z$
axis\cite{ButtsR99}, but shows small clusters of vortices with local
crystalline order.

Rotational symmetry breaking may be seen in the properties of the
exact energy spectrum in the limit $N\to \infty$ with $N_{\rm v}\sim L/N$ constant.
Specifically, in this limit, the exact energy per particle (\ref{eq:tote}) of the groundstate
tends to the form 
\be
\label{eq:epp}
\frac{E_\Omega(L,N)}{N}
\stackrel{N\to\infty}{\longrightarrow} \mbox{const.}+
V_0 N f(L/N) + \hbar(\omega_\perp -\Omega)(L/N) \ee for a certain subset of
angular momentum states $L$, discussed further below. In this equation, $V_0$
is the energy scale (\ref{eq:v0}) and $f(L/N) \equiv \mbox{lim}_{N\to \infty}
\epsilon_I(L,N)/N^2$ is a dimensionless function
which is set by the form of the interactions. In the thermodynamic limit, we
require that the energy per particle (and hence the chemical potential)
remains finite. For example, to remain within the 2D LLL (\ref{eq:weak})
requires the chemical potential to remain finite, $\mu \lesssim
\hbar\omega_\perp$, $\hbar\omega_\parallel$. Since $\mu \sim V_0N/N_{\rm v}$, this
requirement sets the condition that $V_0 N \to \mbox{constant}$. Under this
condition, the function $f$ has an overall scale of order 1.

At a general value of $\Omega$, the exact groundstate angular momentum $L^*$
is found by minimizing the energy per particle (\ref{eq:epp}) over the allowed
integer values of $L$. Provided that $f(L/N)$ is a (locally) smooth function such that the total energy has a simple minimum at $L^*$ -- which is found to be the generic case -- all
states which have $L$ that is within of order $\sqrt{N}$ of the optimal value
$L^*$ have an excitation energy per particle that is of order $V_0$, and so
vanishes in the appropriate thermodynamic limit, $N\to\infty$ with $V_0 N \to
\mbox{const.}$. That is, in the thermodynamic limit there is a degeneracy
between of order $\sqrt{N}$ states of angular momentum close the optimal value
$L^*$. The emergence of a degeneracy between a large number of states of
different angular momenta 
 is a signature of
spontaneous rotational symmetry breaking in the thermodynamic limit\cite{anderson}.

Evidence for the appearance of rotational symmetry breaking can be found in
the exact spectra at $L>N$, computed numerically for small numbers of
particles. 

\begin{figure}
\center\includegraphics[width=9cm]{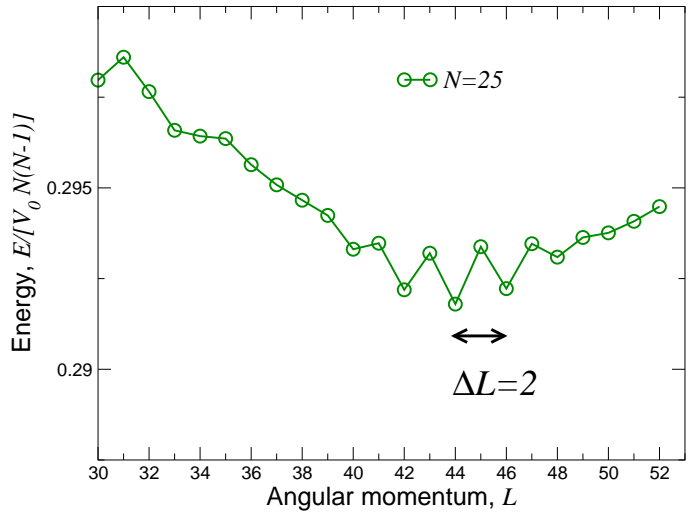}
\caption{Groundstate energy as a function of the angular momentum,
computed from exact diagonalization studies of rotating bosons in the
2D LLL with contact repulsion (\protect\ref{eq:contact}), at rotation
frequency $\Omega$ set to a value for which the mean-field groundstate
shows two vortices. The groundstate is at $L^*=44$, the spectrum shows
the appearance of low-lying excited states at angular momenta that differ from
this by multiples of 2.
In the thermodynamic limit, $N\to\infty$, the
 exact spectrum shows the appearance of
 quasi-degeneracies
 between states spaced in angular momentum by $\Delta L
=2$. This is a signature that in the thermodynamic limit the
groundstate breaks rotational symmetry partially, leaving a two-fold
symmetry axis\protect\cite{cmmp}.}
\label{fig:l2}
\end{figure}

Fig.~\ref{fig:l2} shows the exact groundstate energy as a function of
total angular momentum $L$ for $N=25$ particles, at a value of
$\Omega$ chosen such that the mean-field state is a pair of vortices,
with a two-fold rotational symmetry.
A notable feature is that the groundstate degeneracies occur between states
which differ in angular momentum by multiples of $\Delta L=2$. This feature is
an indication that in the thermodynamic limit the groundstate does not break
rotational symmetry completely, but retains a two-fold symmetry. This is
consistent with the mean-field state (see Ref.\cite{ButtsR99} at $L/N=1.75$).
For a state that retains a two-fold rotation symmetry, the values of $L$ for
which the limiting expression (\ref{eq:epp}) applies are those that are spaced
by $\Delta L=2$ from the true groundstate value $L^*$; the function $f(L/N)$
describes the smooth envelope of the dependence of the energies of the states at these angular
momenta.

\begin{figure}
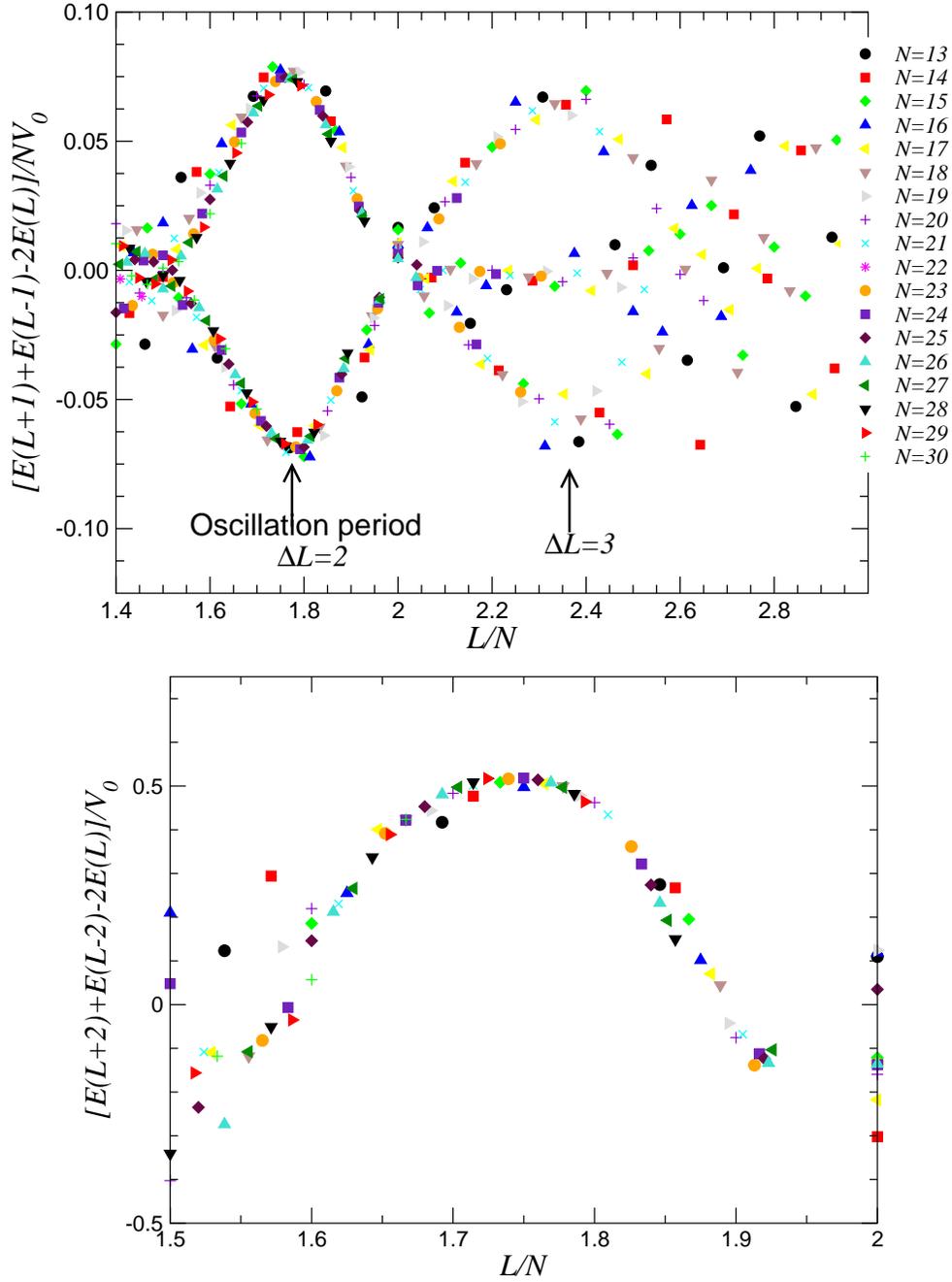

\center\includegraphics[width=13cm]{eaaf_scurve1_editmore.eps}
\center\includegraphics[width=11cm]{eaaf_curve2_edit.eps}
\caption{Scaling plots of the (numerical second derivatives of) the
groundstate energy as a function of the angular momentum, computed
from exact diagonalization studies of rotating bosons in the 2D LLL
with contact repulsion (\protect\ref{eq:contact}). The data collapse indicates
that the large $N$ asymptotic behaviour has been reached.  These plots
determine the scaling of the energies with $N$ 
discussed in the text, which 
lead to the prediction of spontaneous rotational symmetry breaking
in the thermodynamic limit $N\to \infty$,  $V_0 N\to$ constant.  There
are regions which show clear two-fold and three-fold periods,
corresponding to mean-field groundstates that break rotational
symmetry partially to retain two-fold and three-fold symmetry axes
respectively\protect\cite{cmmp}.}
\label{fig:scaling}
\end{figure}

The appearance of the quasi-degeneracies with increasing $N$ is
conveniently analysed by constructing scaling plots of $[E(L+1)+E(L-1)-2E(L)]$, as a function of the scaled angular
momentum, $L/N$, see Fig.~\ref{fig:scaling}. This plot shows regions
with quasi-degeneracies spaced by $\Delta L =2$ ($1.5\lesssim L/N \lesssim 2$)
and by $\Delta L =3$ ($2\lesssim L/N \lesssim 2.5$), which indicate rotational
symmetry breaking to states with residual 2-fold and 3-fold rotation axes.
Analysis of the second derivatives $[E(L+\Delta L)-E(L-\Delta L) -2 E(L)]$ for $\Delta
L=2,3$ in these regions (see Fig.~\ref{fig:scaling} for the case $\Delta L=2$)
allows the construction of scaling plots that confirm the above behaviour of
the function $f(L/N)$ at large $N$.

For $L/N < 1$,  rotational symmetry breaking can be deduced directly from
considerations of the exact analytic wavefunctions of \S\ref{sec:exactlowl}. For $L\leq N$, the
exact energy spectrum for contact interactions (\ref{eq:exacte}) leads to
$f(x) = -x/4$. Since the function $f(x)$ has vanishing second derivative, this
case is special from the point of view of rotational symmetry breaking. At one
special value of the rotation frequency, $1 -\Omega/\omega_\perp = V_0
N/(4\hbar\omega_\perp)$, {\it all} states with $L\leq N$ are degenerate in the
thermodynamic limit. There are therefore $\sim N$ degenerate states, rather
than $\sim \sqrt{N}$. This enhanced degeneracy reflects the fact that, at this
rotation frequency, the mean field state (\ref{eq:onevortex}) is itself degenerate with respect to
the position of the vortex, that is with respect to all possible angular
momenta $0\leq L/N \leq 1$.  

Evidence of rotational symmetry breaking can also be found in the exact
spectrum of the excitations above the lowest energy state at a given $L$. The
relationship of the exact spectrum to the expected properties of Bogoliubov
excitations of the mean-field groundstate has been discussed for $L\leq N$ in
Ref.\cite{ueda:043603}.

The condensate fractions and condensed wavefunctions for small numbers of
rotating bosons have been investigated in Ref.\cite{dagnino:013625}. These are
deduced from the exact groundstates, allowing for rotational symmetry
breaking. The positions of the vortices in the condensate wavefunction
(\ref{eq:condensate}) thus obtained show patterns similar to those from
mean-field theory\cite{ButtsR99}. The condensate fraction is large for a
single vortex ($L=N$) for $N\gtrsim 9$. For the system size studied ($N=6$)
the condensate fraction falls quickly to small values as the angular momentum
is increased beyond the case of a single vortex ($L>N$)\cite{dagnino:013625}.

{\it Anisotropic Traps}

For a rotating trap with a quadrupolar
deformation\cite{linn,PhysRevA.69.023618,sinha:150401,sanchez-lotero:043613,fetter:013620},
which can represent the stirred atomic BEC, angular momentum is not conserved.
However, the wavefunction {\it parity} is preserved: the deformation does not
couple states with odd and even $L$. Close to the rotation frequency at which
the first vortex enters the system, the exact spectra show close degeneracy
of levels of opposite parity for large numbers of
particles\cite{parke:110401}. This quasi-degeneracy of the levels of the exact
spectrum is an indication that the mean-field state spontaneously breaks the
reflection symmetry of the trap: for $N\to \infty$ there are two degenerate
condensed states in which a vortex sits either on one side or on the other
side of the trap (in the reflection symmetry related position). The lifting of
the degeneracy of these two states, at finite $N$, has been explained in terms
of the rate of tunnelling of the vortex between these two locations
\cite{parke:110401}.

{\it Restoration of Rotational Symmetry}

Noticing that the GP theory breaks rotational symmetry, an improved mean-field
theory for rotating Bose gases has been proposed and analysed in
Ref.\cite{romanovsky:011606}. In this theory the GP wavefunction is projected
onto an eigenstate of angular momentum. This restores the property that the
exact groundstate for a finite $N$ must have definite angular momentum. The
resulting ``rotating vortex cluster'' state has a lower energy than the GP
state, which may be important in situations when $N$ is sufficiently small.

\subsubsection{Signatures of Strong Correlations}

\label{sec:cfdots}

From the first studies of rapidly rotating atomic Bose gases\cite{wgs} it has
been recognized that the groundstates include strongly correlated phases.
In particular, the exact
groundstate of $N$ bosons in the 2D LLL at high angular momentum, $L=N(N-1)$\cite{wgs}, is
the bosonic Laughlin state\cite{laughlinstates}. This state
(\ref{eq:laughlin}) and its properties are described in detail in
\S\ref{sec:stronglycorrelated}. For now we simply note that this is a
strongly correlated liquid phase of the bosons. It has a vanishing condensate
fraction, and does not have any long-range crystalline order (in either the
particle or vortex density).

Evidence for other strongly correlated states was found in exact
diagonalization studies of small numbers of particles\cite{WilkinG00}. As the
rotation frequency is swept from $\Omega=0$ to the centrifugal limit
$\Omega=\omega_\perp$, the groundstate undergoes a series of transitions,
between states with different values of angular momentum, $L$. The angular
momentum $L$ does not increase uniformly, but undergoes steps between certain
``magic angular momenta'' in a sequence that depends on the total number
of particles $N$. The states at these magic angular momenta were shown to be
strongly correlated, in the sense of small condensate fraction. Furthermore,
these states were shown to include the Moore-Read, or Pfaffian, state at
$L=N(N-2)/2$ (see \S\ref{sec:stronglycorrelated}), and a set of other
uncondensed states could be accounted for in terms of condensates of
``composite bosons'' rather than of the underlying bosons\cite{WilkinG00}.

\begin{figure}
\center\includegraphics[width=12cm]{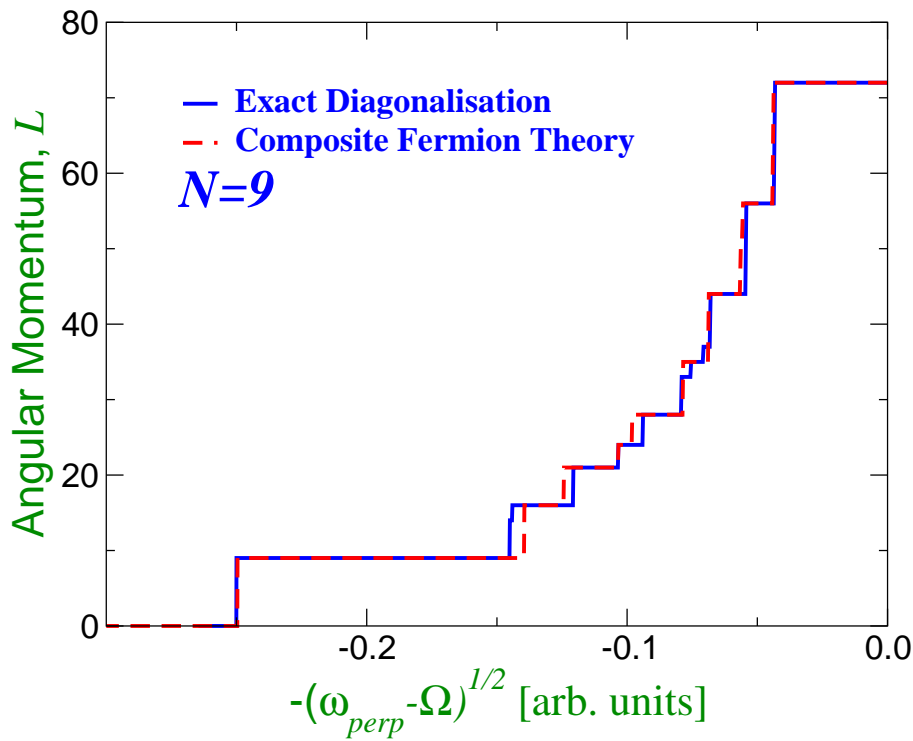}
\caption{Groundstate angular momentum as a function of rotation rate
$\Omega$ for $N=9$ bosons in the 2D LLL with repulsive contact
interactions.  The solid line shows the results of exact
diagonalization studies, with plateaus at the ``magic angular
momenta''.  The dashed line is the parameter-free prediction from the
CF theory\protect\cite{CooperW99}. This is obtained by finding the
groundstate angular momentum using just the set of ``compact states''
of the CFs. These compact states account for all of the magic angular
momenta that appear in the exact spectrum; discrepancies arise only in the
positions and widths of the plateaus, which depend very sensitively on
small energy differences.}
\label{fig:cftheory}
\end{figure}

An alternative description of the rotating groundstates was introduced in
Ref.\cite{CooperW99}. This work proposed a generalization of the
``composite fermion'' construction used in the
FQHE\cite{jainoriginal,dassarmapinczuk,ole} to the case of rotating bosons. See
\S\ref{sec:stronglycorrelated} for a definition and discussion of these
states. In this theory, one parameterizes the $N$-boson wavefunction in terms
of a Slater determinant of $N$ single-particle wavefunctions, which represents
a set of $N$ non-interacting composite fermions (\ref{eq:cf}). For a composite
fermion state with total angular momentum $L_{CF}$ the resulting boson
wavefunction has angular momentum\cite{CooperW99}  $L = L_{CF} + 
N(N-1)/2$.  Treating the composite fermions simply as non-interacting
particles, with a Landau level spectrum, leads to the prediction of a set of
special values of the angular momentum $L_{CF}^*$ at which one expects the
non-interacting composite fermions to have stable groundstates. These
so-called ``compact states''\cite{jainkawamura,CooperW99} identify a set of
special values of the angular momentum for the bosons $L^* = L_{CF}^* +
N(N-1)/2$ and provide a parameter-free trial many-boson wavefunction at each
angular momentum. The theory is very successful at describing the exact
groundstates of rotating bosons for the systems sizes that were studied (up to
$N=10$): the values $L^*$ deduced from the CF theory account for almost all of
the magic angular found in the exact studies; the overlaps of the trial CF
wavefunctions with the exact groundstates are very large\cite{CooperW99}. An
example of the accuracy of the CF theory is shown in Fig.~\ref{fig:cftheory}.

Ref.\cite{barberancrystals} has investigated the pair correlations of
the exact groundstates for small numbers of bosons. The
results show evidence for local crystalline order. These
crystalline correlations appear even for the Laughlin state for the
system sizes studied ($N\leq 8$).  Since the Laughlin state is the
exact groundstate for all $N$, and it is not crystalline in the
thermodynamic limit, these crystalline correlations are
interpreted as a feature of the states of small numbers of
particles\cite{barberancrystals}.

The exact groundstates of small numbers of particles have been related
to ``rotating boson molecule'' states\cite{PhysRevLett.93.230405}, in
studies of up to $N=11$ particles\cite{baksmaty:023620}. This
description is based on the idea of an underlying crystalline order,
combined with the projection of these cluster states onto states of
definite angular momentum, leading to angular averaging of the
particle density. This interpretation provides an alternative
viewpoint on the magic angular momenta. It can account for the values
of the magic angular momenta in terms of the shell structure of
ordered crystalline arrangements of the bosons, most clearly for $N=6$
and $N=9$.  It would be of interest to see the results of calculations
of the overlaps of the variational rotating boson molecule
wavefunctions with the exact groundstates.

For small numbers of particles it is perhaps moot to discuss which of a set of
approximate theories should be used to describe the system.\footnote{When the
  exact groundstate is known (as for the Laughlin state at $L=N(N-1)$),
  discussions reduce to considerations of the properties of this state.}  Descriptions that
are apparently very different (liquids of composite fermions, rotating vortex
clusters or rotating boson molecules) could, in principle, all capture aspects
of the groundstate and each could give useful physical insight.
The relative merits can only be judged in their quantitative successes. 
However, qualitative distinctions  between different
descriptions can arise in the thermodynamic limit $N\to \infty$.

\subsection{Strongly Correlated Phases}

\label{sec:stronglycorrelated}

\subsubsection{The Filling Factor}

In order to make sharp statements about {\it phases of matter}, one
must consider the thermodynamic limit, $N\to \infty$.  There are
different ways in which to take this limit for a rotating gas of
bosons, depending on how one chooses to scale the angular momentum $L$ with 
the number of particles $N$.

The simplest situation to consider is $N\to \infty$ keeping the number of
vortices fixed. Since $N_{\rm v}\sim L/N$ this requires $L\to \infty$ such
that $L/N \to\mbox{const}$.  The particle density then grows as $\bar{n}
\propto N$, so for the mean-field interaction $g \bar{n}$ to remain finite
(for example, to stay in the 2D LLL regime), one should take $g N \to
\mbox{const}$. With this set of scalings, the Gross-Pitaevskii theory has been shown to be
  exact\cite{LiebS06}: the groundstate is a pure condensate in this
  thermodynamic limit. We have shown
explicit examples of how this limit emerges from the exact spectra
at finite $N$ in \S\ref{sec:symmetry}.

Although this limit is appropriate for the description of atomic BECs in
many experimental situations, from the perspective of many-body physics it
is not the most interesting limit, leading as it does to states that are fully
condensed. A much more interesting limit can be found by exploiting analogies
with the FQHE. This regime may be of importance in future experiments at high
vortex density.

The FQHE refers to a set of strongly correlated quantum phases
of electrons in the 2D LLL states\cite{prangeandgirvin,dassarmapinczuk}. These states are characterized by the
electron {\it filling factor}
\begin{equation}
\nu_{\rm e} \equiv n_{\rm e}\frac{h}{eB}\,,
\end{equation}
where $n_{\rm e}$ is the 2D number density of electrons.
From the above mapping of the rotating atomic gas (\ref{eq:qB}), 
the analogous quantity is\cite{cwg}
\begin{eqnarray}
\nu \equiv n_{\rm 2d}\,\frac{h}{q^*\!B^*}  & = &  n_{\rm 2d}\,\frac{h}{2M \Omega}
\label{eq:fillingfactor}
\end{eqnarray}
where the 2D density $n_{\rm 2d}$ is related 
to the 3D density by the integral along the rotation axis, $n_{\rm 2d} \equiv
\int n({\bm r}) dz$.
From (\ref{eq:feynman}), the filling factor can be written in terms of the
vortex density as
\begin{equation}
\nu=\frac{n_{\rm 2d}}{n_{\rm v}} \,.
\label{eq:nulocal}
\end{equation}
In an inhomogeneous trap, Eqn.(\ref{eq:nulocal}) should be interpreted as a local relation,
defined on lengthscales larger than the mean vortex spacing.  Considering the
uniform limit, in which 
the particles are uniformly distributed (on average) over an area containing
$N_{\rm v}$ vortices, the relation takes the simple form
\begin{equation}
\nu=\frac{N}{N_{\rm v}} \,.
\label{eq:fillingfactorN}
\end{equation}
Note that, for system with a large number of vortices $N_{\rm v}$, which shall be the
focus of the studies described below, the rotation frequency is $\Omega\simeq
\omega_\perp$ (\ref{eq:close}), and so $n_{\rm v} \simeq 1/(\pi a_\perp^2)$
in (\ref{eq:nulocal}).

It is interesting to study the nature of the groundstate of a rotating atomic
gas for $N\to \infty$, $N_{\rm v}\to\infty$ such that $\nu\to
\mbox{const}$. A sharp question can be formulated in this thermodynamic limit:
what is the phase diagram of a rapidly rotating atomic gas as a function of
the filling factor $\nu$? This is the question that was first raised in
Ref.\cite{cwg}, and that we shall address in the remainder of the section.

Note that for {\it bosons} in the 2D LLL, the filling factor can take any
value, including values larger than one. (This contrasts with the case of
fermions where the Pauli exclusion principle limits the occupation.) The only
limitation to the theory as presented is that, for the single particle states
to remain in the 2D LLL we require that interactions (and hence chemical
potential) remain sufficiently small (\ref{eq:weak}). Since $\mu \sim
\nu V_0$, if
 $V_0$ is
small compared to the trap level spacings ($\hbar\omega_\perp,
\hbar\omega_\parallel$) the filling factor can be very large and the system
still be
in the 2D LLL. This is the situation in the experiments of
Ref.\cite{SchweikhardCEMC92}, where the LLL condition is achieved $\mu
\lesssim 2\hbar\omega_\perp$ with a large filling factor $\nu\simeq 500$.

It has been argued in Refs.\cite{cwg,shm1,baym} that for filling factor above
a critical value $\nu_{\rm c}$ the groundstate is  a triangular vortex
lattice, so GP mean-field theory is at least qualitatively accurate. Indeed,
as described above, the GP theory is exact in the limit $\nu\to \infty$ (this
follows from the thermodynamic limit in which $N\to \infty$ with $N_{\rm v}$
finite). However, for small values of $\nu$, corrections to mean-field theory
are large. These corrections cause mean-field theory to fail qualitatively for
$\nu <\nu_{\rm c}$, and the groundstate is replaced by strongly correlated
phases very different from the vortex lattice\cite{cwg}. A very instructive
way in which to understand the importance of the filling factor on the
mean-field groundstate is to evaluate the {\it quantum fluctuations} of the
vortices.

{\it Quantum Fluctuations of Vortices}

Consider the dynamics of a single vortex line in a 2D fluid (i.e. a
straight vortex line).  
We shall describe the quantum dynamics of the vortex line using a canonical quantization approach, but
the results agree with those found in other ways\cite{Fetter67,HaldaneWu}.
The classical dynamics of a 2D vortex, at a position $X$ and $Y$ in an
external potential $V(X,Y)$, follows from the standard Magnus force dynamics
of a vortex line in a classical fluid \begin{eqnarray}
    -\rho_s \kappa \dot{Y}  +F_{X}^{\rm ext} & = & 0\\
    +\rho_s \kappa \dot{X} +F_{Y}^{\rm ext} & = & 0
\end{eqnarray}
where $\rho_s$ is the mass density (per unit area) of the fluid
and $\kappa$ the circulation of the vortex.
The only amendment for a quantized vortex in a superfluid is that
the circulation is quantized, $\kappa= h/M$, so we may write
\begin{equation}
\rho_s\kappa = (n_{\rm 2d} M) \frac{h}{M} = n_{\rm 2d}h \,.
\end{equation}
A Lagrangian that reproduces this classical dynamics is\footnote{This is easily
   checked by constructing the Euler-Lagrange equations for (\ref{eq:lagrange}).}
\be
L   =  n_{\rm 2d} h \;\dot{X}Y  - V(X,Y)  \hskip2cm 
  \left[\vec{F}^{\rm ext} = -\vec{\nabla}V\right] \,.
\label{eq:lagrange}
\ee
Constructing the momentum conjugate to the particle
co-ordinate $X$ and applying canonical quantization, leads to
\begin{eqnarray} 
\Pi_X  & \equiv & \frac{\partial L}{\partial \dot{X}} = n_{\rm 2d}h\; Y \\
 {[\hat{X},\hat{\Pi}_X]} = i\hbar & \Rightarrow &  [\hat{X},\hat{Y}]  =
 \frac{i}{2\pi n_{\rm 2d}} \,.
\end{eqnarray}
The $X$ and $Y$ co-ordinates are conjugate,\footnote{These are the
  guiding-centre co-ordinates of a particle in a single Landau level.} so
obey the generalized uncertainty relation
\begin{equation}
\Delta X \Delta Y  \geq \frac{1}{4\pi n_{\rm 2d}}
\end{equation}
which implies
\begin{equation}
\Delta X^2 + \Delta Y ^2  \geq       \frac{1}{2\pi n_{\rm 2d}} \,.
\label{eq:fluct}
\end{equation}

The result (\ref{eq:fluct}) makes physical sense: one cannot locate
the vortex line to a distance less than the mean 2D separation between
the particles. It is interesting to note that this result has an
entirely classically interpretation, but emerges from a quantum
calculation due to the cancellation of Planck's constant in the
circulation with Planck's constant in the commutator.  Furthermore,
the result (\ref{eq:fluct}) is consistent with the calculation of
fluctuations of the vortex (\ref{eq:zetafluct}) for the exact
one-vortex wavefunction (\ref{eq:leqn}). Eqn. (\ref{eq:zetafluct})
implies $\Delta X^2 + \Delta Y^2 \sim \ell^2/N \sim 1/n_{\rm 2d}$,
where we take the typical particle density $n_{\rm 2d} \sim N/\ell^2$
noting that the $N$ particles are within an area of $\sim \ell^2$ in
this inhomogeneous state (\ref{eq:leqn}).

The importance of the filling factor for the properties of a large vortex
lattice becomes clear if one applies a form of Lindemann criterion and asserts that the
vortex lattice will become unstable if the (rms) quantum
fluctuations in vortex position are larger than some multiple $\alpha_L$ of the
vortex spacing
\begin{eqnarray}
\sqrt{\Delta X^2 + \Delta Y ^2} =
    \frac{1}{\sqrt{2\pi n_{\rm 2d}}} & \geq & \alpha_L \times  a_{\rm v} =
    \alpha_L \sqrt{\frac{2}{\sqrt{3}n_{\rm v}}} \\
\nu \equiv \frac{n_{\rm 2d}}{n_{\rm v}} \geq \nu_{\rm c} =
\frac{\sqrt{3}}{4\pi \alpha_L^2}
 \,.
\end{eqnarray}
Putting in a typical value for the Lindemann parameter,
$\alpha_L^2\simeq {0.02}$\cite{RozhkovS96},  one
finds $\nu_c \simeq 7$.  

The theory described above for a single vortex neglects the kinetic energy of
the vortex. It amounts to the evaluation of the quantum fluctuations of only
the guiding centre co-ordinate.  Including the inertia of the
vortex leads, in addition, to a zero-point cyclotron motion of the
vortex position. This zero point cyclotron motion gives an equal contribution to the
quantum fluctuations of position as that of the guiding centre.  Combining
these fluctuations, one finds \be \Delta X^2+\Delta Y^2 \geq \frac{1}{\pi
  n_{\rm 2d}} 
\label{eq:fluct2}
\ee that is, {\it twice} the value (\ref{eq:fluct}) from the
guiding centre fluctuations alone. Using this expression (\ref{eq:fluct2})
 within the Lindemann analysis leads to a critical filling
 factor
that
is twice as large,
$\nu_{\rm c}= \sqrt{3}/(2\pi \alpha_L^2)$,
which evaluates to $\nu_{\rm c}\simeq 14$ using the same value
for the Lindemann parameter.
This is the estimate that was given
in Refs.\protect\cite{RozhkovS96,cwg}.

A more controlled calculation of the quantum fluctuations of a vortex lattice
involves the consideration of the collective modes of the entire array, rather
than just the motion of a single vortex. This calculation has been performed
for a uniform vortex lattice array, by determining the low frequency
``Tkachenko'' modes\cite{tkachenko1,tkachenko2,tkachenko3} and evaluating
their contributions to the quantum fluctuations of an individual
vortex\cite{shm1}. Applying the same Lindemann criterion as above, leads to a
critical filling factor of $\nu_{\rm c} \sim 8$. A more complete calculation that includes,
in addition to the Tkachenko modes, the gapped ``inertial modes'' of the
vortex lattice\cite{baym} leads to a critical filling factor of $\nu_{\rm
  c}\sim 17$.  The difference between these two results, of roughly a factor
of two, is again due to the neglect or inclusion of the inertial
contributions to the fluctuations of the vortices\cite{baym}.  That the
results of calculations based on the collective
modes are so close to those found from single-vortex calculations is an
indication that the fluctuations are dominated by the short-wavelength
modes.

\subsubsection{Numerical Evidence for Crystalline Phases}

\label{subsec:crystals}

While instructive, considerations based on the Lindemann
criterion are hardly predictive, depending very sensitively on
$\alpha_L$ which, itself, is estimated\cite{RozhkovS96} from the
thermal melting of 3D crystals!  For this reason it is useful to have
a direct determination of the transition. 

\begin{figure}
\center\includegraphics[width=12cm]{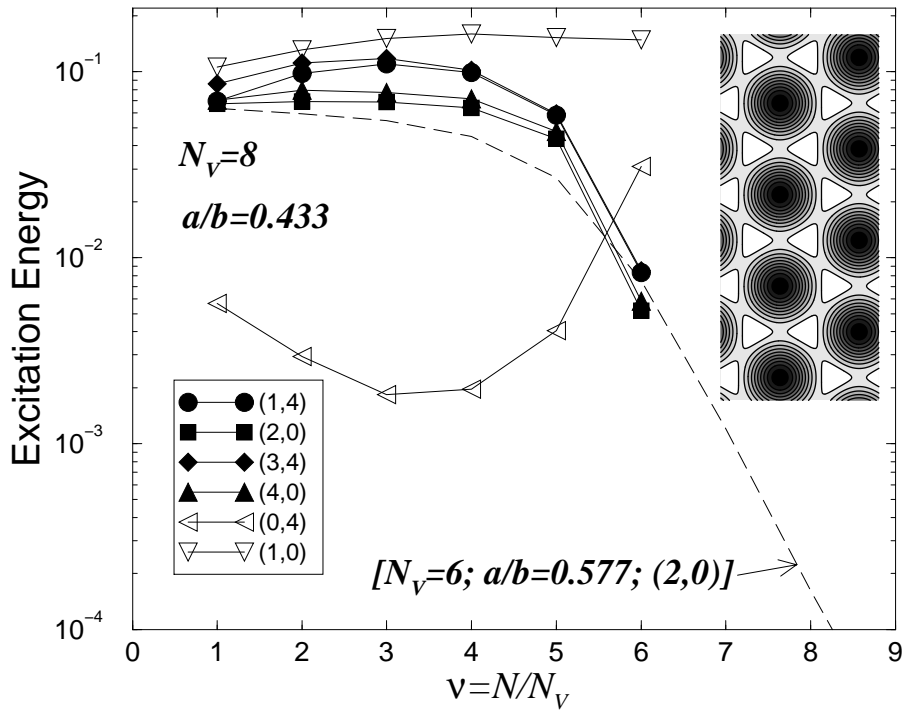}
\caption{Numerical evidence for the development of short-range crystalline
  order 
at filling factor $\nu=N/N_{\rm
  v}\gtrsim 6$\protect\cite{cwg},
for bosons in the 2D LLL  on the torus geometry with aspect ratio chosen to
 be commensurate with the triangular vortex lattice.
The graph shows the excitation energies 
at a set of different points $(K_x,K_y)$ in the Brillouin zone
 as a function filling factor $\nu$. For $\nu\gtrsim 6$ the excitations
at reciprocal lattice vectors of the triangular vortex lattice (filled symbols) become very
small.
This is a signal, in the finite-size system, of a tendency to translational
symmetry breaking. (For $\nu\lesssim 6$ there is a low energy excitation which
is located at a position in the Brillouin zone that is consistent both with
the Read-Rezayi states and with the  ``smectic'' phase.)
[Reprinted figure from: N.R. Cooper,  N.K. Wilkin, and J.M.F. Gunn, 
 Phys. Rev. Lett {\bf 87}, 120405 (2001). Copyright (2001) by the American
 Physical Society. 
]
}
\label{fig:translationsymmetry}
\end{figure}

The transition to the vortex lattice phase at large $\nu$ has been studied in
large scale exact diagonalization studies\cite{cwg,CooperR07}.  The strategy
is to work on a geometry with periodic boundary conditions (the torus geometry, \S\ref{sec:torus}),
which is consistent with the formation of a vortex lattice.  The signal of
crystallization is the collapse to very low energies (above the groundstate
energy) of a set of excitations at momenta that are reciprocal lattice vectors
of the vortex lattice. (This is directly analogous to the signal of
rotational symmetry breaking in a trap, leading to quasi-degeneracies of
energy levels at angular momenta spaced by $\Delta L$, which was discussed in
\S\ref{sec:symmetry}.)
By looking for the emergence of broken translational
symmetry, it was estimated that the transition to the triangular vortex lattice occurs at
{$\nu_{\rm c}\simeq 6$}\cite{cwg}. Numerical evidence for the appearance of
a triangular vortex lattice is shown in Fig.~\ref{fig:translationsymmetry}.
At $\nu=6$ the largest system that could be studied by exact diagonalization
was $N= 48, N_{\rm v}=8$. Since the lattice contains only $8$ vortices, there
are surely significant finite size effects in the  estimate of the location of the phase transition.

\begin{figure}
\includegraphics[width=6cm]{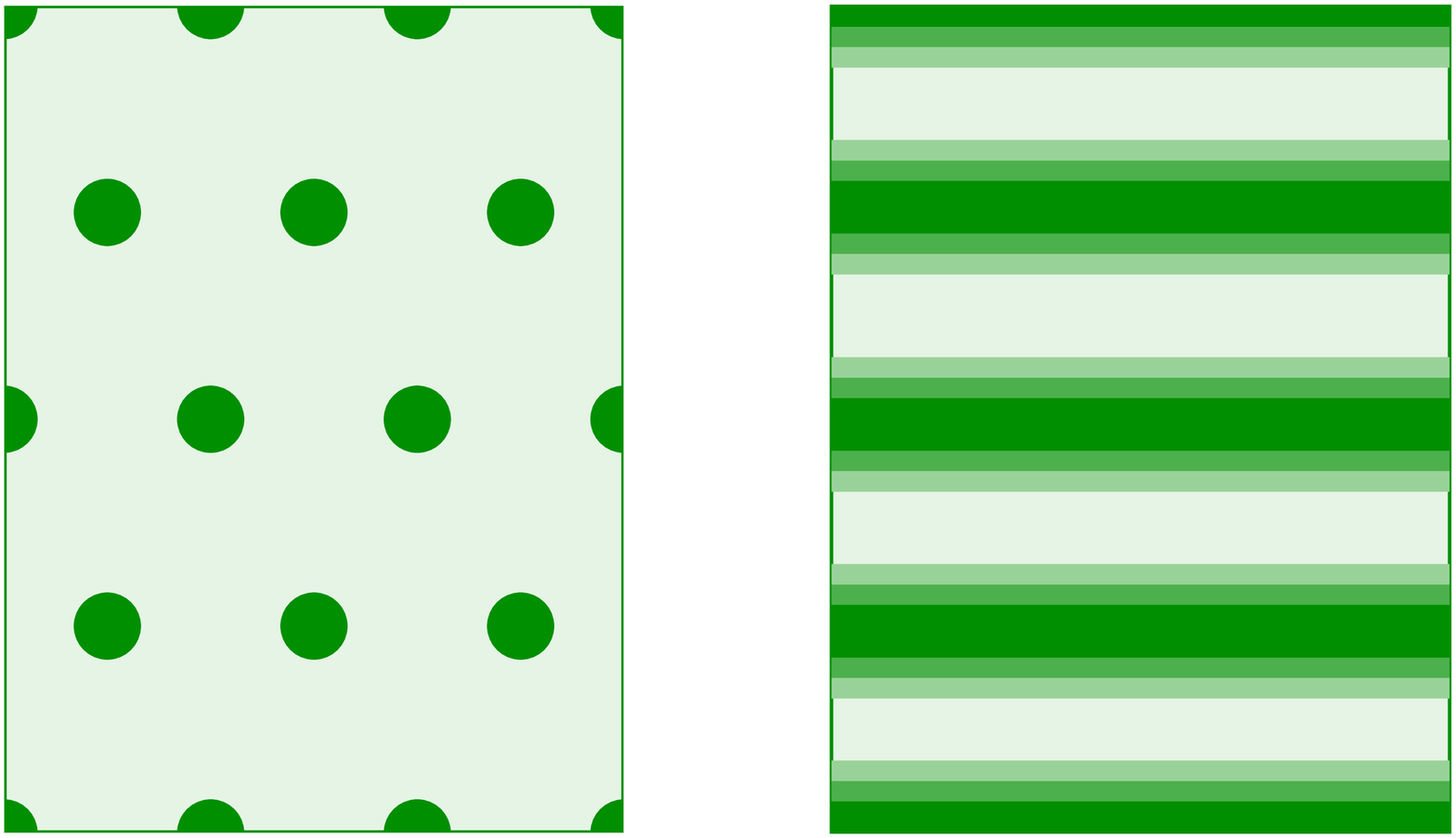} 
\vskip-3.2cm\hfill
\includegraphics[width=6cm]{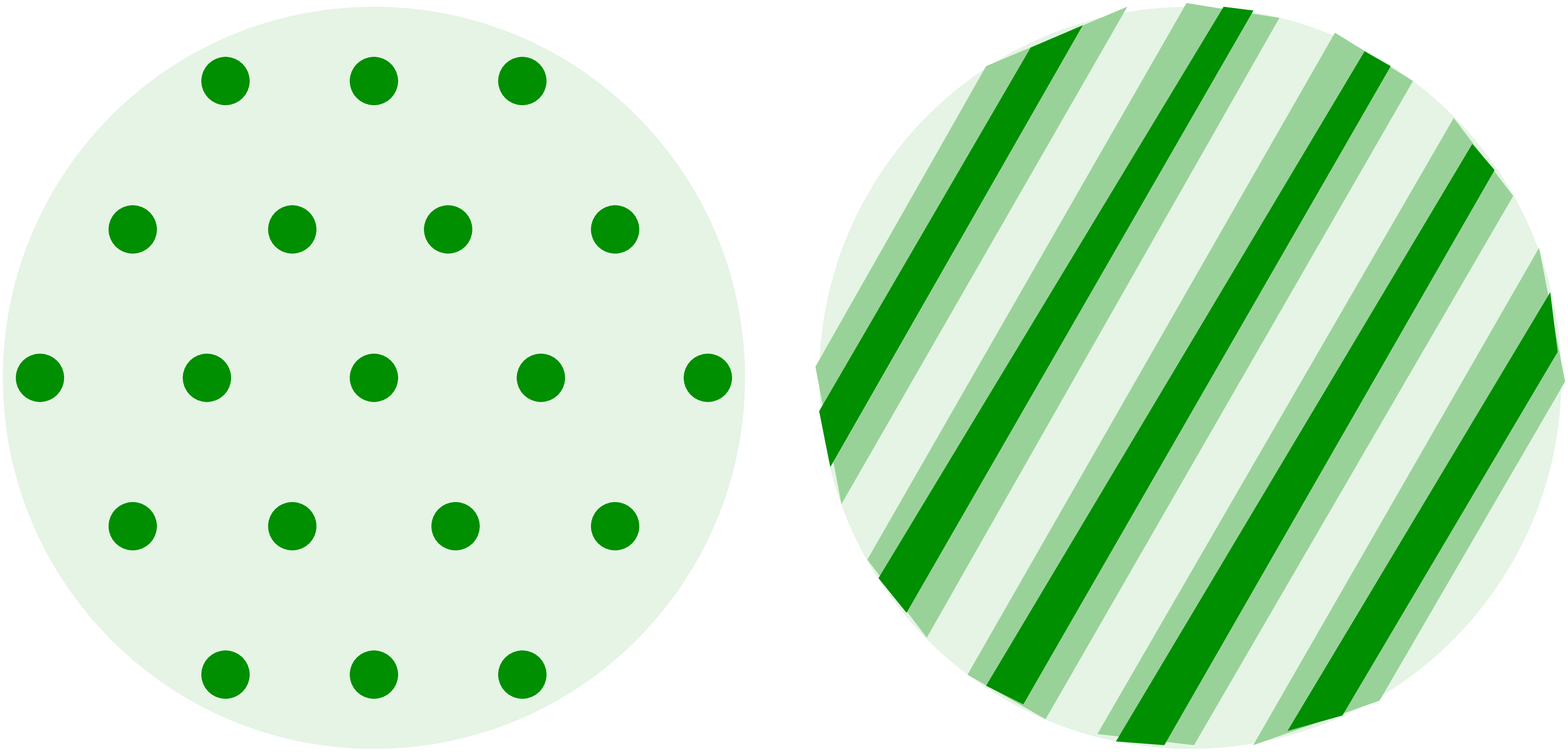}
\caption{Schematic diagrams of the ``smectic'' phase that appears in 
exact diagonalization studies on the torus (left), and its possible 
effect in disordering 
the vortex lattice in a trapped gas (right). The vortex lattice is
quantum-disordered along lines of the vortices, giving a phase with broken
translational symmetry in just one direction\protect\cite{CooperR07}.
}
\label{fig:smectic}
\end{figure}

More recent work\cite{CooperR07} focusing on $\nu=2$ has shed more light on
this issue. In the regime $\nu=2\to 6$ the
numerical results  were shown to be consistent with the existence of a
competing crystalline phase with ``smectic'' order -- that is, a stripe state
with broken translational order in only one direction. This can be viewed as a
phase in which the vortex lattice is quantum melted due to fluctuations of
vortices along lines, leading to averaged density patterns of the form shown
in Fig.~\ref{fig:smectic}. 
While this state appears to describe the groundstate in the system sizes that
can be studied numerically ($N_{\rm v}\leq 12$), it is possible that
finite-size effects may favour this over competing phases.  Estimates suggest that the
finite-size effects in experiments on cold gases 
will mean that the smectic phases are likely to be important in
practice\cite{CooperR07}. The first indications of quantum disordering of
vortex lattices as filling factor is reduced is likely to be the appearance
of (local) stripe ordering, perhaps as in Fig.~\ref{fig:smectic}.

In the range of filling factor $2\lesssim \nu \lesssim 6$ close to the
appearance of the short-range vortex-lattice ordering at $\nu\gtrsim 6$, the
usefulness of exact diagonalization studies becomes limited with respect to
determining the groundstates in the thermodynamic limit, owing to the
influence of finite-size effects.  This is apparent from the strong dependence
of the numerical spectrum on the boundary conditions (aspect ratio of the torus),
which shows there to be  competition between different phases: the
vortex lattice, Read-Rezayi states and the smectic state. Understanding the nature
of the groundstates in this regime is a very difficult theoretical challenge,
involving a strongly interacting many-body system close to a quantum phase
transition. This  would be a  particularly interesting regime to
explore experimentally.

\subsubsection{Incompressible Liquid Phases}

\label{sec:incompressible}

For $\nu< \nu_{\rm c}$ the vortex lattice phase (a triangular vortex lattice)
is unstable to quantum fluctuations, and is replaced by a series of
strongly correlated phases. These phases are best understood at small filling
factors, far from the transition to the vortex lattice phase. As the filling
factor approaches this transition, our understanding becomes much poorer (see
the discussion in \S\ref{subsec:crystals}).

{\it Laughlin State}

The strongly correlated phase that is best
understood is the bosonic Laughlin state\cite{laughlinstates}, at $\nu=1/2$.
The wavefunction describing this state is 
\begin{equation}
\Psi_{\rm Laughlin}(\{\zeta_i\}) \propto \prod_{i<
  j}^{N}(\zeta_i-\zeta_j)^2 \,.
\label{eq:laughlin}
\end{equation}
This is the {\it exact} groundstate for contact repulsion
(\ref{eq:contact}) at total angular momentum {$L=N(N-1)$}\cite{wgs}.
This may be noted from the fact that this is the unique symmetric
polynomial function of $N$ variables that has the property that it
vanishes when any two co-ordinates coincide -- it is therefore the
unique zero energy eigenstate of the contact interaction
(\ref{eq:contact}) in the 2D LLL.

The state has the property that its average density is
uniform\cite{laughlinstates}, up to a radius $\simeq 2\sqrt{N}\ell$
beyond which the density falls to zero.
Although the particle density of this state is uniform, there are still
vortices in the system. However, unlike the vortex lattice phase
(\ref{eq:condensate}), these vortices are not localized in space
(translational symmetry is not broken). Rather, the vortices can be viewed as
being {\it bound to the particles}\cite{girvinmacdonald,readoperator}. This can be seen by noting that the
wavefunction (\ref{eq:laughlin}) changes phase by $2\times 2\pi$ each time the
position $\zeta_i$ of any particle $i$ encircles the position of any
other particle $\zeta_{j\neq i}$. Each particle therefore experiences two vortices bound
to the position of every other particle, so the total number of vortices
experienced by any one particle is $N_{\rm v} =
2(N-1)$. Thus, (\ref{eq:laughlin}) describes a phase that
has filling factor $\nu = N/N_{\rm v} = 1/2$ in the thermodynamic
limit.\footnote{The exact relation $N_{\rm v} = 2N-2$ should be interpreted as
  $N_{\rm v} = N/\nu-{\mathcal S}$ where ${\mathcal S}$ is the ``shift'' of
  the state on the sphere, see \S\ref{sec:sphere}.}

The Laughlin state for bosons shows all of the characteristics familiar from
the conventional FQHE\cite{prangeandgirvin,dassarmapinczuk}: it is an
incompressible fluid, with gapped collective excitations in the bulk, and
gapless edge modes; the particle-like excitations have fractional particle
number and fractional statistics (they are ``abelian
anyons'')\cite{prangeandgirvin}.

\begin{figure}
\center\includegraphics[width=9cm]{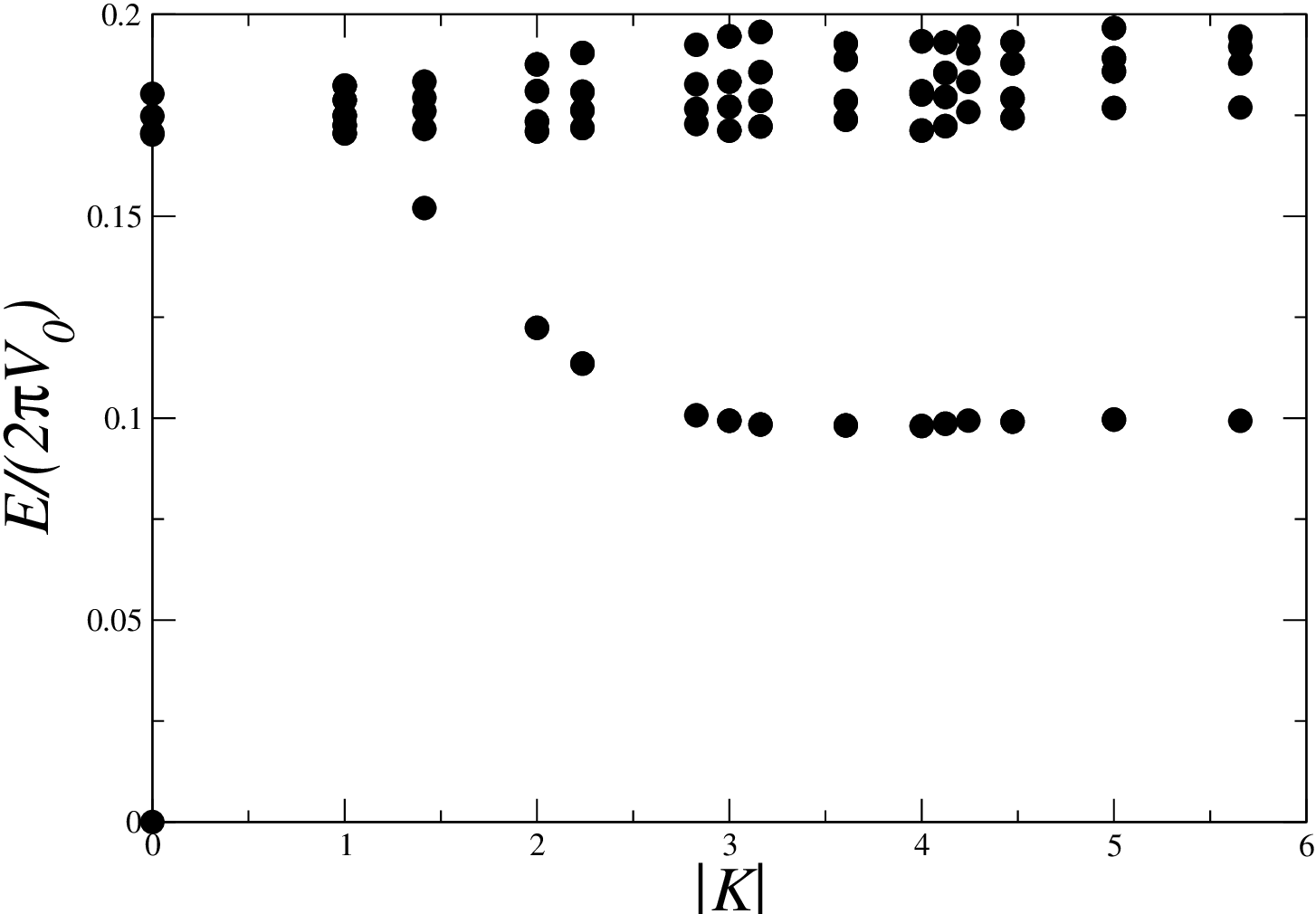}
\caption{Excitation spectrum of the Laughlin state at $\nu=1/2$, for
  bosons interacting with contact interactions, 
obtained by exact diagonalization studies for $N=8$ and $N_{\rm v}=16$
on a torus with square geometry,
$a/b=1$. The states are labelled by the magnitude of the
conserved wavevector in units of $2\pi/a$ (see \S\ref{sec:torus}). The zero
energy state at $|\bm{K}|=0$ is the Laughlin groundstate; the excitations
are gapped at all wavevectors, characteristic of an incompressible liquid.}
\label{fig:nu0.5}
\end{figure}

Incompressibility of the bosonic Laughlin state for rotating bosons is evident
from exact diagonalization studies of the collective excitation spectrum. The
spectrum shows the characteristics of an incompressible liquid (see
Fig.~\ref{fig:nu0.5}) in which the excitations are gapped at all momenta rather
than having the linear sound mode of a compressible liquid.  Calculations of
the energy gap on finite size systems, show convergence to the thermodynamic
limit, with a collective mode gap of  $0.095(5) \times (2\pi
V_0)$\cite{RegnaultJ03,regnault:235309}.\footnote{The numerical results in
  Refs.\cite{RegnaultJ03,regnault:235309} are stated to be in units of 
$\sqrt{32\pi} \hbar\omega_\perp a_{\rm s}/a_\parallel = 4 \pi V_0$. For
consistency with the numerical calculations of Ref.\cite{cwg} and here, I assume that the results of
Refs.\cite{RegnaultJ03,regnault:235309} are, in fact, in units of $2\pi V_0$.}

\label{rj}

For $L>N(N-1)$, the groundstate of the contact interaction becomes
degenerate, with a subspace spanned by quasi-hole excitations of
(\ref{eq:laughlin})\cite{Viefers}.
These states may be written
\begin{equation}
\Psi (\{\zeta_i\}, \{w_\alpha\}) \propto \prod_{i=1}^N\prod_{\alpha=1}^{N_{\rm
    qh}}
\left(\zeta_i - w_\alpha\right) \times \Psi_{\rm Laughlin}
 \,,
\label{eq:laughlinqh}
\end{equation}
where $w_\alpha$ are the co-ordinates of $N_{\rm qh}$
quasiholes.\footnote{As written the states (\ref{eq:laughlinqh}) are not
eigenstates of total angular momentum. They form an over-complete basis
for the zero-energy states with $L>N(N-1)$.}  From general properties
of the Laughlin states\cite{prangeandgirvin}, these quasiholes have a
fractional particle number, giving rise to a density depletion the
integral of which is exactly $1/2$ of an atom; they behave under
adiabatic exchange as particles with fractional exchange
statistics\cite{halperinhierarchy} (``anyons'') with statistics parameter
$1/2$, corresponding to ``semion'' statistics.

Incompressibility implies that the chemical potential has a discontinuity: the
chemical potential for adding particles to a bulk system, $\mu_+$, is not the
same as the chemical potential for removing particles, $\mu_-$. Since the
chemical potential for a bulk system is the derivative of the energy density
with respect to the number density, this implies that the energy density has a
cusp for an incompressible state. One can relate these thermodynamic
quantities to the creation energies of quasiparticles, $\Delta_{\rm qp}$, and
quasiholes, $\Delta_{\rm qh}$, in the incompressible state. For the $\nu=1/2$
Laughlin state, quasiparticle and quasihole excitations have fractional
particle number of $1/2$. Thus, adding a single particle to the bulk of the
system involves the creation of two quasiparticles, and an energy increase of
$\mu_+ = 2\Delta_{\rm qp}$; removing a particle involves the creation of two
quasiholes, and an energy increase of $-\mu_- = 2\Delta_{\rm qh}$. Hence, the
discontinuity in chemical potential is $\mu_+-\mu_- = 2(\Delta_{\rm qp}+
\Delta_{\rm qh})$. For the Laughlin state with contact interactions, the
energy to create a quasihole vanishes, $\Delta_{\rm qh} =0$. The quasiparticle
creation energy is non-zero, and is available only from numerical
studies\cite{cwg,RegnaultJ03,regnault:235309}. The numerical results are
consistent with the expectation that the quasiparticle energy is equal to the
collective mode gap\cite{RegnaultJ03,regnault:235309}. [The large wavevector
collective mode can be viewed as a widely separated quasiparticle quasihole
pair, so is a measure of $(\Delta_{\rm qp}+ \Delta_{\rm qh})$. From Fig.\ref{fig:nu0.5} this is  $\Delta_{\rm qp}+ \Delta_{\rm qh}  = \Delta_{\rm qp} \simeq
0.1 (2\pi V_0)$.]

{\it Composite Fermion States}

At certain higher filling factors, $\nu > 1/2$, there appear
strongly correlated phases that are accurately described in terms of
non-interacting ``composite fermions''\cite{CooperW99,RegnaultJ03,regnault:235309,chang:013611}.  As described in \S\ref{sec:cfdots},
the states formed from non-interacting composite fermions provide an accurate
description of the groundstates of small numbers of rotating bosons in a
trap. These are the small-system signatures of bulk incompressible
liquid phases (in the thermodynamic limit).

Composite fermions are formed from binding vortices to particles. Since the
underlying particles are bosons, one can form a composite particle which has
Fermi statistics by binding a single vortex to the location of each particle.
This may be made explicit within the Jain construction\cite{jainoriginal}, by
writing the many-particle wavefunction for the bosons as\cite{CooperW99}
\begin{equation}
\Psi(\{\bm{r}_k\}) = \hat{P}_{\rm LLL} {\prod_{i<j} (\zeta_i-\zeta_j)}\; \psi_{\rm
CF}(\{\bm{r}_k\}) \,.
\label{eq:cf}
\end{equation}
The Jastrow factor $\prod_{i<j} (\zeta_i-\zeta_j)$ causes any particle
$i$ to experience a single vortex at the location of any other
particle $j\neq i$. Since this factor is completely antisymmetric under
particle exchange, to obtain a bosonic wavefunction the function
$\psi_{\rm CF}$ must also be antisymmetric: this is the wavefunction
of the composite fermions.  In general terms, one can appreciate why
this is a useful variational state: the Jastrow factor suppresses the
amplitude for two particles to approach each other, as does the
composite fermion wavefunction. Taken alone, these two factors would
give a bosonic wavefunction that is a zero energy eigenstate of the contact
interactions. However, for $L< N(N-1)$, this construction requires
$\psi_{\rm CF}$ to include basis states that are not in the lowest
Landau level.  To recover a bosonic state within the LLL, one must project the
single particle states into these states, as represented by the operator $\hat{P}_{\rm LLL}$\cite{jainoriginal}.  For
$L<N(N-1)$ the projected wavefunction does not vanish as two particles
approach each other, so it is a state with a non-zero contact
interaction energy.

Since one vortex is bound to each particle,
in a large (uniform) system the composite fermions experience
\be
n_{\rm v}^{\rm CF} = n_{\rm v} - n_{\rm 2d}
\ee
vortices per unit area, so have filling factor (\ref{eq:fillingfactor})
\be
\nu^{\rm CF} = \frac{n_{\rm 2d}}{n_{\rm v}^{\rm CF}} \,.
\ee
Treating the composite fermions as {\it non-interacting} particles which 
completely fill $p$ Landau levels, {\it i.e} $\nu^{\rm CF} = \pm p$, 
one is led to the bosonic version of the
``Jain'' sequence
\begin{equation}
\nu = \frac{p}{p\pm 1} \;.
\end{equation}
At these filling factors, the composite fermion theory predicts the appearance
of an incompressible liquid state of the bosons, and provides a trial
wavefunction (\ref{eq:cf}).

The states constructed in this way have been shown to successfully  describe
the results of exact diagonalization studies in a variety of ways.

For trapped systems (in the disk geometry):
(i) The prediction of the ``magic angular momentum'' for states
of small numbers of rotating clusters\cite{CooperW99}.
(ii) Large overlaps of the groundstate wavefunctions with the composite fermion wavefunctions\cite{CooperW99}.
(iii) The form of the edge excitation spectrum for a bulk incompressible 
liquid of this kind is known\cite{wen,cazalilla03}. The edge modes found in exact
diagonalization studies are consistent with these expectations for 
bulk regions at  $\nu=1/2, 2/3, 3/4$\cite{cazalilla:121303}.

For uniform systems, the evidence of composite fermion states at $\nu=1/2, 2/3$
and $3/4$ includes:
(i) The existence of a sequence of states for different particle number $N$ at
the expected ``shift'' on the sphere [see Eqn.(\ref{eq:shift})], which have a uniform groundstate charge
density and an energy gap (i.e. consistent with incompressible
liquids)\cite{RegnaultJ03,regnault:235309}.
(ii) The prediction of the quantum numbers of the low-lying excited
states\cite{regnault:235309}.
(iii) Large overlaps of the exact groundstate and of the low-lying excited
states with the parameter-free wavefunctions formed from the
above composite fermion wavefunctions\cite{chang:013611}.

Extrapolation of the energy gap at $\nu=2/3$ shows a value of about $0.05\times (2\pi
V_0)$
in the thermodynamic limit\cite{regnault:235309};\footnote{See footnote 9 on
  page~\protect\pageref{rj} regarding the units in Ref.\protect\cite{RegnaultJ03,regnault:235309}.} a similar extrapolation for
$\nu=3/4$ is unreliable owing to the small number of system sizes available.

{\it Moore-Read and Read-Rezayi States}

One of the most interesting aspects of the physics of rapidly rotating Bose
gases is the prediction\cite{wgs,cwg,rrc,regnault:235324,CooperR07} of the
appearance of {\it non-abelian} phases: incompressible liquid phases whose
quasiparticle excitations obey ``non-abelian exchange
statistics''.

Non-abelian phases of matter are currently attracting intense theoretical and
experimental interest in the condensed matter physics community\cite{nayak:1083}. In part, this
is due to the possibility to explore experimentally this most exotic
consequence of quantum many-body theory.  In part, these studies are motivated
by the possibility ultimately to build a ``topological quantum computer'',
which makes use of the properties of a non-abelian phase to form a
quantum register with topological protection against decoherence.
Experimental work is largely centered on studies of the $\nu=5/2$ FQH state of
electrons in semiconductors. This state is thought likely to be a non-abelian
phase of matter described by the Moore-Read, or Pfaffian,
state\cite{MooreR91} or a close relative thereof. Experiments are working towards the detection of the
expected unconventional exchange statistics of the quasiparticle excitations
of this state.  Unfortunately, the creation of a universal topological quantum computer
would not be possible with the Moore-Read state, as its properties do now
allow arbitrary unitary transformations of the quantum register. Other
non-abelian phases, known as the Read-Rezayi states\cite{ReadR99}, would allow universal topological quantum computation. There is some theoretical
evidence that these states might appear in semiconductor systems, but this is
still very preliminary\cite{RezayiR06}.

With this background, it is striking that theoretical studies
have shown very convincing evidence for the appearance of the Moore-Read
state\cite{MooreR91} and the Read-Rezayi states\cite{ReadR99} for rotating
Bose gases with realistic two-body interactions. One of the major goals of the
creation of rapidly rotating atomic gases would be to allow (possibly the
first) measurements of non-abelian phases of matter.

The construction of Moore-Read and Read-Rezayi states for bosons may be viewed
as a generalization of the bosonic Laughlin state. 
The Laughlin state for bosons is
the densest exact zero energy eigenstate of the 2-body contact interaction
(\ref{eq:contact}) within the 2D LLL.  Similarly, the Moore-Read and
Read-Rezayi states are the densest exact zero energy eigenstates of a $k+1$-body
contact interaction
\begin{equation}
\sum_{i_1< i_2 < \ldots i_{k+1}=1}^N
\delta (\bbox{r}_{i_1}-\bbox{r}_{i_2})
\delta (\bbox{r}_{i_2}-\bbox{r}_{i_3})
\ldots \delta (\bbox{r}_{i_{k}}-\bbox{r}_{i_{k+1}}) \,.
\label{eq:k+1}
\end{equation}
For $N$ divisible by $k$, the groundstate wavefunctions of (\ref{eq:k+1}) may be written in a simple way
\begin{equation}
\Psi_{\rm RR}^{(k)}(\{\zeta_i\}) \propto {\mathcal S}\left[\prod_{i< j\in
    A}^{N/k}\!\!(\zeta_i-\zeta_j)^2\prod_{l< m\in
    B}^{N/k}\!\!(\zeta_l-\zeta_m)^2\ldots\right]
\label{eq:psik}
\end{equation}
where ${\mathcal S}$ symmetrizes over all possible 
ways of dividing the $N$ particles into the $k$ groups ($A,B\ldots$)
of $N/k$ particles each\cite{cappelli}. It is straightforward to
convince oneself
that (\ref{eq:psik}) vanishes when the positions of $k+1$ particles
coincide, as required.
By counting the degree of the polynomial $L=N(N/k-1)$, or the number of vortices
$N_{\rm v} = 2(N/k-1)$ experienced by each particle, one sees that these states describe bosons at
filling factor 
\begin{equation}
\nu^{(k)}= \frac{k}{2}\,.
\end{equation}
$k=1$ is the Laughlin state;
$k=2$ is the Moore-Read state;
$k\geq 3$ are the Read-Rezayi states.
Although these states have a similar structure, it should be
remembered that the physics of these phases is very different -- with
quasiparticle excitations that obey abelian ($k=1$) or non-abelian ($k \geq
2$) exchange statistics.

\begin{figure}
\center\includegraphics[width=9cm]{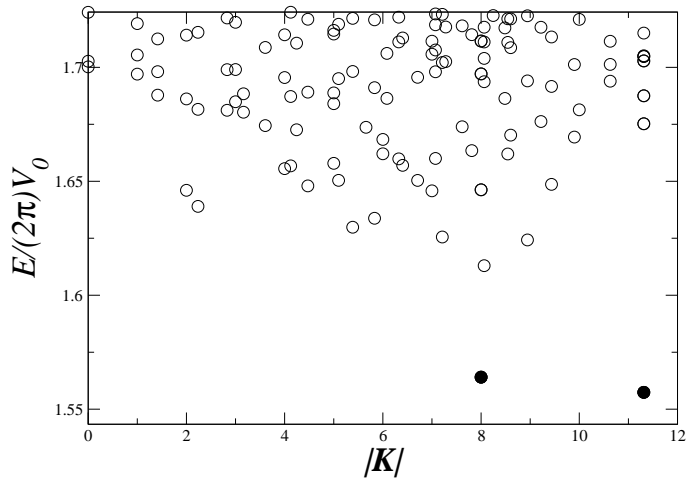}
\caption{Low energy spectrum for the $\nu=1$ state with contact interactions,
  obtained by exact diagonalization studies on a torus with square geometry
  for $N=16$ and $N_{\rm v}=16$. The states are labelled by the magnitude of
  the conserved wavevector on this periodic geometry. The appearance of
  low-energy states (shown as filled circles) at $\bm{K}=(8,8)$, $\bm{K}=(8,0)$ and $\bm{K}=(0,8)$ (the last two are indistinguishable
  in this plot by symmetry) is consistent with the expected three-fold
  degeneracy of the Moore-Read state.  [$\bm{K}$ is measured in units of $2\pi/a$ where $a$ is the side of the (square) torus.]}
\label{fig:nu1}
\end{figure}

Numerical evidence for  the Moore-Read state ($k=2$) for bosons
interacting with contact interactions has been reported on the disc\cite{wg},
torus\cite{cwg} and spherical\cite{RegnaultJ03,chang:013611} geometries. Large overlaps of
the exact wavefunctions with the model states (\ref{eq:psik}) are found.
While a large wavefunction overlap is encouraging, this is not necessarily the
best way to characterize a phase of matter. In the thermodynamic limit, 
owing to the
exponential increase in the size of the Hilbert space, the
wavefunction overlap with any trial state will surely vanish,
 even if the two
wavefunctions describe the same topological phase. A robust characterization
of the topological phase is provided by the groundstate degeneracy on the
torus\cite{oshikawa}; this degeneracy is expected to survive (and even improve) in the
thermodynamic limit, provided the wavefunction is in the same topological
phase as the trial state (\ref{eq:psik}). For the Moore-Read state, this
degeneracy appears clearly in the spectrum for the system sizes that can
be studied numerically\cite{cwg,chung:043608}, see Fig.~\ref{fig:nu1}.

Evidence for the appearance of phases of matter that are described by the Read-Rezayi states ($k\geq 3$) at $\nu=
k/2$ for contact repulsion (\ref{eq:contact}) has been found on the torus
geometry\cite{cwg}. While wavefunction overlaps are large, at least for $k$
that is not too large, the expected groundstate degeneracy is less clearly
resolved than for the Moore-Read state ($k=2$)\cite{rrc}.  Studies of the
groundstates in the spherical geometry at $\nu=3/2,2$, at the shifts
[see Eqn.(\ref{eq:shift})] appropriate for the $k=3$ and $k=4$ Read-Rezayi states, were determined to be
inconclusive\cite{RegnaultJ03,rjreview} with no clear sign that the
thermodynamic limit had been reached. It thus appears that, for contact
interactions, the correlation lengths of the groundstates at $\nu=3/2$ and
$\nu=2$ are somewhat larger than the available system sizes.  (The origin of
this lack of convergence is related to the competing ``smectic'' state at
intermediate filling factors\cite{CooperR07}, \S\ref{subsec:crystals}.)
However, it has been found that by changing the inter-particle interactions to
introduce an additional small non-local repulsion, the spectra at $\nu=3/2$ and $2$ smoothly evolve into
spectra with very clear groundstate degeneracies which show excellent evidence of
being in phases described by the $k=3$ and $k=4$ Read-Rezayi states\cite{rrc,regnault:235324,CooperR07}. It
appears that the small non-local interaction makes the correlation lengths
smaller than the system sizes available. These results are described further in
\S\ref{sec:dipolarcorrelated}.

\subsubsection{Feshbach Resonance}

For a model of rotating bosons interacting by a Feshbach resonance, it
can be shown that (for certain parameters) the {\it exact}
groundstates are the Moore-Read and Read-Rezayi
states\cite{CooperFesh04,cooperleshouches}. The hybridization of atoms into molecules
leads to a suppression of the repulsive inter-atomic interactions.
Tuning the resonance to the point where the net two-body repulsion
vanishes leads to an effective theory for dressed atoms (which are
resonating into molecules) in the 2D LLL which involves effective
three- and four-body contact interactions. These arise from processes
in which two atoms hybridize into a molecule which then collides with
an atom, or another molecule.

\subsubsection{Dipolar Interactions}

\label{sec:dipolarcorrelated}

The effects of dipolar interactions on the strongly correlated phases
of rotating bosons have been investigated in
Refs.\cite{crs,rrc,CooperR07,seki:063602,chung:043608}.  Qualitatively
similar effects are found from studies of a model interaction in which
a non-zero $V_2$ Haldane pseudo-potential is
introduced\cite{crsproc,regnault:235324,seki:063602} in addition to the contact
$V_0$. In this section we will focus on the case of dipolar
interactions, with pseudo-potentials (\ref{eq:dipolar}).

The strength of the dipolar interaction relative to the contact
interaction is characterized by the ratio $\alpha=V_2/V_0$ (\ref{eq:gamma}).
One expects the nature of the groundstates to vary with $\alpha$.

\begin{figure}
\center\includegraphics[width=9cm]{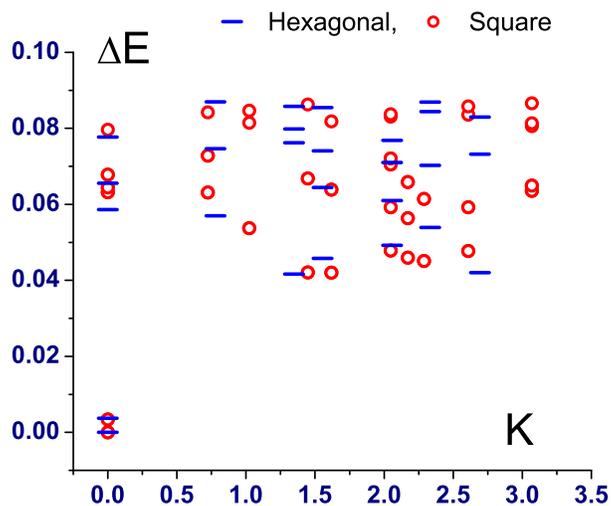}
\caption{
Numerical evidence for the appearance of the $k=3$ Read-Rezayi state for
bosons at
$\nu=3/2$. ($N=18, N_{\rm v} =12$ on a torus with two different
units cells, for 
 contact interactions with a small amount of dipolar interaction,
$\alpha = V_2/V_0=0.380$.)
The spectrum depends weakly on the unit cell,
so the results are representative of the thermodynamic limit. There is a very clear $2$-fold
degeneracy of the groundstate, exactly as expected for 
the $k=3$ Read-Rezayi state. (The overall $4$-fold degeneracy
is recovered
by the additional two-fold centre-of-mass
degeneracy
due to the half-integer filling factor.) From Ref.\protect\cite{rrc}.
[Reprinted figure from: E.H. Rezayi, N. Read, and N.R. Cooper,
 Phys. Rev. Lett {\bf 95}, 160404 (2005). Copyright (2005) by the American
 Physical Society. 
]
}
\label{fig:k3}
\end{figure}

For small positive values of $\alpha$ the nature of the groundstate does not
change significantly. However, it has been shown that this can lead to
improvements in the overlaps of the groundstates (as computed in exact
diagonalization studies) with the Moore-Read state at $\nu=1$ and the
Read-Rezayi states at
$\nu=3/2$ and $2$\cite{rrc,CooperR07,seki:063602,chung:043608}. It appears that the
effect of a small amount of additional long-range repulsion is to reduce the
correlation length of the groundstate sufficiently that it is less than the
available system sizes. Numerical results then show convergence to the
thermodynamic limit. Evidence for the $k=3$ Read-Rezayi state at $\nu=3/2$ is shown in,
Fig.~\ref{fig:k3}. Currently, these results provide by far the most convincing theoretical
evidence for the existence of a Read-Rezayi topological phase in a system with
realistic two-body interactions.

For larger values of the non-local interaction, the nature of the
groundstate changes character.

The change in the nature of the groundstate at $\nu=1/2$ was described
in Ref.\cite{crs}, based on the results of exact diagonalization
studies on the torus. For $\alpha =0$ the exact groundstate is the
Laughlin state. For $\alpha\gtrsim 0.5$ the groundstate changes
character abruptly (the overlap with the Laughlin state becomes very
small).  The states for $\alpha\gtrsim 0.5$ were found to be compressible
crystalline phases. Numerical evidence was given for a stripe phase at
$\alpha = 0.528$, and for the $q=4$ bubble crystal phase at $\alpha = 0.758$.
These results are consistent with expectations of mean-field theory,
Fig.~\ref{fig:dipolargs}, which indicates stripe and bubble crystal
phases at these values of $\alpha$. Mean-field theory is expected
to be reliable only for sufficiently large $\nu$. Why does it
work so well even for $\nu=1/2$ in this regime of $\alpha\gtrsim 0.5$?
The answer can be found by noting that the mean-field phases at these
values of $\alpha$ involve clustering of particles, so the quantum
fluctuations are reduced. 
For example, the
$q=4$ bubble crystal phase at $\alpha = 0.758$ has $\nu q =2$ particles per bubble, so can be viewed
as a crystal of pairs of particles.
A calculation of the Lindemann criterion for quantum melting of the
$q=4$ bubble crystal phase 
leads to a critical filling factor of $\nu_c^{q=4} \simeq 0.4$, using
a Lindemann criterion that gives $\nu_c \simeq 6$ for the
critical filling factor at which the triangular vortex lattice is
unstable to quantum fluctuations\cite{crs}. This estimate is
consistent with the observation of crystalline order even down to
$\nu=1/2$.

The collapse of the incompressible states for large values of $\alpha$
has been reported at filling factors
$\nu=1$\cite{seki:063602,chung:043608}, $\nu=3/2$\cite{rrc} and
$\nu=2$\cite{CooperR07}. In all cases the transition is into compressible states.  At
$\nu=1$ the Moore-Read state is destroyed for $\alpha\gtrsim 0.4$, and
evidence has been given for the appearance first of a stripe
phase\cite{seki:063602,chung:043608} and then of bubble crystal
phases\cite{seki:063602}. These results are consistent with the
expected increased stability of the stripe and bubble crystal phases at
$\nu=1/2$ as the filling factor is increased\cite{rrc}. (The $q=4$ bubble crystal
phase is a crystal of clusters of $\nu q = 4, 6$ and $8$ particles for
$\nu=1,3/2$ and $2$ respectively.)

As compared to the case of contact interactions, one qualitatively new
effect of small non-zero value of $\alpha$ is that it allows the emergence of
other strongly correlated phases for $\nu<1/2$. At $\nu=1/3$ numerical
evidence  has been given\cite{chung:043608} for the
appearance of the composite fermion liquid\cite{hlr}
state.  One further anticipates the appearance of 
other incompressible states (e.g. the Laughlin states at $\nu=1/m$
with $m\geq 4$) and the transition, at sufficiently small $\nu$, to a
Wigner crystal phase of the individual atoms, analogous to that for
electrons in FQHE systems or dipolar interacting
fermions\cite{baranov:200402}. The properties of rotating dipolar
bosons in this regime remain to be studied in detail.

\subsection{Density Distribution in a Trap}

\label{sec:lda}

In much of the above, we have treated the rapidly rotating gas in the uniform
limit. This applies when $\Omega=\omega_\perp$ and the residual
transverse trapping potential in the rotating frame \be V^\perp_\Omega(r_\perp) = \frac{1}{2}
M\left(\omega_\perp^2-\Omega^2\right) r_\perp^2 \ee
 vanishes. In this formula $r_\perp\equiv \sqrt{x^2+y^2}$ is the radial distance
 from the rotation axis.
 Clearly $\Omega = \omega_\perp$ requires the numbers of vortices $N_{\rm v}$
and of particles $N$  to be  infinite. In any finite
  system one has $\Omega< \omega_\perp$ so there will remain a weak harmonic
  trapping potential. This trapping leads to inhomogeneity of the
  density, which we now discuss.

The first corrections to the particle density at large but finite $N_{\rm v}$
can be found by using the local density approximation (LDA). This  amounts
to the assumption that at each point in space there is a locally defined
chemical potential
\be
\mu(r_\perp) = \mu - V^\perp_\Omega(r_\perp) \,.
\ee
Then, the 2D particle density is found from the
local chemical potential, as
\be
n_{\rm 2d}(r_\perp) = n_{\rm 2d}[\mu(r_\perp)] \,,
\label{eq:nmu}
\ee
where $n_{\rm 2d}[\mu]$ is the 2D density of a {\it uniform system} at
chemical
potential $\mu$.
Within a reasonable assumption of locality in the physics
determining the equation of state, one expects the LDA to be accurate provided
the particle density varies slowly on a lengthscale of the correlation length of
the relevant phases; for the situations we have been discussing this correlation length
is of order the mean
vortex spacing $a_{\rm v} \simeq a_\perp$.

Through (\ref{eq:nmu}), the spatial density profile $n_{\rm 2d}(r_\perp)$ depends
on the equation of state, $n_{\rm 2d}[\mu]$. The profile therefore takes different forms in the vortex-lattice
($\nu > \nu_{\rm c}$) and strongly correlated ($\nu < \nu_{\rm c}$) regimes.

A useful general result exists. For a Bose gas with repulsive
interactions, the density falls to zero for $\mu < 0$.
Therefore, the edge of the cloud is at a radius \be R_\perp =
\sqrt{\frac{2\mu(0)}{M(\omega_\perp^2-\Omega^2)}} \ee where $\mu(0)$ is the
chemical potential at the centre of the cloud. Since the average vortex density is
$n_{\rm v} = 2M\Omega/h$, 
the total number of vortices within the Thomas-Fermi radius $R_\perp$ is 
\be 
N_{\rm
  v} = \pi R_\perp^2 n_{\rm v} = \frac{2\pi
  M\Omega}{h}\frac{2\mu(0)}{M(\omega_\perp^2-\Omega^2)} \,. \ee 
Inverting this relation, and taking the limit $N_{\rm v}\gg 1$ in which the LDA
is valid, shows that the 
rotation frequency is
\be 
\label{eq:close}
\frac{\Omega}{\omega_\perp} \simeq 1 - 
\frac{\mu(0)}{\hbar\omega_\perp}\frac{1}{N_{\rm v}} + {\cal O}(1/N_{\rm v}^2)
\,.
\ee
Thus, for $N_{\rm v}$ sufficiently large (larger than
$\hbar\omega_\perp/\mu(0)$), the rotation frequency is close to the
trap frequency $\omega_\perp$.

Deep in the vortex lattice phase $\nu\gg 1$, the equation of state can
be found from mean-field theory, Eqn. (\ref{eq:egplattice}), giving  $\mu =
2\beta_A \nu V_0$ where $\nu$ is the local filling factor (\ref{eq:nulocal}).
The local density approximation therefore predicts a
density distribution of the form\cite{watanabe} 
\be n_{\rm 2d}(r_\perp) = \nu(r_\perp) n_{\rm v} = \frac{\mu(0)}{2\beta_A V_0}
n_{\rm v}\left(1-\frac{r_\perp^2}{R_\perp^2}\right)\,.
\ee
This is
the inverted parabola familiar from the Thomas-Fermi approximation for a
trapped BEC. This been shown to provide an excellent description
of vortex lattice phase ($\nu\gg1$) within the 2D
LLL\cite{watanabe,CooperKR}, showing that the LDA is accurate in this
regime. 
Further applications of the LDA allow one to relate $\mu(0)$ to the total
number of particles $N$, and $\Omega$ to the angular momentum
$L$\cite{watanabe,CooperKR}. This leads to the result that the number of
vortices inside the Thomas-Fermi radius scales as
\be N_{\rm v} \simeq
3\frac{L}{N} \ee
for large $N_{\rm v}$.
Detailed theoretical studies of the particle density and vortex distributions
of finite vortex lattices in the mean-field LLL regime have been reported
in Refs.\cite{watanabe,CooperKR,aftalion:023611,aftalion:011601}. Among the motivations for these
considerations was the observation that if the vortices adopt a uniform
triangular array the mean particle density follows a Gaussian profile on
average\cite{Ho01}, rather than the more conventional Thomas-Fermi profile.
The reconciliation of these views led to the conclusion that the vortex
lattice distorts slightly towards the edge of the
trap\cite{watanabe,aftalion:023611,CooperKR}, in such a way that the above Thomas-Fermi
profile is achieved. The arrangement of vortices outside the
Thomas-Fermi radius has been discussed in Ref.\cite{aftalion:011601}.

In the strongly correlated regime, the equation of state has a jump in the
chemical potential at the density of each incompressible phase. As a consequence, within the
LDA the filling factor does not fall smoothly with radius. The density
distribution shows a
``wedding cake''
structure\cite{cooper:063622} characteristic of incompressible phases, with regions in which the 2D particle density
is pinned to values $\nu^* n_{\rm v}$ where $\nu^*$ are filling factors at
which the groundstate is incompressible. This is expected to give clear
experimental evidence for the appearance of incompressible phases in
images of the density distribution following
expansion\cite{cooper:063622}.

\subsection{Quantum Melting in the 3D LLL Regime}

\label{sec:3dmft}

\begin{figure}
\center\includegraphics[width=13cm]{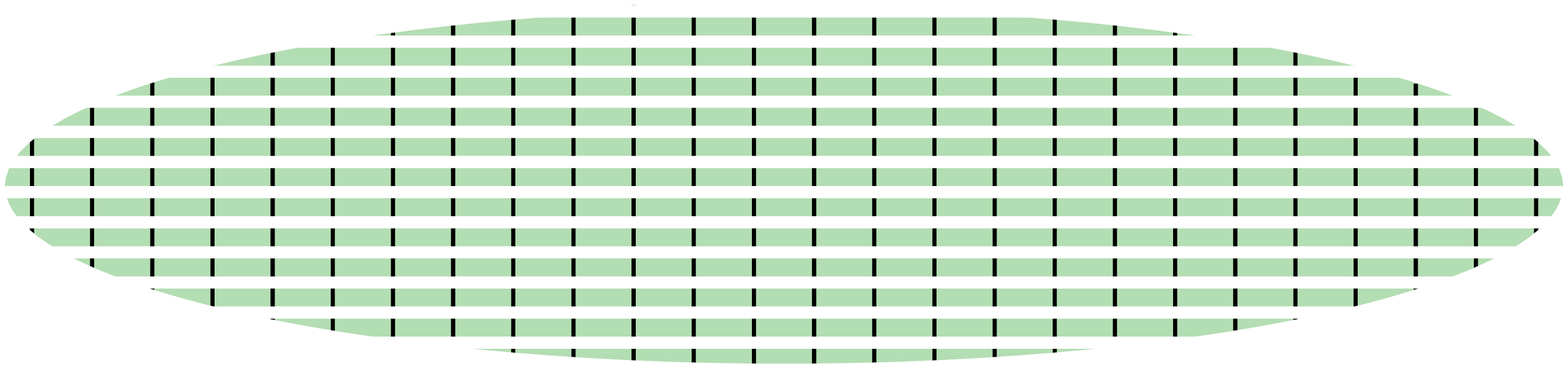}
\caption{Schematic diagram illustrating the use of an optical lattice,
  directed along the rotation axis, to slice the rotating condensate into many
  parallel layers. Each layer has a reduced filling factor, $\nu_{\rm layer} \simeq
  \nu/N_{\rm L}$, where $N_{\rm L}$ is the number of layers. In addition, the tight axial confinement of each layer ($a^{\rm layer}_\parallel\lesssim d_\parallel$) leads to an increased energy scale for interactions
  (\protect\ref{eq:v0}).}
\label{fig:slice}
\end{figure}

The 3D LLL regime
 (\ref{eq:q3dregime}) can be relevant for a
rapidly rotating Bose gas in an anisotropic trap $\omega_\parallel \ll
\omega_\perp$\cite{Ho01}.
The regime is also of importance for the situation\cite{dalibardcornell} in
which the rotating gas is sliced into many parallel layers by a 1D optical
lattice with wavevector parallel to the rotation axis, see
Fig.~\ref{fig:slice}.
The optical lattice  leads to an increased
effective mass $M_\parallel$ for motion along the rotation axis, $M_\parallel
> M$. 
For a deep lattice in the
tight binding limit, with lattice period $d_\parallel$ and 
single particle tunneling amplitude  $J$, the effective
mass for motion along the rotation axis is 
\be
M_\parallel =
\frac{\hbar^2}{2Jd_\parallel^2}  \,,
\ee
which diverges as $J\to 0$. 
Hence, the oscillator energy $\omega_\parallel \propto 1/M_\parallel$
is reduced, leading to an anisotropic confinement $\omega_\parallel \ll
\omega_\perp$.

The use of a 1D optical lattice has been suggested as a means of increasing
the quantum fluctuations of the vortices and allowing entry into the
strongly correlated regime\cite{dalibardcornell,cooper:063622,snoek1,snoek2}.
This also has the advantage of increasing the interaction energy scale
of the strongly correlated phases (see \S\ref{sec:status}).

The origin of the enhanced quantum fluctuations of the vortices is simple to
understand in the limit of a very deep lattice in which the layers become
decoupled. Then, the relevant quantity controlling the quantum fluctuations is
the filling factor {\it per layer} \be \nu_{\rm layer} \simeq \frac{N}{N_{\rm
    v} N_{\rm L}} = \frac{\nu}{N_{\rm L}} \ee where $N_{\rm L}\simeq
a_\parallel/d_\parallel$ is the number of layers that the optical lattice
forms, which could be of order $N_{\rm layer}\sim 50-100$. The vortex lattice
in each layer will disorder due to quantum fluctuations for $\nu_{\rm layer} <
\nu_{c}$ (where $\nu_{\rm c}\sim 6$ is the 2D result\cite{cwg,shm1,baym}),
leading to quantum melting at a very high {\it total} filling factor $\nu <
 \nu_{\rm c}N_{\rm L}$.

The quantum fluctuations of a uniform vortex lattice ($N_{\rm v}$ large) in
the
3D LLL regime of a layered geometry have been investigated theoretically
in Ref.\cite{cooper:063622}. The critical filling factor for a uniform system
was estimated  as a function of the coupling between the layers
both by applying the Lindemann criterion based on the quantum
fluctuations of the vortices, and by evaluating the condensate
depletion.
The degree of interlayer coupling is
controlled by the dimensionless ratio of the interlayer tunneling energy $J$
and the mean interaction energy in a single layer, $V_0\nu_{\rm layer}$ with
$V_0$ (\ref{eq:v0}) evaluated for the subband thickness $a_{\parallel}^{\rm
  layer} < d_\parallel$ of each individual layer. The results show the
crossover from the regime of strongly coupled layers $J\gg V_0\nu_{\rm layer}$
where the 2D LLL regime is valid and quantum melting occurs for $\nu \lesssim
\nu_{\rm c}$, to the regime of decoupled layers for $J\ll V_0\nu_{\rm layer}$
and quantum melting requires only $\nu \lesssim \nu_{\rm c} N_{\rm L}$.

The criterion for quantum melting in the  intermediate  3D regime
can
be
understood in simple geometric terms. In this 3D geometry, the vortices do not
simply fluctuate as straight lines. Quantum fluctuations involve also bending
of the vortex lines -- the Kelvin modes of the
vortices\cite{PhysRevLett.91.240403}. The relevant Kelvin modes are those with
wavelengths longer than a certain minimum wavelength,
which one can think of as the lengthscale over which the vortices remain
straight.  This minimum wavelength  is set by the healing length
(\ref{eq:healinglength}) along the rotation axis\cite{cooper:063622} \be
\xi_\parallel \simeq \sqrt{\frac{\hbar^2}{2 M_\parallel \mu}}
\label{eq:xiparallel} \, , \ee 
or, if the healing length becomes less than the lattice period
$\xi_\parallel < d_\parallel$, then one should use $d_\parallel$ as the minimum wavelength.
 Note that, for a system in
the 3D LLL (\ref{eq:q3dregime}), the healing length satisfies $\xi_\parallel
\ll a_\parallel$, so the vortex fluctuations do involve flexural motion.

Applying a Lindemann criterion\cite{cooper:063622} to these fluctuations of
the vortex lines leads to the conclusion that quantum melting of the 3D vortex
lattice occurs when the filling factor in a slab of thickness
$\xi_\parallel$ is less than a critical value of order 1. We
take this critical value to be $\nu_{\rm c}\sim 6$ for consistency with the 2D
studies\cite{cwg,shm1,baym}. Then the condition for quantum melting of the 3D
vortex lattice is
\be
\frac{\bar{n}\xi_\parallel}{n_{\rm v}} < \nu_{\rm c} \,,
\label{eq:3dcriterion}
\ee
where $\bar{n}$ is the 3D particle density.
With this criterion in mind, and using (\ref{eq:xiparallel}),
the condition for validity of mean-field theory in the
3D LLL regime (\ref{eq:q3dregime}) is
\be
\bar{n}\gg a_{\rm s} n_{\rm v}^4 \sim \frac{a_{\rm s}}{a_\perp^4} \,.
\label{eq:3dmft}
\ee
It is only in this limit (\ref{eq:3dmft}) that one can  model the
3D LLL regime by a Thomas-Fermi profile for the particle density
(\ref{eq:tfsubband}) along
the rotation axis.

The quantum fluctuations of a confined vortex lattice ($N_{\rm v}$ finite) in
a layered geometry have been studied in Ref.\cite{snoek1,snoek2}. This work
provides a detailed study of the Tkachenko modes for the finite-size vortex
lattice. The quantum melting of clusters of vortices is found to be
inhomogeneous, due to the inhomogeneous particle density. Melting can lead to
a fluid of vortices, or to a
shell structure in which shells of vortices can decouple from their
neighbours but maintain inter-vortex correlations within a shell. A critical
comparison is made with the expected location of the vortex-lattice/vortex
liquid boundary in the inhomogeneous system as predicted by the local density
approximation. For small vortex arrays ($N_{\rm v}$ small) in the limit of
weak interlayer coupling, one expects a transition to quantum Hall ``Mott''
phases in which number fluctuations between layers are suppressed\cite{cooper:063622}.

\section{Atomic Bose Gases in Rotating Optical Lattices}

\label{sec:lattices}

The effects of a periodic pinning array on vortices in superconducting
materials is a matter of significant experimental and theoretical
interest\cite{fiory:73,PhysRevB.57.7937,PhysRevB.67.014532,PhysRevLett.78.2648}. At
low temperatures, the vortex lattice can undergo a transition from the
triangular Abrikosov lattice (in the absence of pinning) into phases
that are pinned to the lattice, with a series of different
ordered structures.  The nature of the groundstate involves the
interplay between the vortex density $n_{\rm v}$ and the density of
periodic pinning sites $n_{\rm site}$, and can involve rich and
interesting commensurability effects\cite{PhysRevB.57.7937} as a
function of the ratio \be f \equiv \frac{n_{\rm v}}{n_{\rm site}}
\,, \ee the mean number of vortices per pinning site.

The interplay between vortex-vortex interactions and a vortex pinning
potential can be conveniently studied in rotating atomic BECs. Regular arrays
of pinning sites can be imposed by a rotating optical lattice
potential\cite{tung}, and allow detailed investigations of the vortex lattice
structure and the dynamics. Furthermore, the extension of experimental studies
into the strongly interacting regime, where the optical lattice is strong and
the (non-rotating) Bose gas is close to the Mott insulating regime, will allow
the study of vortex-commensurability effects in a regime of strong
correlations.

\subsection{Rotating Optical Lattices}

Conceptually, the simplest way to investigate the physics of vortices in a
static periodic potential is to impose a rotating optical lattice on a trapped
BEC in an otherwise cylindrically symmetric potential well. The optical
lattice is at rest in the rotating frame. In the steady state, the gas will
come to equilibrium in this frame of reference.

The coupling of a BEC to rotating lattices has been explored in
experiments in which a rotating mask impose a rotating optical lattice
of square or triangular symmetry\cite{tung}. Owing to the small angle
of the light passing through the mask relative to the rotation axis,
the periods of these lattices ($\sim 8\mu\mbox{m}$) are large compared
to the optical wavelength. An alternative geometry has been proposed
that could allow rotating optical lattices with smaller lattice
constants, of order half the optical wavelength\cite{hafezi-2007}.

For a strong optical lattice, in which atoms are restricted to a single
Wannier orbital at each site\cite{jakschbh,greiner}, schemes have been
proposed by which to imprint the required Peierls phases to simulate the
motion of a charged particle in a uniform magnetic field. The uniform magnetic
field has the same effects as the Coriolis forces of uniform rotation; we shall
therefore use ``rotation'' and ``magnetic field'' interchangeably, assuming
that a harmonic confinement potential is applied to cancel the centrifugal
effects of rotation. One class of method makes use of optically induced
transitions to other hyperfine levels of the atom to imprint geometric
potentials on the tunnelling matrix elements\cite{jaksch,mueller}. In another
proposal, time-dependent potentials are imposed to modulate the lattice
potentials and tunnelling amplitudes in register. Provided this modulation is
sufficiently fast compared to the timescales relevant for the physics of the
cold atomic gas, these have the effect of imprinting the necessary phase
modulations to simulate a uniform magnetic field\cite{sorensen:086803}.

The results of these studies\cite{jaksch,mueller,sorensen:086803} are
proposals for generating systems that are well-described by the Bose
Hubbard model in a uniform magnetic field \be
\label{eq:bh}
\hat{H} = -J \sum_{\langle i,j\rangle } \left[\hat{a}_i^\dag\hat{a}_j
e^{i A_{ij}} + \hat{a}_j^\dag\hat{a}_i e^{i A_{ji}}\right] +
\frac{1}{2}U\sum_i \hat{n}_i(\hat{n}_i-1) - \mu\sum_i \hat{n}_i \ee
where $\hat{a}_i^{(\dag)}$ are the field operators for boson
creation/destruction at the lattice site $i$, and $\hat{n}_i =
\hat{a}_i^\dag\hat{a}_i$ is the number operator. $J$ is the tunnelling
energy and $U$ is the on-site boson interaction. The
ratio size of $J/U$ can be varied by changing the strength of
the optical lattice\cite{jakschbh,greiner}.  The sites $i$ are defined
on a 2D lattice, which, for convenience, we shall consider to be a
Bravais lattice (all sites are equivalent, with one site per unit
cell). To simulate a uniform magnetic field (or uniform rotation), the
link phases $A_{ij}= -A_{ji}$ must have the gauge-independent property
that their sum around any unit cell of the
lattice (the discrete version of the line integral $\oint \bm{A}\cdot
d\bm{r}$) is a constant \be \sum_{\rm unit-cell} A_{ij} = 2\pi f \ee
where $0\leq f < 1$ is the mean number of vortices (flux quanta) per
unit cell of the lattice.

A related scheme has been proposed in which ``non-abelian'' gauge fields can
be generated in optical lattice systems\cite{osterloh:010403}. We will
restrict attention to the physics of the above model with an abelian gauge
field (\ref{eq:bh}).

\subsection{Weakly Interacting Regime: Vortex Pinning}

For a weakly interacting Bose gas, and in the regime where the 2D
density of particles is sufficiently large, $n_{\rm 2d}\gg n_{\rm v},
n_{\rm site}$ it is appropriate to treat the system within mean-field
theory.

This was first discussed for a uniform system, within an effective model for
the vortex interactions in the presence of a square lattice pinning
potential\cite{reijndersprl}. The maxima of the potential act as pinning
centres for the vortices, owing to the particle depletion at
the vortex cores. At a vortex density of $f=1$ it was found that there exist
three phases: the unpinned triangular Abrikosov lattice; the strongly pinned
square lattice with one vortex per lattice site; and an intermediate phase in
which half of the vortices are pinned. The phase diagram is a function of  the strength and period of the pinning potential relative to, respectively,
the chemical potential and the healing length of the BEC. Extensions to other
vortex densities $f< 1$ and calculations of the collective mode dynamics are
reported in Ref.\cite{reijnderspra}.

A study of a rotating square lattice over a wide range of vortex
fillings in a harmonically trapped atomic BEC is reported in
Ref.\cite{pu:190401}. A direct numerical solution of the GP equation
in a trap containing $\sim 100$ lattice sites was performed.  A very
rich range of configurations was found, including states with
doubly quantized vortices pinned to certain lattice sites, and the
appearance of domain walls between regions of (locally) different
order. Owing to the many competing phases, the inhomogeneity of a
trapped gas leads to a very much more complex situation than the
uniform case.

The experiments reported in Ref.\cite{tung} studied the effects of
rotating triangular\cite{sato:053628} and square lattices. The locking
of a BEC (initially rotating at a different rate to the optical
lattice) to a rotating triangular lattice was investigated. No clear
enhancement in the locking was found at the commensurability point $f
=1$.  For a rotating square lattice, the structural phase transition
was observed from the triangular vortex lattice (at weak pinning) to a
square lattice (strong pinning), at a pinning strength in rough
agreement with the predictions of
Refs.\cite{reijndersprl,pu:190401}. Incommensuration effects were
found to be difficult to observe in experiment\cite{tung}, as the
finite number of vortices can compress or expand to stay close to
commensurate conditions.

Another approach to the physics of this mean-field regime is
applicable in the case of a very strong optical lattice, for which the
Bose-Hubbard model (\ref{eq:bh}) can be used. Under conditions of weak
interactions $U\ll J$ and when the mean number of particles per
lattice site is large,\cite{polini:010401,kasamatsu} one can replace
the Hamiltonian (\ref{eq:bh}) by the phase representation \be 
\label{eq:qphase}
\hat{H}
= -\sum_{\langle i,j\rangle} \tilde{J}_{ij}
\cos\left(\theta_i-\theta_j+A_{ij}\right) - \frac{U}{2} \sum_j
\frac{\partial^2}{\partial\theta_j^2} \ee where $\tilde{J}_{i,j} =
2\sqrt{n_i n_j} J$ with $n_i$ the mean particle number per site. This
model is familiar from studies of ``frustrated Josephson junction
arrays''. In the limit $U\ll \tilde{J}$ it reduces to the classical
uniformly frustrated XY model, the groundstates of which have been
found for infinite uniform lattices at certain rational values of the
vortex density
$f$\cite{PhysRevLett.51.1999,PhysRevB.31.5728,PhysRevB.48.3309}.  The
groundstate at $f=1/2$ is two-fold degenerate.  A Gutzwiller
mean-field study\cite{palmer:013609} of the Bose-Hubbard model in this
regime $J\ll U \ll \tilde{J}$ for $f$ close to $1/2$ shows the
appearance of two competing groundstates.  In Ref.\cite{kasamatsu} the
influence on the classical groundstates of the inhomogeneity in
$\tilde{J}$ due to the harmonic trap has been studied.

\subsection{Strongly Interacting Regime}

Understanding the nature of the groundstate of the Bose-Hubbard model
in a magnetic field (\ref{eq:bh}) poses a very difficult theoretical
problem. It involves the interplay of several very interesting and
complex aspects of physics.

\begin{itemize}

\item As discussed above for weak interactions, there is the interplay
of commensurability of the vortex density $n_{\rm v}$ with the lattice
site density $n_{\rm site}$, controlled by $f =n_{\rm v}/ n_{\rm
site}$.  This same commensurability is responsible for a  complex
form of the energy spectrum of a single particle\cite{harper}, which
has a fractal structure referred to as the ``Hofstadter
butterfly''\cite{hofstadter}.

\item In the continuum limit (in the absence of any pinning potential), there
  can appear strongly correlated phases related to FQH states when
  the 2D particle density $n_{\rm 2d}$ is comparable to the vortex density
  $n_{\rm v}$, controlled by the filling factor $\nu = n_{\rm 2d}/n_{\rm v}$. (See
  \S\ref{sec:incompressible}.)

\item In the absence of rotation, $f=0$, the interplay between interactions
$U$ and kinetic energy $J$ leads to the physics of the Mott-Hubbard
transition for bosons\cite{jakschbh,greiner}.

\end{itemize}

The relevant dimensionless parameters controlling the properties of
the system are the vortex filling factor $f$, the particle filling
factor $n$, and the ratio $U/J$ of particle interactions to the
bandwidth.  In addition the properties can depend on the geometry of
the lattice on which the system is defined. (Unless stated otherwise,
it shall be assumed that the lattice is square.)

\subsubsection{Mean-Field Theory}

The groundstates of the Hamiltonian (\ref{eq:bh}) have been studied within
mean-field theory by several authors, focusing on different aspects of
the very rich physics that can arise.

The nature of the vortices and vortex lattices close to the
Mott-insulating states ($n$ is integer) were studied in
Ref.\cite{PhysRevA.69.043609}. It was found that the vortex cores
tend to have a local particle density that is equal to that of the nearest
Mott state, which can be larger or smaller than the average background
density.

In Ref.\cite{polini:010401} a mean-field treatment of the quantum
phase model (\ref{eq:qphase}) has been used to study the transition at
$f=1/2$ into the Mott insulating phase.  The boundary of the Mott phase
has been studied in detail in Ref.\cite{oktel:045133} as a function of
the vortex filling $f$. It was shown to follow a dependence that is
set by the extremal energy of the corresponding one-particle state in
the Hofstadter spectrum. That the phase boundary of a
strongly interacting phase (the Mott insulator) can be described in
terms of a single particle property is a surprise; the connection
arises from the fact that the boundary is determined by
the appearance of dilute particles, the properties of which are
governed by the single-particle spectrum\cite{goldbaum:033629}.

Ref.\cite{goldbaum:033629} has studied in detail the vortex lattice
structures close to the Mott-insulating state. It is shown that, in
contrast to the case of {\it weakly} interacting Bose
gases\cite{reijndersprl,pu:190401} the vortex cores do not necessarily get
pinned to the maxima of the potential (i.e. the plaquettes of the
lattice); vortex lattices are found in which the vortices are pinned
to the minima of the potential (the sites of the lattice). This is
related to the presence of Mott phases at vortex
cores\cite{PhysRevA.69.043609}, which can act to increase the particle
density at the vortex core as compared to the average particle
density, making it energetically favourable to locate the core in the
potential minima.

\subsubsection{Strongly Correlated Phases}

Going beyond mean-field theory involves considerations of the
effects of quantum fluctuations of the vortices.

The quantum fluctuations of a single vortex have been described in
Ref.\cite{vignolo:023616}, for $\tilde{J}\gg U$, when the interaction
term in Eqn.(\ref{eq:qphase}) is small.  Fluctuations lead to a finite
vortex mass, which is predicted to have experimental consequences in
Bragg spectroscopy measurements.

For a dense array of vortices forming a vortex lattice, one can expect
that, if sufficiently strong, quantum fluctuations will disorder the
vortex lattice. A novel form of disordered vortex phase was proposed
for the ``dice'' lattice, at vortex density
$f=1/3$\cite{burkov:180406}. It is argued that in this situation, the
first effect of quantum fluctuations of the vortices is to lead to a
so-called ``vortex Peierls'' phase: in this phase the vortices are
partially delocalized, tunnelling between pairs of nearest-neighbour
sites, while retaining the broken translational symmetry expected for
a conventional vortex lattice.

For very strong quantum fluctuations, one expects the vortex
lattice states to disorder completely and form strongly correlated
liquid states, analogous to the incompressible liquid states discussed
for rotating Bose gas in the continuum, see \S\ref{sec:incompressible}.
Indeed, the Bose-Hubbard model in a magnetic field (\ref{eq:bh})
contains all of the physics of the continuum (as discussed in \S\ref{sec:rapid}) in the corner of parameter space with
vortex filling and particle filling are small $n,f\ll 1$: in this
continuum limit, the groundstate depends only on the filling factor
$\nu\equiv n/f$ and includes the correlated phases described in
\S\ref{sec:incompressible}.

The influence of a lattice potential on the bosonic Laughlin state
(\ref{eq:laughlin}) has been studied in large-scale exact
diagonalization studies on  periodic lattice
geometries\cite{sorensen:086803,hafezi-2007}, intended to represent
the bulk region of a system with a large number of lattice sites.
Fixing the filling factor (\ref{eq:fillingfactor}) to  $\nu\equiv n/f=1/2$, the influence of the lattice
was varied by increasing the vortex density $f$ from small values
(when a continuum approximation is valid), to large values (when the
lattice structure is important). It is found that the groundstate is
well described by the Laughlin state for sufficiently small
$f$. Calculations of the wavefunction overlap indicate that the
(continuum) Laughlin wavefunction accurately describes the exact
groundstate for $f\lesssim 0.25$\cite{sorensen:086803,hafezi-epl}; a
computation of the Chern number associated with the topological order
show that this takes the 
value expected for the $\nu=1/2$ Laughlin phase for
$f\lesssim 0.4$\cite{hafezi-2007,hafezi-epl}. The excitation gap of
the Laughlin state is estimated from the numerics as a function of $f$
and $U/J$.  In the hard-core limit $U\gg J$, the gap above the
$\nu=1/2$ Laughlin state is found to be as large as $\simeq 0.25J$ at
$f\simeq 0.11$\cite{hafezi-2007}.

Exact diagonalization studies have been performed to investigate the
properties of small clusters of sites with open boundaries,
corresponding to rotating lattices in a bounded trap
geometry\cite{bhat:060405,bhat:063606,bhat:043601}. As a function of
the rotation rate, the groundstate shows transitions between states of
differing ``quasi-angular momentum'' -- the conserved quantum number
associated with the point-group symmetry of the lattice in the
rotating frame. These transitions represent the entry of vortices
into the system\cite{bhat:063606}.

Numerical constraints for the Bose-Hubbard model limit  
exact diagonalization studies to very small system sizes (typically the
number of particles is $N\leq 5$ in
Refs.\cite{sorensen:086803,hafezi-2007,bhat:060405,bhat:063606,bhat:043601}). This
makes the investigation of possible strongly correlated phases of
bosons on rotating lattices very challenging.  An analytic proposal
for strongly correlated quantum Hall states of bosons has been given
in Refs.\cite{palmer:180407,palmer:013609}. The construction is based
on the observation that the single-particle energy spectrum close to
rational fractions $f=l/q+ \delta$ ($l$, $q$ integer, $\delta\ll
1$) appears to form a set of $q$ states (which differ strongly on the
scale of the lattice constant $a$), but each of which is modulated by
the same Landau level wavefunctions over a large lengthscale of order
$\sqrt{1/\delta}$ times the lattice constant. For $\delta$ sufficiently small, one
can treat the system within a continuum theory, but now for $q$
species of bosons. The interspecies interactions are determined by the
microscopic wavefunctions.  At $f=1/2+\delta$ it is argued that in
the continuum limit ($n, \delta\ll 1$) one can construct exact
groundstates of the model with on-site interaction (\ref{eq:bh});
these are formed from the ``(221)'' quantum Hall
state\cite{Halperin83}  [see Eqn.(\ref{eq:221})] at the effective filling factor $\nu' \equiv
n/\delta = 2/3$.

The physics of strongly interacting bosons in rotating lattices is a
very rich and interesting topic, with the potential for exotic
strongly correlated phases. Much of the phase diagram remains to be
explored. It is an area where experiment is likely to raise surprises.

\section{Rotating Multi-Component Bose Gases}

\label{sec:multi-component}

The physics of ultra-cold atomic gases is very much enriched by the
possibility to trap and cool more than one atomic species. Mixtures of
degenerate atomic gases offer a vast array of different situations and
parameter regimes in which new and interesting phenomena can emerge.
Consistent with this wide range of parameters, there have been many
theoretical studies of vortices and vortex lattices in multi-component Bose
gases. Here, I describe only those studies that are most closely related to
the topics discussed in \S\ref{sec:rapid} for the one-component Bose gas.
Specifically, I shall focus on results in the 2D LLL regime (\ref{eq:weak})
appropriate for conditions of rapid rotation, and shall limit attention to the
two-component Bose gas and the spin-1 Bose gas.

\subsection{Two-Component Bose Gases}

\label{sec:twocomponent}

The bulk properties of an ultra-cold two-component Bose gas, with the
components labelled by $i=1,2$, are characterized by the two masses $M_i$ and
the set of all mutual two-body $s$-wave scattering lengths,
$a_{ij}$\cite{HoS96}. A two-component BEC is described by the
condensate wavefunction
\be \left(\begin{array}{c} \psi_1(\bm{r}) \\
    \psi_2(\bm{r})\end{array}\right) \,,
\label{eq:bec2}
\ee 
and the mean-field interaction
energy is \be E_I^{\rm GP} = \frac{1}{2} \int \sum_{i,j=1}^2 g_{ij}
|\psi_i|^2|\psi_j|^2 \;d^3\bm{r} \ee where $g_{ij} \equiv 2\pi\hbar^2
a_{ij}/\mu_{ij}$ with $\mu_{ij} \equiv M_iM_j/(M_i+M_j)$ the reduced mass\cite{esry}.

\label{values}

For the case of equal interactions $g_{ij}=g$ there is an exact symmetry of
the interaction energy under SU(2) rotations of the condensate wavefunction
(\ref{eq:bec2}). It is then convenient to parameterize the condensate
wavefunction in a manner that makes this invariance manifest, writing
\be
\label{eq:spinor}
 \left(\begin{array}{c} \psi_1(\bm{r}) \\
    \psi_2(\bm{r})\end{array}\right) = \sqrt{n}
e^{i\chi}\left(\begin{array}{c} \sin(\theta/2)\, e^{i\phi/2} \\
    \cos(\theta/2)\,e^{-i\phi/2} \end{array}\right) \ee where $n$ is the particle density,
$\chi$ a global phase, and $\theta$ and $\phi$ parameterize the spinor
co-ordinates by the polar and azimuthal angles on the surface of the sphere.
For SU(2) invariant interactions, $g_{ij}=g$, the interaction energy is
invariant under rotations of $\theta, \phi$, as well as of the global phase
$\chi$. For experimentally relevant situations, the interaction parameters can
be close to the SU(2) invariant condition.\footnote{For the
  $|F=1,m_F=-1\rangle$ and $|F=2,m_F=1\rangle$ states of $^{87}$Rb, the ratios
  are $g_{11}:g_{22}:g_{12} = 1.024:0.973:1.0$\cite{kempen,harber}. For the
  $|F=1,m_F=0\rangle$ and $|F=1,m_F=1\rangle$ states of $^{23}$Na these are
  $g_{11}:g_{22}:g_{12} = 1:1.035:1.035$\cite{StengerISMCK98}.} Departures from the SU(2) 
lead to weak anisotropies in the energy as a function of $\theta$ and $\phi$\cite{kasamatsu:043611}.

For a uniform 3D system, stability of the individual condensates requires
$g_{11}, g_{22} \geq 0$. Within this stable regime, depending on the
interactions the two components can be miscible or immiscible. The condition
for immiscibility, in which case the BEC phase separates into
spatially separated regions of component-1 and component-2, is
\cite{HoS96,PhysRevA.58.4836} \be g_{11}g_{22} < g_{12}^2 \,.
\label{eq:phasesep}
\ee

Two-component BECs have been formed from the trapping of two hyperfine states
of the same atomic species (for $^{87}$Rb\cite{PhysRevLett.78.586}, and for
$^{23}$Na\cite{PhysRevLett.82.2228}), and from the trapping of mixtures of two
different atomic species (for
$^{41}$K-$^{87}$Rb\cite{modugno,thalhammer:210402} and
$^{85}$Rb-$^{87}$Rb\cite{papp:040402}). Experiments on these two-component
condensates show evidence for both miscible and immiscible regimes compatible
with (\ref{eq:phasesep}), and have investigated the transition between these
regimes as the scattering lengths are
varied\cite{thalhammer:210402,papp:040402}.

When the two components are two hyperfine states of a single atomic species,
the masses are identical $M_1=M_2\equiv M$; when they correspond to two
different atomic species, then $M_1\neq M_2$. The rotational properties of the
two-component gas are very different in these two cases, which we therefore
discuss in turn.

\subsubsection{Rotating two-component BEC: Equal masses}

{\it Single Vortex States}

Experimental studies of quantized vortices in atomic BECs were initiated by
experiments on a two-component BEC\cite{MatthewsAHHWC99}. Coherent
excitation of a $^{87}$Rb BEC in a single hyperfine state (which could be
either one of the two trapped states) was performed in such a way as to
convert atoms into the other trapped hyperfine level with one unit of
circulation. The imprinted configuration therefore had a single quantized
vortex for the particles in the excited component. The phase of the rotating
component was directly imaged in an interference technique, to confirm the
quantized circulation. The subsequent evolution of the two-component system
was studied, showing evidence for differences in the dynamics and stability
depending on which of the two hyperfine levels was excited into a circulating
state.

The structure of the axisymmetric vortex state created in these experiments
is  easily written in the 2D LLL regime, for which the condensate
wavefunction takes the form\cite{muellertexture}
\be \left(\begin{array}{c} \psi_1 \\
    \psi_2\end{array}\right)
=  \left(\begin{array}{c} a \\
    b \zeta \end{array}\right) \times e^{-|\zeta|^2/4} e^{-z^2/2a_\parallel^2}
\,,
\label{eq:2componentvortex}
\ee where $a$ and $b$ are amplitudes whose ratio determines the relative
populations of the two components $N_1/N_2$. This wavefunction describes a state in which component-2 has one unit of angular momentum while
component-1 is at rest. Note that, despite the density suppression in the core
of the vortex of component-2, there remains non-zero particle density for
component-1 in the core region. This is a general feature of the vortices in
two-component BECs with repulsive interspecies interactions
$g_{12}>0$\cite{HoS96,chuivortex,jezek,leonhardt,muellertexture,kasamatsu:043611}.
Since the density does not vanish, one can use the representation
(\ref{eq:spinor}) to describe this vortex as a spin-texture
in the variables $(\theta,\phi,\chi)$. 
In terms of this description,
 the vortex configuration (\ref{eq:2componentvortex})
has a non-trivial topological classification.
It is referred to as a ``half quantum vortex''\cite{leonhardt} or
(two-dimensional) ``Skyrmion''\cite{muellertexture}. The properties of these
spin textures have been investigated in detail for interactions that (weakly)
break the SU(2) symmetry\cite{kasamatsu:043611}. The results show the
appearance also of vortex states which break the axial symmetry of the trap.

Beyond mean-field theory, there exist exact results for the case of SU(2)
invariant interactions, $g_{ij} = g$, in the 2D LLL limit. For small values of
angular momentum, $L\leq\mbox{min}(N_1,N_2)$, the results of exact
diagonalization studies on $N= N_1+N_2 \leq 8$ particles have been used to motivate a
conjecture for the exact (analytic) groundstates\cite{bargi:130403}. The
numerical studies show that the groundstate contains only single particle
states with angular momenta $m\leq 1$; such states can be shown to be
exact eigenstates of the SU(2) interactions. Similar 
conclusions hold also for the rotating spin-1 Bose gas with SU(3)
invariant interactions\cite{reijnderss1}, discussed in \S\ref{sec:spinone}.

{\it Many Vortex States}

\begin{figure}
\center\includegraphics[width=12cm]{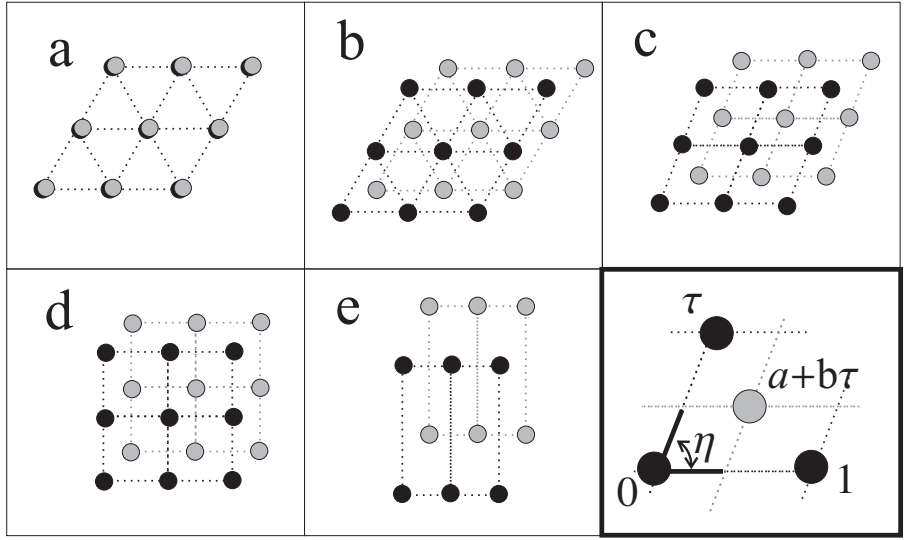}
\caption{
Vortex lattices in the mean-field LLL regime of a
two-component BEC for $g_{11}=g_{22}\equiv g$, and at
different values of $g_{12}/g$. The filled and shaded circles
show the locations of vortices in each of the components.
For $g_{12}/g< 0$ the interspecies interactions are attractive, so the vortex
cores of the two lattices attract each other. The vortex lattices of the two
components each form a triangular vortex lattice, with coincident vortices
(a).
For $g_{12}/g>0$ the interspecies interactions are repulsive, so the vortex
cores repel each other. A series of lattices is found which appear as
interlaced triangular lattices (b); interlaced rhombic lattices (c); and finally
interlaced square (d) or rectangular (e) lattices. At the SU(2) point $g_{12}/g=1$, the interlaced
rectangular lattices of the two components can be viewed as a triangular
lattice if the differences of the two components is ignored. The final panel
illustrates the variational parameters used in the study. 
[Reprinted figure from: E.J. Mueller and T.L. Ho,
 Phys. Rev. Lett {\bf 88}, 180403 (2002). Copyright (2002) by the American
 Physical Society. 
]
}
\label{fig:muellerho}
\end{figure}

Theoretical studies of lattices of many vortices were performed
by Mueller and Ho\cite{muellerho} within the mean-field LLL
regime.\footnote{Within mean-field theory the results are identical for the
  quasi-2D and 3D LLL regimes.} As for the case of the one-component BEC, in
the LLL limit the condensate wavefunction may be expressed in terms of the
positions of the vortices. However, for the two-component condensate
wavefunction there is a set of vortex locations for each of the two
components. For the case $g_{11}=g_{22} \equiv g\neq g_{12}$ studied in
Ref.\cite{muellerho} the two components are identical, so one expects each to have
the same vortex structure up to a relative translation.
 The groundstates were
determined as a function of $g_{12}/g$ for an infinite lattice.
A series of vortex lattice phases was found, with structures depicted
in
Fig.~\ref{fig:muellerho}. These results illustrate the very rich physics that
is possible in a rotating two-component BEC.
For $g_{12}/g> 1$, corresponding to the regime of phase separation
(\ref{eq:phasesep}) for the non-rotating gas, the groundstate consists of interlaced rectangular arrays of vortices.
For large $g_{12}/g$ the vortices become closely
spaced along one direction, leading to vortex sheets; the particle
density then tends towards an array of parallel stripes in which one component
has low density and the other a high density. 

The vortex lattice states have been studied beyond the LLL and including the
effects of the trap\cite{kasamatsuprl,woo:031604}. The results show
qualitatively similar phases to those described above. Stripe-like states
(vortex sheets) appear in the regime of phase separation. Dynamical modes have
been studied in the uniform system in the LLL\cite{kecceli:023611} and beyond
the LLL in the trap geometry\cite{woo:031604}.

Experimental studies of the vortex lattices in rotating two-component
$^{87}$Rb BECs have been reported in Ref.\cite{schweikhardspin}. The
interspecies interactions are repulsive $g_{12}>0$ so the vortices in the two
components repel each other. The interactions are close to the SU(2) invariant
values $g_{11}\simeq g_{12}\simeq g_{22}$ (see footnote 1 on page
\pageref{values}), but just inside the regime of phase separation
(\ref{eq:phasesep}). The experimental results showed evidence for the
appearance of a stable vortex lattice phase with two inter-penetrating lattices
of approximate square symmetry, consistent with the phases found in
Refs.\cite{muellerho,kasamatsuprl}.

Exact results for the interacting two-component system with many vortices are
very limited.

In the 2D LLL regime, the groundstate at high angular momentum can be determined analytically for repulsive contact
interactions. Recall that, for spinless bosons with repulsive contact
interactions, the exact groundstate at high angular momentum is the Laughlin state (\ref{eq:laughlin})
with total angular momentum $L=N(N-1)$ corresponding to filling factor
$\nu=1/2$. For the two-component case with $g_{ij}> 0$, and for equal numbers
of atoms in the two components $N_1=N_2=N/2$, the exact zero-energy groundstate is 
 an incompressible liquid  described by the
``(221)'' bosonic Halperin state\cite{Halperin83}
\begin{equation}
\Psi_{\rm Halperin}^{(221)}(\{\zeta_i\}) \propto 
\prod_{i<
  j}(\zeta^{(1)}_i-\zeta^{(1)}_j)^2
\prod_{i<
  j}(\zeta^{(2)}_i-\zeta^{(2)}_j)^2
\prod_{i,j}(\zeta^{(1)}_i-\zeta^{(2)}_j)
 \,,
\label{eq:221}
\end{equation}
where $\zeta_i^{(1)}$ are the co-ordinates of the $N/2$ particles in
component-1 and $\zeta_i^{(2)}$ the co-ordinates of the $N/2$
particles in component-2. 
This state can be viewed as a
fluid of pairs of particles forming spin singlets.
This function vanishes when the co-ordinates of any two particles coincide, so
it is a zero-energy eigenstate of
the contact interaction Hamiltonian; it is the groundstate provided the interactions are
repulsive $g_{ij}>0$. The total angular momentum is $L=3N^2/4-1$; which
corresponds to a uniform fluid state with a total filling factor $\nu_1+\nu_2 =
2/3$, therefore denser than the Laughlin state. 

A series of non-abelian states that can describe two-component Bose systems in
the 2D LLL was proposed in Ref.\cite{nass}. These ``non-abelian spin-singlet''
(NASS) states form a sequence labelled by $k=1,2,\ldots$, and are the exact
groundstates of the two-component Bose gas for a $k+1$-body contact
interaction (\ref{eq:k+1}), at filling factors $\nu = 2k/3$.\footnote{The more
  general form is $\nu = 2k/(2kq+3)$ with $q$ an even/odd integer for
  bosons/fermions\protect\cite{nass}.} For $k=1$ the state in this sequence is
the above Halperin (221) state; for $k\geq 2$ the NASS states are {\it
  non-abelian} phases which are spin-singlet generalizations of the Moore-Read
and Read-Rezayi states. Owing to the importance of the Moore-Read and
Read-Rezayi phases in describing the phases of rotating spinless bosons with
realistic two-body interactions (see \S\ref{sec:stronglycorrelated}), it would
be interesting to explore the relevance of NASS states for a two-component
Bose gas with realistic two-body interactions.

\subsubsection{Rotating two-component BEC: Unequal masses}

The rotational properties of a two-component BEC formed from two atomic
species raises additional interesting features. Since the atomic masses
differ, $M_1\neq M_2$, if the two components were to rotate (on average) at
the same rotation frequency $\Omega$, then by the Feynman criterion (\ref{eq:feynman}) the mean
vortex densities in the two components would be unequal \be \frac {n_{\rm
    v}^{(1)}} {n_{\rm v}^{(2)}} = \frac{M_1}{M_2}\,.
\label{eq:unequal}
 \ee This imbalance of the average vortex
densities competes with an energetic gain from interspecies interactions which
is obtained by making the vortex densities equal (or commensurate). 

The multi-vortex states of a two-component BEC with unequal masses have been
studied theoretically in Ref.\cite{barnettrefael}. The interspecies
interaction is assumed to be attractive $g_{12}<0$, such that the cores of the
vortices in the two components attract. 

For strong attractive interactions, the vortices pair, and the system forms a
locked state in which the average vortex densities are equal. How does one
reconcile the equality of the vortex densities in this phase with the above
relation (\ref{eq:unequal})? Consider a system that has come to equilibrium at
a drive frequency $\Omega$, and that forms a locked vortex lattice of pairs of
vortices. The locations of the vortex cores must rotate uniformly at the drive
frequency $\Omega$, and they will have some density $n_{\rm v}$ that is
determined by the energetics (including the inter-vortex interactions). Since
the vortex density is fixed to $n_{\rm v}$, the individual fluid components
are flowing with average flow fields that lead to rotation of the two
fluids with rates $\Omega_i = h n_{\rm v}/(2M_i)$. Clearly $\Omega_1\neq
\Omega_2$, and, in addition, neither $\Omega_1$ nor $\Omega_2$ need match the drive frequency
$\Omega$. Thus, there is a relative motion of the fluids with respect to the
vortex cores. This relative motion leads to Magnus forces on the vortices. One
reconciles these features by noting that 
the Magnus forces are exactly balanced by the inter-particle interactions that act
to pin the vortices to each other\cite{barnettrefael}, as they must be for
consistency with the statement that the vortex lattice is at equilibrium at
the drive frequency $\Omega$.

For weak attractive interspecies interactions the vortices separate and the
vortex densities are set by those expected for independent condensates, with
$n_{\rm v}^{(1)}\neq n_{\rm v}^{(2)}$ (\ref{eq:unequal}). The transition
between these two regimes is controlled by the parameter
$|g_{12}|\bar{n}/(\hbar\Omega)$ where $\bar{n}$ is the mean particle density.
Thus, in a trap geometry, the inhomogeneity of density leads to spatial
variation: a region of locked vortex lattice at centre of the trap, and 
an unlocked region at the outer edge\cite{barnettrefael}.

Related issues arise in models of rotating Bose gases close to a Feshbach
resonance\cite{CooperFesh04}. The atoms and molecules (formed from resonant
association of pairs of atoms) have unequal masses $M$ and $2M$, leading to
different mean vortex densities for the molecular and atomic
condensates\cite{woo:120403}. It is found that the coherent coupling can lead
to the formation of states in which the atomic and molecular vortices form
units resembling a carbon dioxide molecule. The arrangement of the
orientations of these units within an approximately triangular array leads to
many near degenerate configurations and glass-like behaviour. A range of other
interesting vortex lattice phases is identified\cite{woo:120403}.

\subsection{Spin-1 Bose Gas}

\label{sec:spinone}

Spinor atomic gases may be realized by trapping a high spin atom in an optical
trap, which does not lift the spin degeneracy. The inter-particle interactions
are spin-rotationally invariant, and for spin $S=1$ must take the
form\cite{ohmimachida,Ho98} \be H_{\rm int} = \sum_{i<j} \delta({\bm
  r}_i-\bm{r}_j)\left[ c_0 + c_2 \bm{S}_i\cdot\bm{S}_j \right]
\label{eq:s1int}
\ee where $c_0$
and $c_2$ are the only two interaction parameters. In terms of
the
two
$s$-wave scattering lengths, denoted $a_{\rm s}^{0}$ and $a_{\rm s}^{2}$ for scattering in the two scattering
channels with total spin $S=0,2$, the coupling constants in
Eqn.(\ref{eq:s1int}) are
$c_0 = (g_0+2g_2)/3$, $c_2 =
(g_2-g_0)/3$, where $g_S = 4\pi \hbar^2 a_{\rm s}^{S}/M$.\footnote{Since the
  components are different hyperfine states of the same atomic species, the mass $M$ is the same for all  components.}

For $c_2<0$, the system acts to maximize the total spin, which is achieved by
aligning all spins; this is referred to as the ``ferromagnetic'' state. It is
characterized by a non-zero value of the expectation value of the local spin
density $\langle \hat{\bm{S}} \rangle$. The breaking of spin-rotational
symmetry leads to a manifold of groundstates related by SO(3)
transformations\cite{Ho98}. For $c_2>0$, the system acts to minimize the total
spin; this is the ``polar'' state. It has vanishing spin density, $\langle
\hat{\bm{S}} \rangle = 0$. Still, it breaks spin-rotational invariance, with a
groundstate manifold described by $S^1\times S^2/{Z}_2$\cite{Ho98,zhou}. Thus,
in contrast to the two-component condensate, which is characterized by a
global phase and a local spinor direction (\ref{eq:spinor}), the topology
of the spin-1 Bose gas is much richer.

Both ferromagnetic and polar spin-1 BECs have been studied experimentally,
with $^{87}$Rb ($c_2<0$)\cite{PhysRevLett.87.010404} and with $^{23}$Na
($c_2>0$)\cite{Stamper-KurnACIMSK98,StengerISMCK98,PhysRevLett.82.2228}. In
both of these experimental systems, the spin-dependent interaction is weak,
with $|c_2/c_0|\ll 1$.
Consequently, in spin-textured configurations such as vortex lattices, it can
be energetically favourable for the condensate to have mixed character, with regions in which the
local order parameter is of ferromagnetic character (with non-zero average
spin) and regions where it is of polar character (with vanishing average spin).

{\it Single Vortex States}

The resulting textures for single vortices of a spin-1 BEC have been studied
theoretically within mean-field
theory\cite{PhysRevLett.83.4677,PhysRevLett.89.030401,PhysRevA.66.053604}.
Within the LLL mean-field regime the description of these states
simplifies\cite{reijnderss1,muellertexture}. However, a wide range of textures
can still appear, with forms that depend on $c_2/c_0$ and on the total angular
momentum. The angular momentum is carried by spin-textures, which can be of
the form of ``skyrmions'' of the ferromagnetic order or of
``$\pi$-disclinations'' of the polar phase.

In the 2D LLL regime, the {\it exact} groundstates at small angular momentum,
$L\leq N$, have been found for the case $c_2=0$ in which case the interactions
have a full SU(3) symmetry\cite{reijnderss1}. There exist exact eigenstates of
the Hamiltonian for which the only occupied single particle states are those
with angular momenta $m\leq 1$ (for $0< L\leq 2N/3$), or those with $m\leq 2$
(for $2N/3<L\leq N$). Exact diagonalization studies confirm the states to be
the exact groundstates for all system sizes studied.

{\it Many Vortex States}

\begin{figure}
\center\includegraphics[width=14cm]{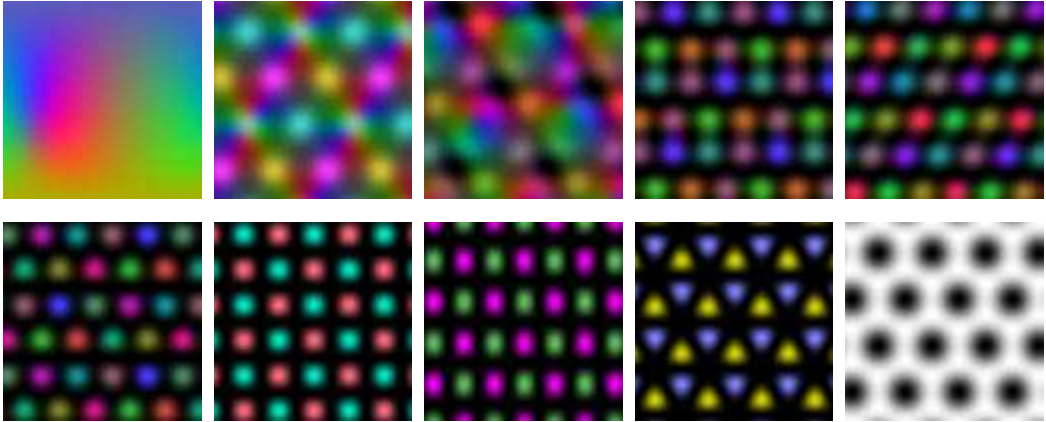}
\caption{(Colour.)
Vortex lattices in the mean-field LLL regime for a spin-1 BEC at different values
of $c_2/c_0$.
 The first panel shows the
Hammer-Aitoff projection of the colours on the spin-sphere, in which the
intensity encodes the size of the spin  $|\langle \bm{\hat{S}}\rangle|$. The images
are the spatial distributions of the expectation values of the local spin density
for the vortex lattices at $c_2/c_0 = -0.1, -0.05, 0.01, 0.016, 0.04, 0.1,
0.54, 0.7, 0.9$. In the last picture, $c_2/c_0=0.9$, the spin vanishes,
and so the intensity shows the particle density, which forms a triangular
Abrikosov lattice. 
[Reprinted figure from: J.W. Reijnders, F.J.M. van Lankvelt, K. Schoutens, and N. Read,
 Phys. Rev. A {\bf 69}, 023612 (2004). Copyright (2004) by the American
 Physical Society. 
]
}
\label{fig:reijnderss1}
\end{figure}

The vortex lattices of a uniform $S=1$ Bose gas have been studied within
mean-field theory for the polar regime $c_2>0$\cite{PhysRevA.66.061601}. 
 A rich series of
phases of five different lattices of textured vortices was found, depending on
$c_2/c_0$. The vortex lattices
were studied  within 
LLL mean-field theory 
in Ref.\cite{reijnderss1}
for both positive and negative $c_2$. Eight different symmetries of
vortex lattice were found, which are illustrated in 
Fig.~\ref{fig:reijnderss1}.
The phases involve arrangements of various spin-textures of the spinor order,
with lattices of Skyrmions, $\pi$-disclinations and conventional vortices\cite{reijnderss1}.

Turning to consider rapidly rotating gases beyond mean-field theory, there
exist some interesting exact results.
In the 2D LLL regime appropriate for rapid rotation, the groundstate at high
rotation rate can again be determined analytically for repulsive contact
interactions, $g_0,g_2>0$. It is an incompressible liquid state that may be
viewed as a generalization of the Halperin (221) state, Eqn.(\ref{eq:221}). The exact (zero-energy) groundstate
can be viewed as a liquid of clusters of triplets of particles in the three
spin projections of spin-1\cite{homuellers1,ParedesZC02}, and is the
SU(4)$|_{1}$ state in the sequence of Ref.\cite{ReijndersvSR02}. For $N$
divisible by three (such that all particles can form complete clusters) the
angular momentum is $L=N(2N/3-1)$, which corresponds to a filling factor
$\nu=3/4$. 
These clustered states may be generalized to an arbitrary number of components
$n$; they form the exact groundstates of repulsive contact interactions at
filling factors $\nu = n/(n+1)$\cite{ParedesZC02,ReijndersvSR02}.

A series of generalized Read-Rezayi states has been proposed for rotating
$S=1$ bosons\cite{ReijndersvSR02}. The states in this sequence, labelled
SU(4)$|_{k}$, have the property that their wavefunctions vanish when the
co-ordinates of $k+1$
particles (in any spin component) coincide; they are therefore exact groundstates
of repulsive $k+1$-body contact interactions (\ref{eq:k+1}) between the atoms.
They describe incompressible liquid phases at $\nu = 3k/4$\footnote{The more
  general forms are $\nu = 3k/(3kq+4)$ for SU(4)$|_{k}$  and $\nu =
  k/(kq+1)$ for SO(5)$|_{k}$
with $q$ an even/odd integer for
  bosons/fermions\protect\cite{ReijndersvSR02}.} involving clusters of $3k$ particles. For $k=1$ this is the
clustered state\cite{homuellers1,ParedesZC02,ReijndersvSR02} described above, which is the exact groundstate for two-body
contact repulsion between all particles. For $k > 1$ the phases have
quasiparticle excitations which obey {\it non-abelian} exchange statistics.
Although not the exact wavefunctions for two-body interactions, these
represent interesting trial wavefunctions for possible quantum Hall phases of
rotating spin-1 Bose gas at $\nu >3/4$. Refs.\cite{ReijndersvSR02,reijnderss1}
report the results of numerical studies of the exact groundstates in this regime. A large
overlap of the exact groundstate (of the two-body interactions) with the
SU(4)$|_{2}$ state is reported at $\nu=3/2$, which is evidence of the
appearance of this non-abelian phase.

A second generalization of the Read-Rezayi states, labelled SO(5)$|_{k}$,
and describing $S=1$ bosons at filling factor $\nu = k$ is proposed in   
Ref.\cite{ReijndersvSR02}. These states are exact zero-energy eigenstates
of a $k+1$-body contact interaction, for a special choice of $c_2/c_0$.  
They involve clustering of groups of $2k$ particles. These phases have the 
additional interesting feature of involving fractionalization both of
atom number or ``charge'' (as is common in FQH states) and also of spin,
their quasiparticle excitations having atom number $1/2$ and spin $1/2$.

\section{Rapidly Rotating Fermi Gases}

\label{sec:fermions}

The possibility to cool Fermi gases to quantum degeneracy
$T<T_F$\cite{B.DeMarco09101999} and to achieve regimes of strongly correlated
quantum phases of two-component Fermi gases\cite{blochdz}, raises interesting
questions concerning the properties of degenerate Fermi gases under rotation.

For a one-component Fermi gas $s$-wave interactions vanish due to the
Pauli principle.  For an ultra-cold atomic gas (with low relative
momentum $\hbar k$ between colliding particles) the dominant
interactions are in the $p$-wave channel. The cross-section
is typically very small, vanishing as $\sigma_1 \sim V^2 k^4$ as $k\to
0$ where the ``scattering volume'' $V$ is set by the inter-atomic
forces and therefore typically $V \ll 1/k^3$ for a cold
gas. Nevertheless, it is possible to introduce significant
interactions between the identical fermions. This can be achieved by
using atoms or molecules that have strong dipolar
interactions\cite{BaranovPS}; or by use of a ``Feshbach resonance'' to
enhance the $p$-wave interactions\cite{PhysRevLett.90.053201}.

For a two-component Fermi gas, there can be $s$-wave interactions between the
two components. Moreover, by use of a Feshbach resonance, these
interactions can be made very large. In this way, experiments have
been able to study the evolution between the BCS paired state (for weak
attraction between the two species) and the BEC state (of small
molecules formed from tightly bound pairs of the atoms),
through a regime of strong correlations\cite{blochdz}.

In this section we discuss theoretical predictions for the effects
of rapid rotation on one- and two-component Fermi gases.

\subsubsection{Non-Interacting Fermi Gas}

The single particle states for a fermion in an axisymmetric harmonic
trap are the same as those discussed in \S\ref{sec:2dlll}. However, we
shall now consider states beyond the lowest Landau level, so need to
generalize the discussion.

Working in the rotating frame, at angular frequency $\Omega$, the
single particle energies of particle in an axisymmetric harmonic well
are \be E^\Omega_{n_\perp,m,n_\parallel} = \hbar\omega_\perp(2n_\perp + |m| +1) - \hbar\Omega
m + \hbar\omega_\parallel (n_\parallel + 1/2) \ee where $n_\perp = 0,1,2\ldots $
and $m= 0, \pm 1, \pm 2$ are the radial and angular momentum
quantum numbers for motion in the plane perpendicular to the rotation
axis, and $n_\parallel = 0,1,2\ldots$ is the subband index for motion parallel
to the rotation axis.  If there are additional internal degrees of
freedom (hyperfine levels), there is one such orbital state for each
level.

In general, one should take account of confinement in all directions,
owing to the inhomogeneity of the trapped gas.  However, as in the
case of bosons, it is useful to consider the limit $\Omega \simeq
\omega_\perp$ in which the system becomes locally homogeneous in the
plane perpendicular to the rotation axis.  The spectrum can then be
written \be E^{\omega_\perp}_{n,m,n_\parallel} = 2 \hbar\omega_\perp(n +1) +
\hbar\omega_\parallel (n_\parallel + 1/2) \ee where now $n=0,1,\ldots$ is the
Landau level index and the angular momentum takes the values $m=
-n,-n+1,-n+2$.  [The relationship is obtained by noting $n=n_\perp +
(|m|-m)/2$.]  Note that the states in a single Landau level (labelled
by $m$) are degenerate.  The particle density sets the highest
occupied Landau level (maximum $n$) and subband (maximum $n_\parallel$).

The properties of the gas depend on the aspect ratio of the trap,
$\omega_\perp/\omega_\parallel$. It is of interest to consider the two
limiting cases: the quasi-2D regime, $\omega_\parallel\gg
\omega_\perp$; and the 3D regime, $\omega_\parallel\ll
\omega_\perp$.

In the quasi-2D regime, the particle density is sufficiently small or
the axial confinement is sufficiently strong that only
$n_\parallel=0$ is occupied, i.e. $\mu < E^{\omega_\perp}_{0,0,1}$. One can then define the
filling factor as before, and as is conventional in the FQHE systems, by \be
\nu \equiv \frac{n_{\rm 2d}}{n_{\rm v}} = \frac{n_{2d}h}{2M\omega_\perp} \,.\ee 
In this quasi-2D regime, even a non-interacting Fermi gas shows a
series of incompressible states. These arise due to the fact that the
spectrum is discrete, with a set of highly degenerate states (over
$m$) at $E^{\omega_\perp}_{n,m,0}$.  These appear each time $\nu$ is an integer and a
Landau level is filled. (At this point the chemical potential has a
discontinuity equal to the Landau level spacing, $2\hbar\omega_\perp$.) If there
are additional internal degrees of freedom these incompressible states
occur at those filling factors for which atoms of all species fill an
integer number of Landau levels.

The 3D regime (\ref{eq:q3dregime}) is relevant for anisotropic traps
$\omega_\parallel \ll \omega_\perp$. Here, for each Landau level state,
the motion along the rotation axis can be viewed as a quasi-1D gas,
and the energy is more conveniently written as \be E^{\omega_\perp}_{n,m,k_\parallel} = 2
\hbar\omega_\perp(n +1) + \frac{\hbar^2k_\parallel^2}{2M} \,,
\label{eq:q3dspectrum}\ee where $k_\parallel$ is the wavevector along the rotation
axis. In this case, the system does not show incompressibility.
However, there is a $1/\sqrt{E}$ divergence in the density of states
at the position of each Landau level, associated with the quasi 1D
motion.

The existence of Landau level quantization is predicted to lead to 
signatures in the density distribution for a trapped Fermi gas, when
the rotation frequency is close to the trap frequency $\Omega \simeq
\omega_\perp$\cite{hociobanu}.  The experimental signatures depend on
the aspect ratio of the trap: they are strongest for the quasi-2D
limit, where there is strict incompressibility leading to plateaus in
the density distribution; but features survive also in the 3D
case due to the singularities in the density of
states\cite{hociobanu}.

\subsection{One Component Fermi Gas}

Under conditions of rapid rotation, and at sufficiently low particle
density, the Fermi gas will be in the quasi-2D LLL regime ($n=n_\parallel=0$).
At the single particle level, the system is then degenerate when the
filling factor is non-integer, $\nu<1$. The groundstate is determined by
minimizing the inter-particle interactions. 

In the following I describe the results of studies of the effects of
two important forms of inter-particle interactions on the properties of
a rapidly rotating one component Fermi gas.

\subsubsection{Dipolar Interactions}

The properties of a Fermi gas in the 2D LLL have been investigated for the case of
dipolar interactions. Antisymmetry of the wavefunction requires the relative
angular momentum of any two particles to be odd, so the inter-particle
interaction is characterized by the Haldane pseudo-potentials
(\ref{eq:haldanedip}) at odd $m$. The relevant energy scale can be large for
fermionic molecules with electric dipole interactions\cite{baranov:070404},
leading to the possibility of stable correlated phases in rapidly rotating
dipolar Fermi gases.

The possibility to use dipolar interactions to stabilize the $\nu=1/3$
Laughlin state was described in Ref.\cite{baranov:070404}.  The energy
gap for creation of a quasi-hole excitation (which  contributes to the
incompressibility of the state, see \S\ref{sec:incompressible})
  was estimated to be $\Delta_{\rm qh} =
(0.9271\pm 0.019)C_d/\ell^3$ for a trap with spherical symmetry
($\omega_\perp=\omega_\parallel)$\cite{baranov:070404}.

The groundstates of small numbers of dipolar-interacting fermions
($N\leq 10$) in the 2D LLL in a harmonic trap have been investigated
numerically\cite{osterloh:160403} over a wide range of different
rotation rates (angular momenta). The results were analysed within the
framework of the composite fermion construction, in which the
interacting Fermi system is represented by non-interacting composite
fermions\cite{jainoriginal,dassarmapinczuk}. This construction works well for
analogous studies of harmonically trapped electrons in the FQHE regime
(i.e. Coulomb forces instead of dipolar forces)\cite{jainkawamura}.
As described in \S\ref{sec:cfdots} a generalization of the composite
fermion method to treat rotating bosons in harmonic traps also works
well\cite{CooperW99}.  For the case of dipolar interacting
fermions, some discrepancies are reported from the apparent
predictions of CF theory\cite{osterloh:160403}. However, these appear to arise from an incomplete application of 
the composite fermion theory for some of the groundstates found in numerics.\footnote{Most of
the discrepancies reported in Ref.\protect\cite{osterloh:160403} arise from the neglect of a set of the
``compact'' CF states that account for the physics in other
situations\cite{jainkawamura,CooperW99}. Specifically, for $N=10$, the
CF theory predicts a set of compact states at $L =
73,80,85,93,95$; these values fill out most of the missing entries in
Table (4) of Ref.\protect\cite{osterloh:160403}. The results reported
therefore appear consistent with CF theory describing the high angular
momentum states accurately.} To check the accuracy of the CF
description of these states, or to determine whether the states are of
a different character\cite{osterloh:160403}, it would be useful to
compute overlaps of the exact groundstates with the trial composite
fermion states\cite{jainkawamura,CooperW99}.

The fact that dipolar interactions are long-ranged allows the
stabilization of a Wigner crystal state at very low filling factor
$\nu\ll 1$. The transition from Laughlin states to the Wigner crystal
phase is predicted to occur at $\nu < 1/7$\cite{baranov:200402}, close to
the value at which the transition is believed to occur in FQHE systems.

In the case of a quasi-2D fermionic gas with filling factor
larger than 1, $\nu > 1$ -- {\it i.e.} when more than one Landau level
is occupied -- the nature of the groundstate of the particles in the
partially occupied Landau level (let us label this level by $n$) are
determined solely by the interactions. However, now  the Haldane
pseudo-potentials should be computed in this $n$th Landau level. Consequently,
the groundstate at a filling factor $\nu = n + \nu'$ ($0\leq \nu' \leq
1$) can differ from that at $\nu=\nu'$.  Studies in the quantum Hall
literature show that Hartree-Fock mean-field theory is accurate when
$n\gg 1$\cite{MoessnerC96}. For large filling factors, the mean-field
states of dipolar atomic Fermi gases are
expected\cite{komineas:023623} to be qualitatively similar to those of
electrons in semiconductor
systems\cite{KoulakovFS96,MoessnerC96}. Depending on the filling of
the partially occupied Landau level, the groundstate will be a
``stripe'' or ``bubble'' phase\cite{KoulakovFS96,MoessnerC96} of similar
translational symmetry to the mean-field states discussed for bosons in
\S\ref{sec:dipolar}.

\subsubsection{$p$-wave Feshbach resonance}

Although weak at low collision energy, identical fermions experience
non-zero $p$-wave interactions. These interactions can be resonantly
enhanced in the vicinity of a Feshbach resonance -- i.e. close to the
threshold for formation of a new bound state in the $p$-wave channel.
Such resonances have been studied in ultra-cold gases of
$^{40}$K\cite{PhysRevLett.90.053201,gunter:230401,gaebler:200403} and
$^6$Li\cite{PhysRevA.70.030702,schunck:045601,inada:100401}.

Close to a $p$-wave resonance, at low collision energy [when the wavevector 
$\mbox{max}(k_F,1/\ell)$ is sufficiently small]
the $p$-wave
scattering
phase shift takes the form\be \delta_1(\bm{k}) = (1/3)k^3 a_1^3 \, .  \ee
This defines the scattering length $a_1$. The quantity $a_1^3$ is
often referred to as the ``scattering volume''\cite{gurariereview},
the modulus of which  diverges at the position of the resonance.

In the lowest Landau level, the interaction may be represented by the
pseudo-potentials.  For the $p$-wave interaction, the only non-zero
pseudo-potential is in the $m=1$ channel, with\cite{regnaultpwavefermions} \be  V_1 =
\sqrt{\frac{2}{\pi}}\frac{\hbar^2}{M} \frac{a_1^3}{m a_\parallel
\ell^4}\,.\ee This two-body interaction is the special case for which the
$\nu=1/3$ Laughlin state is exact\cite{haldanehierarchy}. For
$\nu<1/3$ the groundstate is highly degenerate, and may be viewed as a
gas of $1/3$ anyons of the Laughlin state.

Analytic results do not exist for $\nu>1/3$,\footnote{One exception is
the trivial statement that $\nu=1$ is a fully filled Landau level.}
or for the excitation spectrum above the $\nu=1/3$
groundstate. Numerical studies on the
sphere\cite{regnaultpwavefermions}  have investigated the groundstates and low-energy excitations above the Laughlin $\nu=1/3$ state and the ``Jain sequence''\cite{jainoriginal} states at $\nu=2/5,3/7$. These studies show
that in the thermodynamic limit $N\to \infty$ the excitation gaps appear to
converge to non-zero values, establishing these states as incompressible
and determining numerical estimates values for the
excitation gaps. The overall energy scale is set by the $V_1$
pseudo-potential. Experimental observation of these states will require
cooling to temperatures less than $\sim V_1/k_B$.  The issue of stability of
a degenerate one-component Fermi gas close to a $p$-wave
resonance\cite{gaebler:200403,levinsen:210402,inada:100401} is an
interesting open question, as is the effect of rapid rotation on this
stability.

\subsection{Two-Component Fermi gas}

\subsubsection{BEC-BCS crossover}

Degenerate gases of two species of fermions may be formed either from
two atomic species, or from two hyperfine states of the same fermionic
atom\cite{varenna-ketterle,varenna-grimm}.  I shall denote the two
components as spin-up and spin-down. For simplicity I shall assume
that the two components have equal densities and equal masses $M$.

Since intra-species $s$-wave interactions are excluded by the Pauli
principle, the dominant interactions at low energy are the inter-species
$s$-wave interactions.  By use of a Feshbach resonance, these
interactions can be made strong. A Feshbach resonance arises when,
close to the energy of the two dissociated atoms, there exists a bound
molecular state of the two atoms in some different hyperfine
levels. Owing to the different hyperfine structure of the bound state
from the dissociated atoms, by varying a magnetic field one can cause
the energy of the bound state to cross the energy of the two
dissociated atoms. Hybridization of the molecular level with the
unbound atoms leads to a Feshbach resonance in the $s$-wave scattering
amplitude.

\begin{figure}
\center\includegraphics[width=6cm]{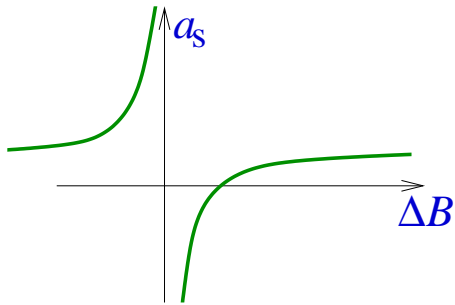}
\caption{Schematic diagram of the variation of the $s$-wave scattering length 
close to a Feshbach resonance as a function of the detuning $\Delta B = (B-B_{\rm res})$.}
\label{fig:feshbach}
\end{figure}

For a detailed discussion of the physics of the resonant scattering,
the reader is referred to Ref.\cite{blochdz}. In brief, the
consequence of the formation of the bound state is that the $s$-wave
scattering length $a_{\rm s}$ passes through a divergence, \be a_{\rm s}(B)
\propto \frac{1}{B_{\rm res}-B} \ee as illustrated in
Fig.~\ref{fig:feshbach}.  On approaching the resonance from the side 
on which the bound state has not formed ($B>B_{\rm
res}$), the scattering length becomes large and negative; approaching
the resonance from the side on which the bound state has
formed ($B<B_{\rm res}$), the scattering length becomes large and
positive.
For short-range interactions, the properties of the (balanced)
two-component Fermi gas close to the Feshbach resonance depend on the
parameter $k_Fa_{\rm s}$. [We denote the Fermi energy $\epsilon_F = \hbar^2
k_F^2/(2M)$.] 

The nature of the groundstate can be understood in the
limiting cases far from the resonance, when $k_F|a_{\rm s}|$ is small.  

When $a_{\rm s}$ is small and positive, there exists a bound-state even for one pair
of atoms. The groundstate of a many particle system consists of tightly bound
pairs of atoms of the two species; the resulting small bosonic molecules
behave as a weakly interacting Bose gas and form a Bose-Einstein condensate at
low temperatures. This is the ``BEC side'' of the resonance. 

When $a_{\rm s}$ is small
and negative, the weak attractive interaction causes a BCS instability of the
Fermi surface, leading to a BCS paired superfluid. There is an energy gap
that, in this BCS limit of weak interaction, is exponentially small \be \Delta
\sim \epsilon_F\; \exp\left(-\frac{\pi}{2k_F |a_{\rm s}|}\right) \,.\ee This is the ``BCS side''
of the resonance.

As the resonance is swept, such that $1/k_Fa_{\rm s}$ varies from the BCS to
BEC regime, it is believed that the groundstate of the Fermi gas
evolves smoothly. This ``BEC to BCS
crossover'' has been confirmed in recent experimental studies of several
groups\cite{blochdz}.

Cold atomic gases allow additional novel physics to be studied  by
imposing an imbalance between the populations of the two
components. In conventional superconducting systems, a population
imbalance can be achieved by using the Zeeman splitting in a magnetic
field.  However, a magnetic field also induces orbital effects,
precluding a study of the effects of the density imbalance alone. In
atomic gases the density imbalance can be achieved without any orbital
effects\cite{Zwierlein01272006,Partridge01272006}.

New physics can also emerge in atomic gases in the complementary
situation: that is, to study purely the {\it orbital} effects of a
magnetic field, without inducing any population imbalance.  This can be
achieved by setting the densities of the two hyperfine levels to be
equal, and introducing the orbital effects of the magnetic field by
rotation. This is the situation that shall be described in
this section.

The effects of moderate rotation are well understood. At all points across the
crossover, the low energy phase is a condensate, and responds as a superfluid
of pairs of fermions. Rotation causes the formation of an array of vortices
with density, ${2 (2M)\Omega}/{h}$, which is the
Feynman result (\ref{eq:feynman}) with the mass, $2M$, of the pair of
particles. The formation of a lattice of quantized vortices has been
demonstrated in experiments described in Ref.\cite{Zwierlein}. The results
reveal the superfluid nature of the system across the whole range of the
crossover regime, and are consistent with vortices having a quantum of
circulation of $h/(2M)$ associated with condensation of pairs of fermions.

\subsubsection{Effects of Rapid Rotation}

One expects the effects of rapid rotation to be rather different in
the BEC and BCS sides of the resonance. 

Far on the BEC side, the effects of rapid rotation are 
similar to those discussed above for rapidly rotating bosons. In the
quasi-2D regime, the mean-field vortex lattice will survive up to
vortex densities at which the filling factor (\ref{eq:fillingfactorN})
is of order one, at which point there is a transition to
strongly correlated states. In the 3D regime, the transition to the
strongly correlated regime occurs at high vortex density according to Eqn.(\ref{eq:3dcriterion}).

Far on the BCS side, the effects of rotation are similar to the
orbital effects of magnetic field on type-II superconductors.
For conventional superconductors, the upper critical field is well
described by the semi-classical approximation of
Gorkov\cite{Gorkov59,Gorkov60,werthamer}.  Translating to the quantities of interest to
us, this predicts the transition to a normal state when\footnote{Note
that, in the BCS limit, this is a much lower rotation rate than that required
to put one flux quantum through one Cooper pair, which would be
$\hbar\Omega\gtrsim \Delta$.}  \be\hbar\Omega \gtrsim 4.24 \Delta^2/\epsilon_F\,.\ee
One can understand this semi-classical result in terms of a balance of
energies. On the one hand the condensation energy per particle on
forming the superfluid is of order $\Delta^2/\epsilon_F$. This should be
compared with the kinetic energy of the superfluid flow. Since the
superfluid cannot rotate as a rigid body, in the rotating frame of
reference the superfluid has a non-zero      kinetic energy.  To estimate
this kinetic energy, note that the vortex spacing $a_{\rm v}\sim
(\frac{2(2M)\Omega}{\hbar})^{-1/2} $ sets a typical velocity
$\hbar/Ma_{\rm v}$, so the kinetic energy per pair is $M
(\hbar/Ma_{\rm v})^2\sim \hbar\Omega$.  The semi-classical formula
states that once this kinetic energy is larger than the condensation
energy, it does not pay to form the superfluid, and a
 rigidly rotating Fermi gas has lower energy.

How the transition between the BEC and BCS regimes occurs in a rapidly rotating gas
is an interesting
question.

The mean-field theory for a rotating Fermi gas in a harmonic trap has
been studied in Ref.\cite{gurarie} across the whole crossover regime.
This theory gives results on the BCS and BEC sides that
match with the above expectations in these limits, and provides a
useful way to connect between these two regimes.  The
approach works explicitly with a trapped gas, and the
mean-field theory has the benefit of being accurate in the limit of a
``narrow'' Feshbach resonance\cite{andreevgl}.  The narrow resonance
limit has the feature that the molecular bosons that form have
vanishing interactions. As a result, for a trapped gas at any
$\Omega\neq \omega_\perp$ the molecular gas on the BEC side is 
confined to the lowest harmonic oscillator
state and carries zero angular momentum. This is an unphysical
situation for rotating atomic Fermi gases which interact, typically, through
``wide'' Feshbach resonances.  However, it is unclear under what
conditions (if any) this might affect the validity of the predictions
for rotating Fermi gases beyond the narrow resonance limit.

{\it BCS Regime: Upper Critical Rotation Rate}

A complementary approach has been described in
Ref.\cite{ZhaiHo,MollerC07}.  These works study mean-field theory in a
situation in which the trapping potential is weak, so the system is
in the uniform 3D regime, \be\Omega \simeq \omega_\perp \gg
\omega_\parallel \,,
\label{eq:q3d}
\ee with single particle energy spectrum
(\ref{eq:q3dspectrum}).  The linearized 
BCS mean-field equations were used to determine
the critical temperature below which the system is unstable to the
formation of a BCS superfluid,  as a function of the rotation rate
$\Omega$.

In Ref.\cite{ZhaiHo}, the effects of a trap are introduced within the local
density approximation (see \S\ref{sec:lda}). This gives rise to inhomogeneity in the
trap, with normal and superfluid regions co-existing at different radii. An
upper critical rotation rate is determined as a function of detuning, at which
the entire cloud becomes normal for a temperature of $k_B T/\hbar\Omega =
10^{-3}$. 

\begin{figure}
\includegraphics[width=14cm]{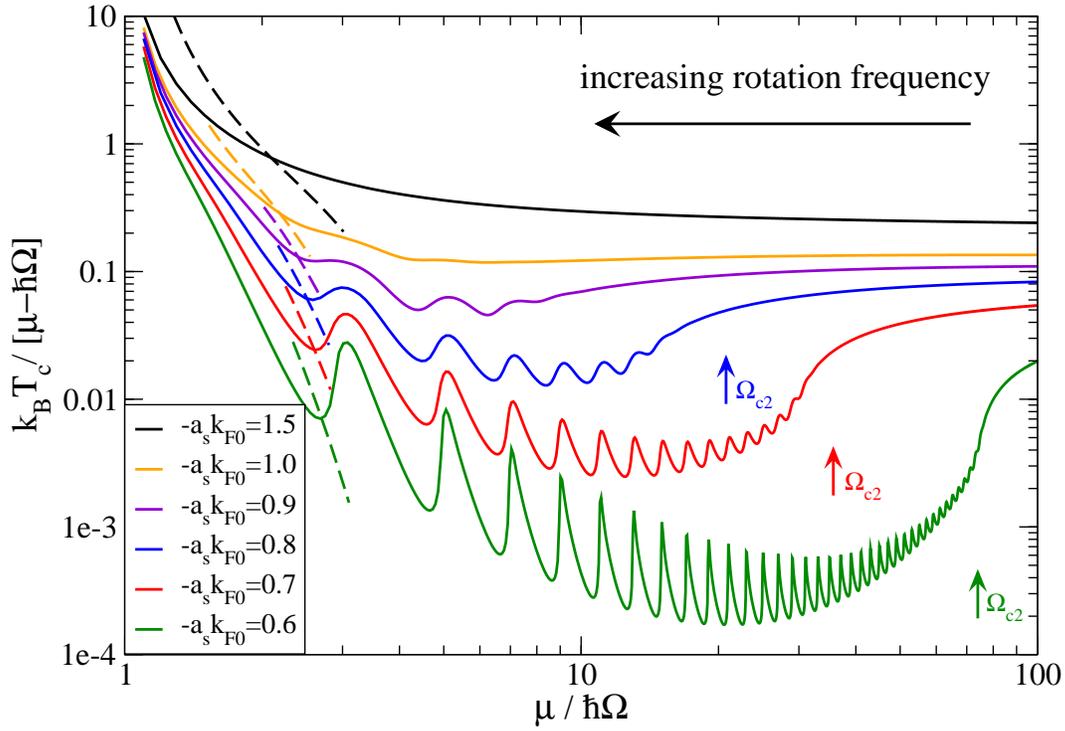}
\caption{Solid lines: Transition temperature for formation of a BCS superfluid
state for a rotating Fermi gas in the uniform 3D limit, as a function
of rotation rate $\Omega$ for different interaction strengths $k_F
a_{\rm s}$. For weak interactions, as the rotation rate increases at fixed
chemical potential $\mu$ (right to left on the graph), the transition
temperature falls abruptly at the semi-classical upper critical rotation rate.
However, it remains non-zero and rises as $\Omega/\mu$ increases. 
For strong interactions the transition temperature becomes a monotonically
increasing function of $\Omega$. Dashed lines: Transition temperature for
formation of a CDW state. 
[Reprinted figure from: G. M\"{o}ller and N.R. Cooper,
 Phys. Rev. Lett {\bf 99}, 190409 (2007). Copyright (2007) by the American
 Physical Society. 
]
}
\label{fig:Figure1}
\end{figure}

In Ref.\cite{MollerC07}, similar mean-field equations were studied for
the uniform 3D limit (\ref{eq:q3d}). These results show that the
critical temperature $T_c(\Omega)$ {\it never} vanishes.  In fact, as
the rotation rate increases, the transition temperature becomes an
{\it increasing} function of rotation rate, see Fig.~\ref{fig:Figure1}.  This is consistent with
studies of BCS theory in the solid state setting beyond the semi-classical Gorkov
approximation\cite{RasoltT92}. The increase in transition temperature
at high rotation rate is related to the enhancement of the density of
states at the Fermi level, due to the $1/\sqrt{E}$ singularities of
the quasi-1D motion along the rotation axis.
These results show that the upper critical rotation rate
found in Ref.\cite{ZhaiHo} is a function of temperature; as $T\to 0$
the BCS mean-field theory predicts that the system remains superfluid
for all $k_Fa_{\rm s}$ and $\Omega$.

\begin{figure}
\center\includegraphics[width=6cm]{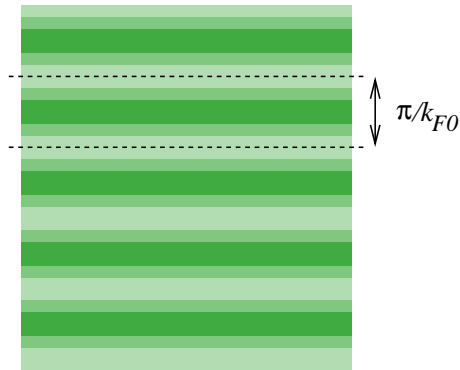}
\caption{Schematic diagram of the charge density wave state that sets
the upper critical rotation rate for destruction of fermionic
superfluidity. The wavevector of the density wave lies along the
rotation axis. It's period is set by the particle density, and is such
that the atomic filling factor of each layer is $\nu_a = n_a/(2\pi
\ell^2) =2$. It is a fully gapped state.}
\label{fig:cdw}
\end{figure}

That BCS mean-field theory predicts the groundstate to be superfluid for all
$k_Fa_{\rm s}$ and $\Omega$ is a statement that the BCS state has a lower (free)
energy than the normal phase for any attractive interaction. Despite
the orbital motion induced by the rotation, the Fermi gas will always undergo a
superconducting (SC) instability at sufficiently low temperature. This does
not rule out the possibility of destruction of the superfluid phase by
transition to another state. In
Ref.\cite{MollerC07} it was shown that the BCS phase competes with a charge
density wave (CDW) state. The CDW is a state in which the atomic density is
spontaneously modulated along the rotation axis with a period such that the
atomic filling factor in each layer is $\nu_a=2$, see Fig.~\ref{fig:cdw}.
This competition was analysed in detail for the case of interacting
fermions in the LLL, allowing fluctuations of both CDW and SC. The
calculations show that the dominant instability is towards the CDW
state. (The transition temperature into the CDW is plotted as a dashed line in
Fig.~\ref{fig:Figure1}.)
 Thus, an upper
critical rotation frequency {\it does} exist on the BCS side of the
transition: it is set by the transition into this CDW\cite{MollerC07}.
At intermediate $\Omega$ it was argued that CDW and SC can
co-exist, leading to a novel form of supersolidity.

{\it Transition from BEC to BCS}

The above considerations are expected to be valid on the BCS side of
the transition.  Understanding how the results connect to the BEC side
is an interesting open problem. One thing that is clear is
that, in the rapid rotation limit, in which all fermions (or bosonic
molecules) are confined to the LLL, the transition from BCS to BEC
side must involve a phase transition.\footnote{This transition cannot
be captured within the theory in the narrow resonance limit\cite{gurarie}, as the
bosonic molecules are non-interacting so the Laughlin state does not
appear.}

In the quasi-2D case this can be seen by noting that states on the two
sides of the transition are topologically distinct. On the BCS side,
where the atoms experience weak attractive interactions, the
groundstate is a compact droplet with $\nu_a=2$ (a full Landau
level for both atomic species); this state has two gapless edge modes,
corresponding to density and spin excitations. On the BEC side, one
anticipates that the groundstate is a Laughlin state of bosonic
molecules, with $\nu_m=1/2$;\footnote{The relation $\nu_m = \nu_a/4$
follows from the fact that there are half as many molecules as atoms,
and the vortex density for a molecule is twice that for an
atom\cite{CooperFesh04}.}  this state has only one gapless edge
mode. The change in edge mode structure shows that the states on the
two sides are topologically distinct; they must be separated by a
(quantum) phase transition\cite{haldanerez,MollerC07,yang:030404}.  An
effective theory has been derived for this phase
transition\cite{yang:030404}, but the nature of
the transition remains unclear.  

The existence of a phase transition in the 3D case was predicted\cite{MollerC07} based on the above 2D argument combined with the
observation that the 3D system forms a CDW of layers of $\nu_a=2$
states at high rotation rate. If the CDW order survives with the same
period on passing to the BEC side, each layer must undergo the above
topological phase transition; the only alternative is that the CDW
does not survive (or changes period), which also signals a phase transition
of another kind.

\section{Experimental Aspects}

Throughout the preceding text, existing experimental studies relevant to the
theoretical results presented have been mentioned. In this section, I focus
only on those aspects of experiment that relate to the strongly correlated
phases at small filling factors. There also remain many interesting
issues to explore experimentally beyond this strong-correlation regime.

\subsection{Experimental Signatures of Strongly Correlated Phases}

Experimental detection of the properties of cold atomic gases is
typically performed following the removal of the trapping potentials
and expansion of the gas. An important aspect of the physics of
rapidly rotating atomic gases is that, provided the interactions are
weak enough that the gas is in the LLL limit (\ref{eq:weak}), the expanded gas
provides a direct map of the wavefunction prior to expansion: the only
effect of expansion is a rotation and rescaling of the
co-ordinates\cite{readcooper}. The use of expansion as a
``wavefunction microscope'' makes probing the rotating gas
particularly straightforward. Contrast the case of a non-rotating gas
where (in the simplest cases) the expanded image is a measure of the
velocity distribution in the trap, and spatial structure is recovered
only after Fourier transformation.

The most dramatic feature of the strongly correlated phases predicted by
theory is that they are incompressible. As a result, the density distribution
in a trapped system will show characteristic ``wedding cake''
structures\cite{cooper:063622}, with regions in which the 2D particle density
is pinned to values $\nu^* n_{\rm v}$ where $\nu^*$ are the filling factors at
which the groundstate is incompressible. This density distribution differs
markedly from the expected density distribution of the vortex lattice phase
which is compressible and hence follows a Thomas-Fermi
distribution\cite{watanabe,CooperKR}. An extreme case is the Laughlin state,
$\nu=1/2$, for which the average particle density is constant out to the
radius $R \sim 2\sqrt{N}\ell$.

An interesting feature of particles in the 2D LLL is that the condensate
fraction of the many-body phases can be determined from measurements of the expectation value of the
particle density\cite{shm2}. In this way, and by repeated measurements to
construct the expectation value of the density, one could directly distinguish
between a condensed state (such as the vortex lattice) and any
strongly correlated phase for which the condensate fraction vanishes.

In order to detect the nature of the correlated phase, it was proposed
that the particle distributions in single-shot measurements of the
expanded wavefunction be used to directly measure the
density-density correlation functions of the states in the
trap\cite{readcooper}. This method -- also referred to as measurements of ``noise
correlations''\cite{altman:013603} -- is particularly powerful in the
present case, where the scaling of the wavefunction allows direct
measurements of spatial correlations in the gas. The statistics of the
density fluctuations can be used to measure the compressibility of the
fluid, and to distinguish between different strongly correlated
phases\cite{readcooper}. Noise correlations are also expected to show
signatures of strongly correlated phases of bosons on rotating
lattices\cite{burkov:180406,palmer:013609}

Further information on the nature of the incompressible phases can be
deduced by studying the properties of their excitations.

The clearest indication of the formation of an incompressible liquid phase in
semiconductor systems is the appearance of a (fractionally) quantized Hall
resistance\cite{prangeandgirvin}. This is a consequence of the pinning of the
quasi-particles by disorder and the dissipationless flow of the incompressible
liquid past these localized states. Proposals have been
made\cite{bhat:043601,palmer:013609} for how analogous transport measurements
might be performed in rotating Bose gases on optical lattices.

Quantum Hall states, while gapped in the bulk, carry gapless edge excitations.
These are collective modes 
(waves of particle density)
 on the
surface of the incompressible fluid. The form of these excitations is fixed by
the nature of the bulk phase\cite{wen}. Direct detection of the spectrum of
edge excitations\cite{baksmaty:063615} could allow identification of the bulk
incompressible state\cite{cazalilla03}.

A direct measurement of fractional statistics of the quasiparticle excitations
of the incompressible states would be a very dramatic confirmation of
theoretical expectations. Indeed, even in conventional electronic FQHE
systems, which are very well understood and much explored experimentally, an
(accepted) experimental measurement of the fractional exchange statistics is
still lacking. An ambitious scheme for measuring the semion statistics of the
quasiparticle excitations of the Laughlin state of rotating bosons has been
proposed in Ref.\cite{Paredes}.

\subsection{Strongly Correlated Phases: Experimental Status \& Outlook }

\label{sec:status}

The reader is referred to Ref.\cite{blochdz} for a recent authoritative review
of experiments on rotating atomic gases. Here I describe the current experimental
status in connection with the strongly correlated phases which are predicted
to occur at low filling factors, and discuss the outlook for future
experimental work in this area.

To date, the highest vortex densities in rotating atomic Bose gases have been
achieved by the group of Eric Cornell. In these experiments, the mean
interaction energy $\mu\sim g\bar{n}$ is sufficiently small that the gases are
in or close to the 2D LLL regime, $\mu\lesssim 2\hbar\omega_\perp,
\hbar\omega_\parallel$. However, the smallest filling factor reported is $\nu
\simeq 500$\cite{SchweikhardCEMC92}. This is deep in the regime $\nu> \nu_c$
where one expects the groundstate to be a vortex lattice\cite{cwg,shm1,baym},
and indeed this is what is observed. Reducing the filling factor further
(reducing $N$ at fixed $N_{\rm v}$) leads to a gas that is so dilute that
collisions are too infrequent for thermalization on the lifetime of the gas.

Clearly further special efforts are require in order to reach the
regime $\nu\lesssim 6$ where one expects quantum fluctuations of
vortices to destabilize the lattice.
There appear to be two main difficulties in reaching this regime:

{\it Weak Interactions}

In the 2D LLL, the typical interaction energy that stabilizes the
strongly correlated phases is small. The relevant energy scale is the
pseudo-potential $V_0$, Eqn.(\ref{eq:v0}). To gain some physical intuition for
the origin of this energy scale, it is useful to note that it sets the
chemical potential (the mean interaction energy) for a rotating gas at a
filling factor $\nu\sim 1$. The centrifugal limit (\ref{eq:centrifugal}) sets
a maximum rotation rate $\Omega = \omega_\perp$, and hence a maximum vortex
density (\ref{eq:feynman}) $n_{\rm v} = 2M\omega_\perp/h$. For $\nu=1$ the
maximum 2D particle density is $n_{\rm 2d} = n_{\rm v} = 2M\omega_\perp/h$ and
so the maximum 3D density is of order $\bar{n} \sim
2M\omega_\perp/(ha_\parallel)$. The chemical potential associated with this
density is
\begin{equation}
\mu \sim g\bar{n} 
\sim \frac{\hbar^2 a_{\rm s}}{M}\frac{1}{a_\perp^2 a_\parallel}
\sim \frac{a_{\rm s}}{a_\parallel}\hbar\omega_\perp \,,
\label{eq:interactionscale}
\end{equation}
which recovers $V_0$ (\ref{eq:v0}) up to numerical factors.

This energy scale sets the size of the incompressibility gap of the
strongly correlated phases. It therefore sets the
required temperature scales in order to observe these phases. In typical
experiments in magnetic traps, $\hbar\omega_\perp \sim 5\mbox{nK}$, while
$a_{\rm s}/a_\parallel\sim 0.01$, leading to gaps that are small fractions of a
nano-Kelvin, much smaller than the temperatures that can currently be reached
in such experiment.

Eqn.(\ref{eq:interactionscale}) highlights ways in which to overcome this
difficulty:
\begin{itemize}
\item
 Increase $a_{\rm s}$. This can be achieved by using a Feshbach resonance
to enter a strongly interacting regime\cite{papp:135301}. For
bosons 
three-body losses can be large
close to a Feshbach resonance,
which may limit the usefulness of this approach. Losses are much
smaller for Fermi gases close to an $s$-wave Feshbach resonance\cite{petrov:090404}. The
FQHE regime for rotating fermions in the BCS to BEC crossover is
therefore a very interesting situation to consider.  It may be easier
to achieve the strong-correlation regime ($\nu\sim 1$) for a
strongly interacting two-component Fermi gas (close to the unitary
limit) than for a  typical Bose gas\cite{antezza:053609}.

\item Decrease $a_\parallel$. The introduction of an optical lattice,
with wavevector parallel to the rotation axis can slice the condensate
up into many parallel layers with a much stronger subband confinement
(much smaller $a^{\rm layer}_\parallel$),
Fig.~\ref{fig:slice} \cite{dalibardcornell}. This can lead to
a significant increase in the interaction energy
scale\cite{cooper:063622}. It is important that the subband
confinement retains rotational symmetry, to prevent loss of angular
momentum or heating (see below).

\item Increase $\omega_\perp$.  A tighter harmonic confinement
 would allow an increased vortex density and hence
stronger interactions at $\nu\sim 1$. Much tighter traps can  be formed by using atom chip traps formed on surface
microstructures\cite{schmiedmayer}, allowing transverse trapping
frequencies on the order of $100\mbox{kHz}$. (Such systems are highly
anisotropic, $\omega_\parallel \ll \omega_\perp$, so could potentially
be used to study the 3D LLL regime (\ref{eq:q3dregime}).)  Very
tight confinement can also be achieved in the wells of an optical
lattice. This might allow studies of rotating clusters of small numbers of
atoms. There are interesting proposals\cite{popp:053612,baur-2008} for
how clusters of small numbers of particles might be transferred
adiabatically from a non-rotating state into the strongly correlated Laughlin
state.

\item Optical lattice systems. 
The energy
  scales that set the correlation energies in optical lattices are large.
  Physical rotation of the optical lattice\cite{tung,hafezi-2007} involves
the same centrifugal limit (\ref{eq:centrifugal}), again limiting
  the vortex density to the values leading to (\ref{eq:interactionscale}). This method will therefore
  likely also need to be supplemented with tight harmonic confinement
in order to achieve a large interaction energy at $\nu\sim 1$. However, the
  proposed schemes for imprinting Peierls phases onto the inter-site
  tunnelling matrix elements\cite{jaksch,mueller,sorensen:086803} appear to
  allow field configurations that mimic uniform rotation without a centrifugal
  limit. If flux densities of order one flux per site can be generated, these
  systems could show strongly correlated states with a large energy scale
(see \S\ref{sec:lattices}).

\end{itemize}

{\it Heating}

An atomic gas that has come to equilibrium at a rotation rate $\Omega$ is at
rest in the rotating frame of reference. For example, in the vortex lattice
phase the positions of the vortices are at rest in the rotating frame. In the
laboratory frame, the fluid a distance $r_\perp$ from the rotation axis
is moving at velocity of $ \Omega
r_\perp$.
As a result, 
the rotating gas will  experience any 
 non-axisymmetric 
perturbation that is static in the
laboratory frame (e.g. a trap that is not circularly symmetric, or any
scattered light from optical traps) as
as a time-dependent potential in the rotating frame. This 
can lead to excitation, or heating, of the rotating gas.

It is instructive to estimate the importance of such excitations on an
incompressible liquid state, by applying a form of Landau criterion. For an
incompressible liquid state, the collective excitations (in the bulk) have a
gap, $\Delta$, that is of order the interaction energy, $V_0$, or of
order the cyclotron energy, $2\hbar \omega_\perp$, whichever is the
smaller.\footnote{The lowest energy gap is proportional to $V_0$ in the 2D
  LLL, $V_0 \ll \hbar \omega_\perp$. For $V_0\gg \hbar \omega_\perp$, there
  will be mixing of Landau levels and the lowest energy gap will not grow
  beyond the cyclotron gap $2\hbar\omega_\perp$.} Therefore, the maximum possible
gap (achievable for $V_0\sim \hbar\omega_\perp$) is $\Delta\sim
\hbar\omega_\perp$. The edge of a rotating cloud of radius $R$ moves with a
linear velocity of $v= \Omega R$, which is close to $v \sim \omega_\perp R$
for a system with a large number of vortices (\ref{eq:close}). Due to this
motion, in this edge region the collective
modes with wavevectors $q \gtrsim \Delta/(\hbar v )\sim 1/R$ have negative
energy in the laboratory frame. These excitations can therefore be spontaneously generated by any
perturbation that is static in the laboratory frame and has Fourier components on the
relevant scale, $q \gtrsim 1/R$. This sets a strict requirement on
the potential: any static perturbation in the laboratory frame that has a
wavelength less than of order the radius of the cloud can efficiently generate
collective modes the rotating gas (i.e. cause heating).

The use of geometric phases to simulate rotation (or a uniform
magnetic field) -- either in a lattice-based
system\cite{jaksch,mueller,sorensen:086803} or in the
continuum\cite{juzeliunas} -- has the advantage that the gas remains at
rest in the laboratory frame.  This removes the relative motion of the gas
with respect to possible perturbations at rest in the laboratory
frame, and eliminates these as sources of heating.  Experimental work
has recently introduced a vector potential\cite{lin-2008} of vanishing
curl using  two-photon induced geometric phases, setting the stage
for extensions to vector potentials with non-zero curl.

\section{Summary}

Rapidly rotating degenerate atomic gases offer the possibility to explore the
physics of quantized vortices and vortex lattices in novel parameter regimes.
This review has described some of the novel equilibrium properties that can
appear: for rotating spinless and multi-component Bose gases; for Bose gases on
rotating lattices; and for rotating Fermi gases. Interesting new physics can
emerge both within the realm where mean-field theories are applicable (as is
dramatically illustrated by the novel vortex lattice phases in
Figs.~\ref{fig:dipolargs}, \ref{fig:muellerho} and \ref{fig:reijnderss1}), and
within regimes where mean-field theory fails (as evidenced by the theoretical
predictions of some very exotic strongly correlated phases). While theoretical
studies have helped to clarify some of the expected properties of these
systems, our understanding is still very much limited. It is hoped that these
theoretical predictions will motivate and inform future experimental
investigations of these novel phenomena. At the very least, these theoretical
studies show that very interesting aspects of many-body quantum physics are at
play in rotating degenerate atomic gases; this is an area where future theory
and experiment are likely to uncover dramatic new phenomena.

\section*{Acknowledgements}

I have benefitted very much from working closely with many people on this
and related topics: Nick Read, Stavros Komineas, Ed Rezayi, Steve Simon,
Kareljan Schoutens, Miguel Cazalilla, Duncan Haldane, Gunnar M{\" o}ller, and,
in particular, Nicola Wilkin and Mike Gunn who introduced me to the subject
and who have been continued sources of help and advice. I also acknowledge
useful discussions with many others, including Misha Baranov, Gordon Baym,
Eric Cornell, Jean Dalibard, Eugene Demler, David Feder, Sandy Fetter, Victor
Gurarie, Jason Ho, Jainendra Jain, Thierry Jolicoeur, Wolfgang Ketterle,
Maciej Lewenstein,
Chris Pethick,
Nicholas Regnault.
Finally, I am grateful to 
Sandy Fetter, Gunnar M{\" o}ller and Kareljan Schoutens for helpful comments on a draft version of this manuscript.

\begin{appendix}

\section{Haldane pseudo-potentials}

\label{sec:haldane}

Within the space of 2D LLL states, any two-body interaction that is
rotationally symmetric in the 2D plane can be conveniently parameterized by
its ``Haldane pseudo-potentials''\cite{haldanehierarchy}. Owing to the
rotational symmetry, the interaction conserves the relative angular momentum
of two particles. Thus, its effects are determined by the expectation value of
the two-body potential in the two-particle wavefunction
\begin{eqnarray}
  \Psi_{m_c,m}(\bm{r}_1,\bm{r}_2) & \propto & \left(\zeta_1 +
    \zeta_2\right)^{m_c} \left(\zeta_1 - \zeta_2\right)^m e^{-|\zeta_1|^2/4}
  e^{-|\zeta_2|^2/4} e^{-z_1^2/2a_\parallel^2}
  e^{-z_2^2/2a_\parallel^2}\\ V_m & = & \int\int V(\bm{r}_1-\bm{r}_2) \;
  |\Psi_{m_c,m}(\bm{r}_1,\bm{r}_2)|^2 \; d^3\bm{r}_1 \,
  d^3\bm{r}_2\end{eqnarray} where $m$ and $m_c$ are the relative and
centre-of-mass angular momenta.  For spinless bosonic (fermionic)
particles, $m$ must be even (odd) so only the subset of even (odd)
Haldane pseudo-potentials of the interaction are important.   
For contact interactions (\ref{eq:contact}), the only non-zero
pseudo-potential is 
\be V_0 = \frac{4\pi\hbar^2 a_{\rm s}}{M} \int d^3\bm{r} 
 \frac{1}{(\pi a_\perp^2)^2}  e^{-2(x^2+y^2)/a_\perp^2} \frac{1}{(\pi
   a_\parallel^2)} e^{-2z^2/a_\parallel^2} = \sqrt{\frac{2}{\pi}}\frac{\hbar^2
   a_{\rm s}}{a_\perp^2 a_\parallel} \,.
\ee
This is the only energy scale entering the low-energy physics of a rapidly rotating
Bose gas in the 2D LLL regime.

For the 3D LLL regime (\ref{eq:q3dregime}), an approximation often
made\cite{Ho01} is that the particle density in the $z$ direction is set by
the Thomas-Fermi distribution. (See \S\ref{sec:3dmft} for a discussion of this
approximation.) One can view this as a change in the subband wavefunction in
the $z$ direction, $\varphi(z)$. In the Thomas-Fermi approximation, the
normalized particle density in the $z$ direction is taken to be
\begin{equation}
|\varphi_{\rm TF}(z)|^2  = \frac{3}{4 W_z}\left(1-\frac{z^2}{W_z^2}\right) \quad\quad (|z|\leq W_z)
\label{eq:tfsubband}
\end{equation}
in which the extension along the $z$ axis, $W_z$, is set by the
chemical potential, $\mu = (1/2) M \omega_\parallel^2 W_z^2$.
The only influence of the subband wavefunction in the $z$-direction
$\varphi(z)$ is to introduce a numerical factor to the Haldane
pseudo-potentials, which for contact interactions is $\int dz
|\varphi(z)|^4$. Using the Thomas-Fermi form for the particle density in the
$z$ direction leads to 
\be
\label{eq:v0tf}
V_0^{\rm TF} =    \frac{6}{5}\frac{\hbar^2 a_{\rm
s}}{Ma_\perp^2 W_z}\,.
\ee

\section{Numerical Techniques}
\label{sec:numerical}

Determining the groundstate of a rapidly rotating atomic gas in a
harmonic well poses an essentially strongly interacting,
non-perturbative theoretical problem.  In the absence of interactions,
the many-particle states at high angular momentum are degenerate,
\S\ref{sec:2dlll}. Interactions lift this degeneracy and select a
groundstate.  To understand the properties of the groundstate (and
the low-lying excitations) it has proved very useful to make use of exact
diagonalization studies.

The restriction of single particle states to the 2D LLL (\ref{eq:2dlll})
limits the size of the Hilbert space. This makes exact diagonalization
feasible for systems that can be large enough to observe clear signatures of
strongly correlated many-particle states. Numerical studies have been
performed in several different geometries, described in detail below. In each
case, the strategy is the use of the ``configuration interaction'' method:

\begin{itemize}

\item One chooses a finite number of (orthonormal) single particle basis
  states, $\phi_\alpha(\bm{r})$, with $\alpha = 1,2\ldots M$ which span the 2D
  LLL for the geometry considered. For example, in the uniform plane, these
  are the 2D LLL wavefunctions of Eqn.(\ref{eq:2dlll}) with $\alpha$
  labelling the angular momentum $m=0,1,\ldots M-1$, where $M$ sets the
  cut-off in total angular momentum. The wavefunctions in other geometries are
  discussed below. The interaction Hamiltonian may then be written in a
  second-quantized form as\cite{fetterwalecka}: \be
\label{eq:2nd}
\hat{V} = \frac{1}{2}\sum_{\alpha,\beta,\gamma,\delta=1}^M
V_{\alpha\beta\gamma\delta} \; \hat{a}^\dag_\alpha \hat{a}^\dag_\beta
\hat{a}_\gamma\hat{a}_\delta \ee where $\hat{a}^{(\dag)}$ are the field
operators for particles (which could be bosons or fermions) in the single
particle state $\phi_\alpha(\bm{r})$. The  matrix elements are \be
V_{\alpha\beta\gamma\delta} \equiv 
\langle \alpha\beta|\hat{V}|\delta \gamma\rangle =
\int d^3\bm{r}\int d^3\bm{r}'
\phi^*_\alpha(\bm{r})\phi^*_\beta(\bm{r}')
\phi_\gamma(\bm{r}')\phi_\delta(\bm{r}) \; V(\bm{r}-\bm{r}')
\ee
where $V(\bm{r})$ is the two-body interaction potential.

\item
One generates a complete set of
(orthogonal) many-particle basis states for a system of $N$ particles occupying
the single particle basis $\phi_\alpha(\bm{r})$, with $\alpha = 1,2\ldots
M$. These states are conveniently taken to be the Fock states, represented by
the occupation numbers of the $M$ orbitals, $|n_1, n_2, \ldots
n_M\rangle$. (The Fock states are symmetric/antisymmetric under particle
exchange for the case of spinless bosons/fermions.)  For bosons,
the total number of such many-particle states is
 \be \frac{(N+M-1)!}{N! (M-1)!}  \stackrel{N,M\to\infty}{\longrightarrow}
 e^{\gamma N }\,, \ee
where the limit $N\to\infty$ is taken with fixed $\nu=N/M$, which sets the value $\gamma $. The number of basis states can be
reduced by the use of symmetries, which depend on the geometry (see
below). Nevertheless, the general feature that the number of basis states
grows exponentially with system size survives, and
sets the ultimate limitation to exact diagonalization studies.

\item
Within the (symmetry reduced) many-particle basis, one constructs the matrix
elements of the Hamiltonian (\ref{eq:2nd}). This is typically a very
sparse matrix.  Efficient numerical solution is possible for a small
number of eigenvalues/eigenvectors using the Lanczos method. This allows
the diagonalization of systems with of order $10^7$ basis states on
a modern desktop PC.

\end{itemize}

The different geometries that have been used in such studies are the ``disc'',
``sphere'' and ``torus'' geometries, and are described in detail in
\S\ref{sec:disk},\ref{sec:sphere},\ref{sec:torus}. They each have their own
benefits and/or drawbacks. 

\begin{itemize}

\item {\bf Disc.} This is appropriate for the direct simulation of
  experimental systems, in which particles are confined in a harmonic well. It
  allows studies of the effects of inhomogeneity, including edge structures.
  Limitations arise in the study of correlated phases which can appear at
  large angular momentum. The inhomogeneity makes it
  difficult to observe emergent bulk phases. For a finite-size system, it is
  unclear how to define the number of vortices $N_{\rm v}$, so the filling
  factor (\ref{eq:fillingfactorN}) is not accurately defined.

\item {\bf Sphere.}  This is useful for studying bulk
incompressible liquid phases. The uniformity of the geometry reduces
the finite size effects associated with numerical studies of the
inhomogeneous system on a disc. The number of particles $N$ and
vortices $N_{\rm v}$ are fixed integers,  allowing a
well-defined filling factor $\nu$ to be studied, via
\be N_{\rm v} =
\frac{1}{\nu} N - {\cal S}\,.
\label{eq:shift}
\ee The offset  ${\cal S}$ is referred to as  the ``shift'', which is a characteristic of a
given incompressible liquid phase\cite{dammorf}. Owing to this offset
(\ref{eq:shift}), different competing incompressible liquid states at
a given filling factor can be studied independently on the sphere if
they have different shifts.  This is an advantage  if the shift of the groundstate is known: consistent
extrapolation to the thermodynamic limit along a sequence with that shift is
strong evidence for this phase describing the thermodynamic
limit. However, it is a disadvantage if the possible groundstates (and
their shifts) are unknown. The sphere is not convenient for the study of
crystalline states.

\item {\bf Torus}. This is useful for studying bulk
incompressible liquid phases, and bulk crystalline phases (states with
broken translational invariance).  Finite size effects are reduced by
the periodic boundary conditions.  The number of particles $N$ and
vortices $N_{\rm v}$ are integers, allowing a well-defined filling
factor $\nu$ to be studied \be \nu = \frac{N}{N_{\rm v}}\,. \ee  
Since there is no shift, all possible groundstates in the
thermodynamic limit can compete within the same finite-size
calculation. This geometry is therefore useful if the nature of the phase in
the thermodynamic limit is unknown.  For crystalline phases, the
possibility to vary the geometry (aspect ratio) of the torus is a
powerful way to investigate the optimal translational symmetry of the
crystal.  Certain incompressible liquid states are characterized by a
groundstate degeneracy on the torus\cite{oshikawa}; the appearance of this
non-trivial degeneracy in the exact spectrum 
 is strong evidence for the groundstate being in this
topological phase. It is somewhat harder to
implement calculations on the torus than on the disk or sphere,
especially when full advantage is made of the translational
symmetries\cite{haldanemtm}.

\end{itemize}

In the following I provide specific details for the different
geometries that are required for the implementation of exact
diagonalization studies.

\subsection{Disk}

\label{sec:disk}

The orthononormal single particle basis states in the 2D LLL are
\begin{equation}
\psi_m(\bm{r})  = \frac{1}{\sqrt{2\pi 2^{m} m!} \ell} 
\left(\frac{x+iy}{\ell}\right)^m
\;  e^{-(x^2+y^2)/4\ell^2}\;\frac{1}{(\pi a_\parallel^2)^{1/4}}e^{-z^2/2a_\parallel^2}
\label{eq:disk}
\end{equation}
with $m=0,1,\ldots$.

For contact interactions (\ref{eq:contact}), the matrix elements of the Hamiltonian are
\be
V_{m_1,m_2,m_3,m_4} = V_0 \frac{(m_1+m_4)!}{2^{m_1+m_2}} \delta_{m_1+m_2,m_3+m_4}\,,
\ee
where $V_0$ is the Haldane pseudo-potential (\ref{eq:v0}).

For a set of $N$ particles, in orbitals $m_{i}$, the total angular momentum
$$L = \sum_{i=1}^N m_i$$ is conserved by the interactions. The
many-particle Hilbert space can therefore be split into subspaces of
fixed $L$ and $N$. There is an additional symmetry reflecting the
independence of the interaction energy on the centre-of-mass
co-ordinate of the particle. This symmetry can be used to further
reduce the size of the Hilbert space, but this is not commonly implemented.

\subsection{Sphere}

\label{sec:sphere}

A monopole of $N_{\rm v} \equiv 2S$ ($N_{\rm v}$ integer) flux quantum
is chosen to be positioned at the centre of a sphere, and to have
uniform flux density over the surface\cite{haldanehierarchy}. The
radius of the sphere is set by the condition that the total flux is $2S = 4\pi
R^2 n_{\rm v}$, i.e. $R = \sqrt{S}\ell$ where $\ell$ is the conventional
magnetic length (\ref{eq:ell}).
 The
lowest-energy single particle states, analogous to the lowest Landau
level states, consist of $2S+1$ degenerate levels. Using a gauge
choice that is symmetric about the polar axis, the single particle
wavefunctions on the sphere may be written\cite{fanoortolani}
\be \phi_m (\theta,\phi) =
\left[\frac{2S+1}{4\pi} \left(\begin{array}{c} 2S \\ S+m \end{array}\right)\right]^{1/2}
u^{S+m} v^{S-m} \ee for $m=-S,-S+1\ldots S$,  where $u \equiv
\cos(\theta/2) e^{i\phi/2}$, $v \equiv \sin(\theta/2) e^{-i\phi/2}$
and
$(\theta,\phi)$ are the polar angles on the sphere.

The matrix elements of the contact interaction  (\ref{eq:contact}) are
\be
V_{m_1,m_2,m_3,m_4} = V_0
\frac{[(2S+1)!]^2 
(2S+m_1+m_2)!(2S-m_1-m_2)!}{S(4S+1)!
\sqrt{\prod_{i=1}^4 (S+m_i)! (S-m_i)!}}
\delta_{m_1+m_2,m_3+m_4}\,,
\ee
where we have chosen a 2D contact interaction of strength $g_{\rm 2d} =
g/(\sqrt{2\pi}a_\parallel)$ to take account of the quasi-2D motion 
along the rotation axis.

These interactions (and any rotationally invariant interaction on the
sphere) preserve the $z$-component of the angular momentum \be L_z =
\sum_{i=1}^N m_i\,.\ee This can be easily imposed to reduce the size of the
many-body  basis. Rotational invariance also conserves the {\it total}
angular momentum.  The Hilbert space can therefore be
further reduced to the many particle basis states of definite
$\hat{L}^2$. In practice it is difficult to construct a basis
of fixed $\hat{L}^2$, so this is
not commonly used.

\subsection{Torus}

\label{sec:torus}

The ``torus'' geometry is defined by a unit cell on the plane on which
opposite faces are identified to impose periodic boundary conditions. It is
specified by two (linearly independent) basis vectors $\bm{a},\bm{b}$ in the
plane.  We shall focus on the simplest case of a
rectangular unit cell, with basis vectors $a\hat{\bm{x}}$ and $b\hat{\bm{y}}$, with
$\hat{\bm{x}}$, $\hat{\bm{y}}$  orthonormal.
A consistent quantum theory on the periodic geometry requires that an integer number of flux quanta
$N_{\rm v}$ pierce the unit cell, with $N_{\rm v} = ab n_{\rm v} = ab/(2\pi\ell^2)$
an integer.
To construct single particle states it is convenient to use the Landau gauge,
with a vector potential directed along $\hat{\bm{y}}$. Then the linear momentum
along  $\hat{\bm{y}}$ is
conserved and spans the $N_{\rm v}$ states in the 2D-LLL.
Expressing the linear momentum along $y$ in units of $\frac{2\pi}{b}$,
it can take the integer values $m=0,1,..N_{\rm v}-1$. The single
particle states in the 2D LLL are\cite{yoshiokaHL83}
\be
\psi_m(\bm{r}) = \frac{\exp(-z^2/2a_\parallel^2)}{(\pi a_\parallel^2)^{1/4}}
\left(\frac{1}{b\pi^{1/2}\ell}\right)^{1/2}\sum_{p=-\infty}^\infty
\exp \left[i\frac{(X_m+pa)y}{\ell^2} - \frac{(X_m+pa-x)^2}{2\ell^2}\right]
\ee
where $X_m \equiv 2\pi \ell^2 m/b$.

The matrix elements of the contact interaction are
\be
V_{m_1,m_2,m_3,m_4} = V_0\sqrt{\frac{8}{\pi}}\frac{\ell}{b}
\sum_{p,q=-\infty}^\infty
e^{-\frac{(X_{m_1}-X_{m_3}+p a)^2}{2\ell^2}-\frac{(X_{m_2}-X_{m_3}+q
    a)^2}{2\ell^2}} \delta^{N_{\rm v}}_{m_1+m_2,m_3+m_4}
\ee
where $\delta^{N_{\rm v}}_{i,j} = 1$ if $i-j=0 \, \mbox{mod} N_{\rm v}$ and
$0$ otherwise.

The total momentum in the $y$ direction
\be
K_y = \sum_i m_i\,
\ee
is conserved modulo $N_{\rm v}$. This provides a convenient way in which to reduce the size of the many-body
basis size (by a factor
order $N_{\rm v}$).
Owing to the symmetry under magnetic translations along $\hat{\bm{x}}$, a second
conserved momentum $K_x$ can be constructed\cite{haldanemtm}. At a filling
factor $N/N_{\rm v} = p/q$ (with $p$ and $q$ co-prime), with number of
particles $N = p\bar{N}$ and number of flux $N_{\rm v} = q \bar{N}$, the
momenta $(K_x,K_y)$ take $\bar{N}^2$ distinct states in a (rectangular)
Brillouin zone. Construction of many-particle basis states that are
eigenstates of both
$K_x$ and $K_y$ allows a significant further reduction in the overall size of the
Hilbert space. 
At particular points in the Brillouin zone there exist additional
point-group symmetries.

\end{appendix}


\begin{thebibliography}{243}
\providecommand{\natexlab}[1]{#1}

\bibitem{donnelly}
R.J. Donnelly {\itshape Quantized Vortices in Helium {I}{I}},    Cambridge
  University Press, Cambridge, 1991.

\bibitem{cornellweimannobel}
E.A. Cornell and C.E. Wieman, {\itshape Nobel Lecture: Bose-Einstein
  condensation in a dilute gas, the first 70 years and some recent
  experiments}, Rev. Mod. Phys. 74 (2002), pp. 875--893.

\bibitem{ketterlenobel}
W. Ketterle, {\itshape Nobel lecture: When atoms behave as waves: Bose-Einstein
  condensation and the atom laser}, Rev. Mod. Phys. 74 (2002), pp. 1131--1151.

\bibitem{blochdz}
I. Bloch, J. Dalibard, and W. Zwerger, {\itshape Many-body physics with
  ultracold gases}, Rev. Mod. Phys. 80 (2008), p. 885.

\bibitem{sokol}
P.E. Sokol, in {\itshape Bose-Einstein Condensation}, A.~Griffin, D.W.~Snoke and S.~Stringari,  eds.,    Cambridge University  Press, 1995, p.~51.

\bibitem{Leggett01}
A.J. Leggett, {\itshape Bose-Einstein condensation in the alkali gases: Some
  fundamental concepts}, Rev. Mod. Phys. 73 (2001), pp. 307--356.

\bibitem{hallvinen}
H.E. Hall and W.F. Vinen, {\itshape The rotation of liquid Helium II.}, Proc.
  Roy. Soc. A238 (1956), pp. 204--14.

\bibitem{cooperleshouches}
N.R. Cooper, {\itshape Les Houches Lecture Notes}  2008 (unpublished).


\bibitem{campbellziff}
L.J. Campbell and R.M. Ziff, {\itshape Vortex patterns and energies in a
  rotating superfluid}, Phys. Rev. B 20 (1979), pp. 1886--1902.

\bibitem{yarm}
E.J. Yarmchuk, M.J.V. Gordon, and R.E. Packard, {\itshape Observation of
  Stationary Vortex Arrays in Rotating Superfluid Helium}, Phys. Rev. Lett. 43
  (1979), pp. 214--217.

\bibitem{yarm1982}
E.J. Yarmchuk and R.E. Packard, {\itshape Photographic studies of quantized
  vortex lines}, J. Low Temp. Phys. 46 (1982), pp. 479--515.

\bibitem{MadisonCWD00}
K.W. Madison, F. Chevy, W. Wohlleben, and J. Dalibard, {\itshape Vortex
  formation in a stirred Bose-Einstein condensate}, Phys. Rev. Lett. 84 (2000),
  pp. 806--809.

\bibitem{abos01}
J.R. Abo-Shaeer, C. Raman, J.M. Vogels, and W. Ketterle, {\itshape Observation
  of vortex lattices in Bose-Einstein condensates}, Science 292 (2001), p. 476.

\bibitem{hodby:010405}
E. Hodby, G. Hechenblaikner, S.A. Hopkins, O.M. Marago, and C.J. Foot,
  {\itshape Vortex Nucleation in Bose-Einstein Condensates in an Oblate, Purely
  Magnetic Potential}, Phys. Rev. Lett. 88 (2002), p. 010405.

\bibitem{PhysRevLett.86.564}
D.L. Feder, A.A. Svidzinsky, A.L. Fetter, and C.W. Clark, {\itshape Anomalous
  Modes Drive Vortex Dynamics in Confined Bose-Einstein Condensates}, Phys.
  Rev. Lett. 86 (2001), pp. 564--567.

\bibitem{PhysRevA.63.011601}
F. Dalfovo and S. Stringari, {\itshape Shape deformations and angular-momentum
  transfer in trapped Bose-Einstein condensates}, Phys. Rev. A 63 (2000), p.
  011601.

\bibitem{madinstab}
K.W. Madison, F. Chevy, V. Bretin, and J. Dalibard, {\itshape Stationary States
  of a Rotating Bose-Einstein Condensate: Routes to Vortex Nucleation}, Phys.
  Rev. Lett. 86 (2001), pp. 4443--4446.

\bibitem{PhysRevLett.92.020403}
C. Lobo, A. Sinatra, and Y. Castin, {\itshape Vortex Lattice Formation in
  Bose-Einstein Condensates}, Phys. Rev. Lett. 92 (2004), p. 020403.

\bibitem{coddington:100402}
I. Coddington, P. Engels, V. Schweikhard, and E.A. Cornell, {\itshape
  Observation of Tkachenko Oscillations in Rapidly Rotating Bose-Einstein
  Condensates}, Phys. Rev. Lett. 91 (2003), p. 100402.

\bibitem{schweikhard:040404}
V. Schweikhard, I. Coddington, P. Engels, V.P. Mogendorff, and E.A. Cornell,
  {\itshape Rapidly Rotating Bose-Einstein Condensates in and near the Lowest
  Landau Level}, Phys. Rev. Lett. 92 (2004), p. 040404.

\bibitem{wgs}
N.K. Wilkin, J.M.F. Gunn, and R.A. Smith, {\itshape Do Attractive Bosons
  Condense?}, Phys. Rev. Lett. 80 (1998), p. 2265.

\bibitem{cwg}
N.R. Cooper, N.K. Wilkin, and J.M.F. Gunn, {\itshape Quantum phases of vortices
  in rotating Bose-Einstein condensates}, Phys. Rev. Lett. 87 (2001), p.
  120405.

\bibitem{prangeandgirvin}
R.E. Prange and S.M. Girvin (eds.)  {\itshape The Quantum {H}all Effect},
  Second Edition   Springer-Verlag, Berlin, 1990.

\bibitem{dassarmapinczuk}
S.D. Sarma and A. Pinczuk (eds.)  {\itshape Perspectives in Quantum Hall
  Effects: Novel Quantum Liquids in Low-Dimensional Semiconductor Structures},
    Wiley, New York, 1997.

\bibitem{tung:240402}
S. Tung, V. Schweikhard, and E.A. Cornell, {\itshape Observation of Vortex
  Pinning in Bose-Einstein Condensates}, Phys. Rev. Lett. 97 (2006), p.
  240402.

\bibitem{hafezi-2007}
M. Hafezi, A.S. S\o{}rensen, E. Demler, and M.D. Lukin, {\itshape Fractional
  quantum Hall effect in optical lattices}, Phys. Rev. A 76 (2007), p.
  023613.

\bibitem{JakschZoller}
D. Jaksch and P. Zoller, {\itshape Creation of effective magnetic fields in
  optical lattices: the Hofstadter butterfly for cold neutral atoms}, New
  Journal of Physics 5 (2003), p.~56.

\bibitem{mueller}
E.J. Mueller, {\itshape Artificial electromagnetism for neutral atoms: Escher
  staircase and Laughlin liquids}, Phys. Rev. A 70 (2004), p. 041603.

\bibitem{sorensen:086803}
A.S. S\o{}rensen, E. Demler, and M.D. Lukin, {\itshape Fractional Quantum Hall
  States of Atoms in Optical Lattices}, Phys. Rev. Lett. 94 (2005),  p.
  086803.

\bibitem{palmer:180407}
R.N. Palmer and D. Jaksch, {\itshape High-Field Fractional Quantum Hall Effect
  in Optical Lattices}, Phys. Rev. Lett. 96 (2006), p. 180407.

\bibitem{palmer:013609}
R.N. Palmer, A. Klein, and D. Jaksch, {\itshape Optical lattice quantum Hall
  effect}, Phys. Rev. A 78 (2008), p. 013609.

\bibitem{fetter}
A.L. Fetter, {\itshape Rotating trapped Bose-Einstein condensates},  (2008),
  arXiv:0801.2952.

\bibitem{LLStatMech}
L.D. Landau and E.M. Lifshitz,  {\itshape \S 26},  {\itshape Statistical
  Physics Pt 1},   Vol. 5,   Butterworth Heinemann, Oxford, 1981.

\bibitem{frohlich1994}
J. Fr{\" o}hlich {\itshape Les Houches Lecture Notes},    Elsevier, Amsterdam,
  1994.

\bibitem{fock}
V. Fock, {\itshape Bemerkung zur Quantelung des harmonischen Oszillators im
  Magnetfeld}, Z. Phys. 47 (1928), p. 446.

\bibitem{darwin}
C.G. Darwin, Proc. Cambridge Philos. Soc. 27 (1930), p.~86.

\bibitem{morris:033605}
A.G. Morris and D.L. Feder, {\itshape Validity of the lowest-Landau-level
  approximation for rotating Bose gases}, Phys. Rev. A 74 (2006), p.
  033605.

\bibitem{Ho01}
T.L. Ho, {\itshape Bose-Einstein condensates with large number of vortices},
  Phys. Rev. Lett. 87 (2001), p. 060403.

\bibitem{Tesanovic91}
Z. Te\u{s}anovi\'{c}, {\itshape Nature of the superconducting transition in the
  presence of a magnetic field}, Phys. Rev. B 44 (1991), pp. 12635--12638.

\bibitem{ButtsR99}
D.A. Butts and D.S. Rokhsar, {\itshape Predicted signatures of rotating
  Bose-Einstein condensates}, Nature 397 (1999), pp. 327--329.

\bibitem{CooperKR}
N.R. Cooper, S. Komineas, and N. Read, {\itshape Vortex lattices in the lowest
  Landau level for confined Bose-Einstein condensates}, Phys. Rev. A 70 (2004),
  p. 033604. 

\bibitem{aftalion:023611}
A. Aftalion, X. Blanc, and J. Dalibard, {\itshape Vortex patterns in a fast
  rotating Bose-Einstein condensate}, Phys. Rev. A 71 (2005), p.
  023611.

\bibitem{watanabe}
G. Watanabe, G. Baym, and C.J. Pethick, {\itshape Landau Levels and the
  Thomas-Fermi Structure of Rapidly Rotating Bose-Einstein Condensates}, Phys.
  Rev. Lett. 93 (2004), p. 190401.

\bibitem{Abrikosov57}
A. Abrikosov, {\itshape On the Magnetic Properties of Superconductors of the
  Second Group}, Zh. Eksp. Teor. Fiz. 32 (1957), p. 1442 [Sov. Phys.
  {J}{E}{T}{P}~{\bf 5}, 1174 (1957)].

\bibitem{KleinerRA64}
W.H. Kleiner, L.M. Roth, and S.H. Autler, {\itshape Bulk Solution of
  Ginzburg-Landau Equations for Type II Superconductors: Upper Critical Field
  Region}, Phys. Rev. 133 (1964), pp. A1226--A1227.

\bibitem{fischerbaym}
U.R. Fischer and G. Baym, {\itshape Vortex States of Rapidly Rotating Dilute
  Bose-Einstein Condensates}, Phys. Rev. Lett. 90 (2003), p. 140402.

\bibitem{cozzini:023615}
M. Cozzini, S. Stringari, and C. Tozzo, {\itshape Vortex lattices in
  Bose-Einstein condensates: From the Thomas-Fermi regime to the
  lowest-Landau-level regime}, Phys. Rev. A 73 (2006), p. 023615.

\bibitem{coddington:063607}
I. Coddington, P.C. Haljan, P. Engels, V. Schweikhard, S. Tung, and E.A.
  Cornell, {\itshape Experimental studies of equilibrium vortex properties in a
  Bose-condensed gas}, Phys. Rev. A 70 (2004), p. 063607.

\bibitem{shm1}
J. Sinova, C.B. Hanna, and A.H. MacDonald, {\itshape Quantum melting and
  absence of Bose-Einstein condensation in two-dimensional vortex matter},
  Phys. Rev. Lett. 89 (2002), p. 030403.

\bibitem{baym}
G. Baym, {\itshape Vortex lattices in rapidly rotating Bose-Einstein
  condensates: Modes and correlation functions}, Phys. Rev. A 69 (2004),
  p. 043618.

\bibitem{schweikhard:210403}
V. Schweikhard, I. Coddington, P. Engels, S. Tung, , and E.A. Cornell,
  {\itshape Vortex-Lattice Dynamics in Rotating Spinor Bose-Einstein
  Condensates}, Phys. Rev. Lett. 93 (2004),  p. 210403.

\bibitem{sonin:021606}
E.B. Sonin, {\itshape Ground state and Tkachenko modes of a rapidly rotating
  Bose-Einstein condensate in the lowest-Landau-level state}, Phys. Rev. A 72
  (2005), p. 021606.

\bibitem{fetter:013620}
A.L. Fetter, {\itshape Lowest-Landau-level description of a Bose-Einstein
  condensate in a rapidly rotating anisotropic trap}, Phys. Rev. A 75 (2007),
  p. 013620.

\bibitem{linn}
M. Linn, M. Niemeyer, and A.L. Fetter, {\itshape Vortex stabilization in a
  small rotating asymmetric Bose-Einstein condensate}, Phys. Rev. A 64 (2001),
  p. 023602.

\bibitem{PhysRevA.69.023618}
M.O. Oktel, {\itshape Vortex lattice of a Bose-Einstein condensate in a
  rotating anisotropic trap}, Phys. Rev. A 69 (2004), p. 023618.

\bibitem{sinha:150401}
S. Sinha and G.V. Shlyapnikov, {\itshape Two-Dimensional Bose-Einstein
  Condensate under Extreme Rotation}, Phys. Rev. Lett. 94 (2005), p.
  150401.

\bibitem{sanchez-lotero:043613}
P. S\'{a}nchez-Lotero and J.J. Palacios, {\itshape Vortices in a rotating
  Bose-Einstein condensate under extreme elongation}, Phys. Rev. A 72 (2005),
  p. 043613.

\bibitem{BaranovPS}
M. Baranov, L. Dobrek, K. Goral, L. Santos, and M. Lewenstein, {\itshape
  Ultracold dipolar gases - a challenge for experiments and theory}, Physica
  Scripta T102 (2002), p.~74.

\bibitem{GriesmaierWHSP05}
A. Griesmaier, J. Werner, S. Hensler, J. Stuhler, and T. Pfau, {\itshape
  Bose-Einstein condensation of chromium}, Phys. Rev. Lett. 94 (2005), p.
  160401.

\bibitem{lahaye2007}
T. Lahaye, T. Koch, B. Fröhlich, M. Fattori, J. Metz, A. Griesmaier, S.
  Giovanazzi, and T. Pfau, {\itshape Strong dipolar effects in a quantum
  ferrofluid}, Nature 448 (2007), p. 672.

\bibitem{ni-2008}
K.K. Ni, S. Ospelkaus, M.H.G. {de Miranda}, A. Pe'er, B. Neyenhuis, J.J.
  Zirbel, S. Kotochigova, P.S. Julienne, D.S. Jin, and J. Ye, {\itshape
A High
  Phase-Space-Density Gas of Polar Molecules};  arXiv:0808.2963.

\bibitem{crs}
N.R. Cooper, E.H. Rezayi, and S.H. Simon, {\itshape Vortex Lattices in Rotating
  Atomic Bose Gases with Dipolar Interactions}, Phys. Rev. Lett. 95 (2005),
  p. 200402.

\bibitem{zhang:200403}
J. Zhang and H. Zhai, {\itshape Vortex Lattices in Planar Bose-Einstein
  Condensates with Dipolar Interactions}, Phys. Rev. Lett. 95 (2005), p. 200403.

\bibitem{KoulakovFS96}
A.A. Koulakov, M.M. Fogler, and B.I. Shklovskii, {\itshape Charge Density Wave
  in Two-Dimensional Electron Liquid in Weak Magnetic Field}, Phys. Rev. Lett.
  76 (1996), pp. 499--502.

\bibitem{MoessnerC96}
R. Moessner and J.T. Chalker, {\itshape Exact results for interacting electrons
  in high Landau levels}, Phys. Rev. B 54 (1996), pp. 5006--5015.

\bibitem{madelunghere}
M. M. Hurley and S. J. Singer, {\itshape Domain Energies of the Dipolar Lattice Gas},
  J. Phys. Chem. {\bf 96} (1992), pp. 1938--1950.

\bibitem{komineas:023623}
S. Komineas and N.R. Cooper, {\itshape Vortex lattices in Bose-Einstein
  condensates with dipolar interactions beyond the weak-interaction limit},
  Phys. Rev. A 75 (2007),  p. 023623.

\bibitem{SmithW00}
R.A. Smith and N.K. Wilkin, {\itshape Exact eigenstates for repulsive bosons in
  two dimensions}, Phys. Rev. A 62 (2000), p. 061602.

\bibitem{BertschP99}
G.F. Bertsch and T. Papenbrock, {\itshape Yrast line for weakly interacting
  trapped bosons}, Phys. Rev. Lett. 83 (1999), pp. 5412--5414.

\bibitem{HusseinV02}
M.S. Hussein and O.K. Vorov, {\itshape Generalized yrast states of a
  Bose-Einstein condensate in a harmonic trap for a universality class of
  interactions}, Phys. Rev. A 65 (2002), p. 035603.

\bibitem{VorovHI03}
O.K. Vorov, M.S. Hussein, and P.V. Isacker, {\itshape Rotating Ground States of
  Trapped Atoms in a Bose-Einstein Condensate with Arbitrary Two-Body
  Interactions}, Phys. Rev. Lett. 90 (2003), p. 200402.

\bibitem{kavoulenergy}
A.D. Jackson and G.M. Kavoulakis, {\itshape Analytical Results for the
  Interaction Energy of a Trapped, Weakly Interacting Bose-Einstein
  Condensate}, Phys. Rev. Lett. 85 (2000), pp. 2854--2856.

\bibitem{LiebS06}
E.H. Lieb and R. Seiringer, {\itshape Derivation of the Gross-Pitaevskii
  Equation for Rotating Bose Gases}, Commun. Math. Phys. 264 (2006), pp.
  505--537.

\bibitem{JacksonKMR01}
A.D. Jackson, G.M. Kavoulakis, B. Mottelson, and S.M. Reimann, {\itshape Weakly
  interacting Bose-Einstein condensates under rotation: Mean-field versus exact
  solutions}, Phys. Rev. Lett. 86 (2001), pp. 945--949.

\bibitem{anderson}
P.W. Anderson,  {\itshape Chap. 2},  {\itshape Basic notions of condensed
  matter physics},  Frontiers in Physics Vol. 55,   The Benjamin Cummings,
  1984.

\bibitem{cmmp}
N.R. Cooper, {\itshape Vortex Liquids and Vortex Lattices in Weakly Interacting
  Bose Gases},  (2000),  CMMP 2000, SSp.P2.29.

\bibitem{ueda:043603}
M. Ueda and T. Nakajima, {\itshape Nambu-Goldstone mode in a rotating dilute
  Bose-Einstein condensate}, Phys. Rev. A 73 (2006), p. 043603.

\bibitem{dagnino:013625}
D. Dagnino, N. Barber\'{a}n, K. Osterloh, A. Riera, and M. Lewenstein,
  {\itshape Symmetry breaking in small rotating clouds of trapped ultracold
  Bose atoms}, Phys. Rev. A 76 (2007), p. 013625.

\bibitem{parke:110401}
M.I. Parke, N.K. Wilkin, J.M.F. Gunn, and A. Bourne, {\itshape Exact Vortex
  Nucleation and Cooperative Vortex Tunneling in Dilute Bose-Einstein
  Condensates}, Phys. Rev. Lett. 101 (2008), p. 110401.

\bibitem{romanovsky:011606}
I. Romanovsky, C. Yannouleas, and U. Landman, {\itshape Symmetry-conserving
  vortex clusters in small rotating clouds of ultracold bosons}, Phys. Rev. A
  78 (2008), p. 011606.

\bibitem{laughlinstates}
R.B. Laughlin, {\itshape Anomalous Quantum {H}all Effect: An Incompressible
  Quantum Fluid with Fractionally Charged Excitations}, Phys. Rev. Lett. 50
  (1983), pp. 1395--1398.

\bibitem{WilkinG00}
N.K. Wilkin and J.M.F. Gunn, {\itshape Condensation of ``composite bosons'' in a
  rotating BEC}, Phys. Rev. Lett. 84 (2000), pp. 6--9.

\bibitem{CooperW99}
N.R. Cooper and N.K. Wilkin, {\itshape Composite fermion description of
  rotating Bose-Einstein condensates}, Phys. Rev. B 60 (1999), pp.
  R16279--R16282.

\bibitem{jainoriginal}
J.K. Jain, {\itshape Composite-Fermion Approach for the Fractional Quantum
  {H}all Effect}, Phys. Rev. Lett. 63 (1989), pp. 199--202.

\bibitem{ole}
O. Heinonen (ed.)  {\itshape Composite Fermions: A Unified View of the Quantum
  Hall Regime},    World Scientific, Singapore, 1998.

\bibitem{jainkawamura}
J.K. Jain and T. Kawamura, {\itshape Composite Fermions in Quantum Dots},
  Europhys. Lett. 29 (1995), pp. 321--326.

\bibitem{barberancrystals}
N. Barber\'{a}n, M. Lewenstein, K. Osterloh, and D. Dagnino, {\itshape Ordered
  structures in rotating ultracold Bose gases}, Phys. Rev. A 73 (2006), p. 063623.

\bibitem{PhysRevLett.93.230405}
I. Romanovsky, C. Yannouleas, and U. Landman, {\itshape Crystalline Boson
  Phases in Harmonic Traps: Beyond the Gross-Pitaevskii Mean Field}, Phys. Rev.
  Lett. 93 (2004), p. 230405.

\bibitem{baksmaty:023620}
L.O. Baksmaty, C. Yannouleas, and U. Landman, {\itshape Rapidly rotating boson
  molecules with long- or short-range repulsion: An exact diagonalization
  study}, Phys. Rev. A 75 (2007),  p. 023620.

\bibitem{SchweikhardCEMC92}
V. Schweikhard, I. Coddington, P. Engels, V.P. Mogendorff, and E.A. Cornell,
  {\itshape Rapidly Rotating Bose-Einstein Condensates in and near the Lowest
  Landau Level}, Phys. Rev. Lett. 92 (2004), p. 040404.

\bibitem{Fetter67}
A.L. Fetter, {\itshape Quantum Theory of Superfluid Vortices. I. Liquid Helium
  II}, Phys. Rev. 162 (1967), pp. 143--153.

\bibitem{HaldaneWu}
F.D.M. Haldane and Y.S. Wu, {\itshape Quantum dynamics and statistics of
  vortices in two-dimensional superfluids}, Phys. Rev. Lett. 55 (1985), pp.
  2887--2890.

\bibitem{RozhkovS96}
A. Rozhkov and D. Stroud, {\itshape Quantum melting of a two-dimensional vortex
  lattice at zero temperature}, Phys. Rev. B 54 (1996), pp. R12697--R12700.

\bibitem{tkachenko1}
V. Tkachenko, Zh. Eksp. Teor. Fiz. 49 (1965), p. 1875 [Sov. Phys.
  {J}{E}{T}{P}~{\bf 22}, 1282 (1966)].

\bibitem{tkachenko2}
---{}---{}---, Zh. Eksp. Teor. Fiz. 50 (1966), p. 1573 [Sov. Phys.
  {J}{E}{T}{P}~{\bf 23}, 1049 (1966)].

\bibitem{tkachenko3}
---{}---{}---, Zh. Eksp. Teor. Fiz. 56 (1969), p. 1763 [Sov. Phys.
  {J}{E}{T}{P}~{\bf 29}, 245 (1969)].

\bibitem{CooperR07}
N.R. Cooper and E.H. Rezayi, {\itshape Competing compressible and
  incompressible phases in rotating atomic Bose gases at filling factor $\nu =
  2$}, Phys. Rev. A 75 (2007), p. 013627.

\bibitem{girvinmacdonald}
S.M. Girvin and A.H. MacDonald, {\itshape Off-diagonal long-range order,
  oblique confinement, and the fractional quantum Hall effect}, Phys. Rev.
  Lett. 58 (1987), pp. 1252--1255.

\bibitem{readoperator}
N. Read, {\itshape Order Parameter and Ginzburg-Landau Theory for the
  Fractional Quantum Hall Effect}, Phys. Rev. Lett. 62 (1989), pp. 86--89.

\bibitem{RegnaultJ03}
N. Regnault and T. Jolicoeur, {\itshape Quantum Hall Fractions in Rotating
  Bose-Einstein Condensates}, Phys. Rev. Lett. 91 (2003), p. 030402.

\bibitem{regnault:235309}
---{}---{}---, {\itshape Quantum Hall fractions for spinless bosons}, Phys.
  Rev. B 69 (2004),  p. 235309.

\bibitem{Viefers}
S. Viefers, T.H. Hansson, and S.M. Reimann, {\itshape Bose condensates at high
  angular momenta}, Phys. Rev. A 62 (2000), p. 053604.

\bibitem{halperinhierarchy}
B.I. Halperin, {\itshape Statistics of Quasiparticles and the Hierarchy of
  Fractional Quantized {H}all States}, Phys. Rev. Lett. 52 (1984), pp.
  1583--1586.

\bibitem{chang:013611}
C.C. Chang, N. Regnault, T. Jolicoeur, and J.K. Jain, {\itshape Composite
  fermionization of bosons in rapidly rotating atomic traps}, Phys. Rev. A 72
  (2005), p. 013611.

\bibitem{wen}
X.G. Wen, {\itshape Topological Orders and Edge Excitations in FQH States},
  Advances in Physics 44 (1995), p. 405.

\bibitem{cazalilla03}
M.A. Cazalilla, {\itshape Surface modes of ultracold atomic clouds with a very
  large number of vortices}, Phys. Rev. A 67 (2003), p. 063613.

\bibitem{cazalilla:121303}
M.A. Cazalilla, N. Barber\'{a}n, and N.R. Cooper, {\itshape Edge excitations
  and topological order in a rotating Bose gas}, Phys. Rev. B 71 (2005),
 p. 121303.

\bibitem{rrc}
E.H. Rezayi, N. Read, and N.R. Cooper, {\itshape Incompressible Liquid State of
  Rapidly Rotating Bosons at Filling Factor 3/2}, Phys. Rev. Lett. 95 (2005),
  p. 160404.

\bibitem{regnault:235324}
N. Regnault and T. Jolicoeur, {\itshape Parafermionic states in rotating
  Bose-Einstein condensates}, Phys. Rev. B 76 (2007), p. 235324.

\bibitem{nayak:1083}
C. Nayak, S.H. Simon, A. Stern, M. Freedman, and S.D. Sarma, {\itshape
  Non-Abelian anyons and topological quantum computation}, Rev. Mod. Phys. 80
  (2008), p. 1083.

\bibitem{MooreR91}
G. Moore and N. Read, {\itshape Nonabelions in the fractional quantum
  hall-effect}, Nucl. Phys. B 360 (1991), pp. 362--396.

\bibitem{ReadR99}
N. Read and E.H. Rezayi, {\itshape Beyond paired quantum Hall states:
  Parafermions and incompressible states in the first excited Landau level},
  Phys. Rev. B 59 (1999), pp. 8084--8092.

\bibitem{RezayiR06}
E.H. Rezayi and N. Read, {\itshape Non-Abelian quantized Hall states of electrons at
  filling factors 12/5 and 13/5 in the first excited Landau level},
  arXiv:0608346.

\bibitem{cappelli}
A. Cappelli, L.S. Georgiev, and I.T. Todorov, {\itshape Parafermion Hall states
  from coset projections of abelian conformal theories}, Nucl. Phys. B 559
  (2001), p. 499.

\bibitem{wg}
N.K. Wilkin and J.M.F. Gunn, {\itshape Condensation of "composite bosons" in a
  rotating BEC}, Phys. Rev. Lett. 84 (2000), pp. 6--9.

\bibitem{oshikawa}
M. Oshikawa, Y.B. Kim, K. Shtengel, C. Nayak, and S. Tewari, {\itshape
  Topological degeneracy of non-Abelian states for dummies}, Annals of Physics
  322 (2007), p. 1477.

\bibitem{chung:043608}
B. Chung and T. Jolicoeur, {\itshape Fermions out of dipolar bosons in the
  lowest Landau level}, Phys. Rev. A 77 (2008), p. 043608.

\bibitem{rjreview}
N. Regnault and T. Jolicoeur, {\itshape Quantum Hall fractions in ultracold
  atomic vapors}, Mod. Phys. Lett. B 51 (2004), p. 1003.

\bibitem{CooperFesh04}
N.R. Cooper, {\itshape Exact Ground States of Rotating Bose Gases Close to a
  Feshbach Resonance}, Phys. Rev. Lett. 92 (2004), p. 220405.

\bibitem{seki:063602}
H. Seki and K. Ino, {\itshape Incompressible liquid, stripes, and bubbles in
  rapidly rotating Bose atoms at $\nu = 1$}, Phys. Rev. A 77 (2008), p.
  063602.

\bibitem{crsproc}
N.R. Cooper, E.H. Rezayi, and S.H. Simon, {\itshape Vortex Lattices in Rotating
  Atomic Bose Gases with Non-Local Interactions}, Solid State Commun. 140
  (2006), p.~61.

\bibitem{hlr}
B.I. Halperin, P.A. Lee, and N. Read, {\itshape Theory of the half-filled
  {L}andau-level}, Phys. Rev. B 47 (1993), pp. 7312--7343.

\bibitem{baranov:200402}
M.A. Baranov, H. Fehrmann, and M. Lewenstein, {\itshape Wigner Crystallization
  in Rapidly Rotating 2D Dipolar Fermi Gases}, Phys. Rev. Lett. 100 (2008),
  p. 200402.

\bibitem{aftalion:011601}
A. Aftalion, X. Blanc, and F. Nier, {\itshape Vortex distribution in the lowest
  Landau level}, Phys. Rev. A 73 (2006), p. 011601.

\bibitem{cooper:063622}
N.R. Cooper, F.J.M. van Lankvelt, J.W. Reijnders, and K. Schoutens, {\itshape
  Quantum Hall states of atomic Bose gases: Density profiles in single-layer
  and multilayer geometries}, Phys. Rev. A 72 (2005), p. 063622.

\bibitem{dalibardcornell}
 Such geometries have been studied in the groups of E. Cornell and J.
  Dalibard.

\bibitem{snoek1}
M. Snoek and H.T.C. Stoof, {\itshape Vortex-Lattice Melting in a
  One-Dimensional Optical Lattice}, Phys. Rev. Lett. 96 (2006), p.
  230402.

\bibitem{snoek2}
---{}---{}---, {\itshape Theory of vortex-lattice melting in a one-dimensional
  optical lattice}, Phys. Rev. A 74 (2006), p. 033615.

\bibitem{PhysRevLett.91.240403}
J.P. Martikainen and H.T.C. Stoof, {\itshape Quantum Fluctuations of a Vortex
  in an Optical Lattice}, Phys. Rev. Lett. 91 (2003), p. 240403.

\bibitem{fiory:73}
A.T. Fiory, A.F. Hebard, and S. Somekh, {\itshape Critical currents associated
  with the interaction of commensurate flux-line sublattices in a perforated Al
  film}, Applied Physics Letters 32 (1978), pp. 73--75.

\bibitem{PhysRevB.57.7937}
C. Reichhardt, C.J. Olson, and F. Nori, {\itshape Commensurate and
  incommensurate vortex states in superconductors with periodic pinning
  arrays}, Phys. Rev. B 57 (1998), pp. 7937--7943.

\bibitem{PhysRevB.67.014532}
W.V. Pogosov, A.L. Rakhmanov, and V.V. Moshchalkov, {\itshape Vortex lattice in
  the presence of a tunable periodic pinning potential}, Phys. Rev. B 67
  (2003), p. 014532.

\bibitem{PhysRevLett.78.2648}
C. Reichhardt, C.J. Olson, and F. Nori, {\itshape Dynamic Phases of Vortices in
  Superconductors with Periodic Pinning}, Phys. Rev. Lett. 78 (1997), pp.
  2648--2651.

\bibitem{tung}
S. Tung, V. Schweikhard, and E.A. Cornell, {\itshape Observation of Vortex
  Pinning in Bose-Einstein Condensates}, Phys. Rev. Lett. 97 (2006), p.
  240402.

\bibitem{jakschbh}
D. Jaksch, C. Bruder, J.I. Cirac, C.W. Gardiner, and P. Zoller, {\itshape Cold
  Bosonic Atoms in Optical Lattices}, Phys. Rev. Lett. 81 (1998), pp.
  3108--3111.

\bibitem{greiner}
M. Greiner, O. Mandel, T. Esslinger, T.W. Hansch, and I. Bloch, {\itshape
  Quantum phase transition from a superfluid to a Mott insulator in a gas of
  ultracold atoms}, Nature 415 (2001), pp. 39--44.

\bibitem{jaksch}
D. Jaksch and P. Zoller, {\itshape Creation of effective magnetic fields in
  optical lattices: the Hofstadter butterfly for cold neutral atoms}, New
  Journal of Physics 5 (2003), p.~56.

\bibitem{osterloh:010403}
K. Osterloh, M. Baig, L. Santos, P. Zoller, and M. Lewenstein, {\itshape Cold
  Atoms in Non-Abelian Gauge Potentials: From the Hofstadter "Moth" to Lattice
  Gauge Theory}, Physical Review Letters 95 (2005), p. 010403.

\bibitem{reijndersprl}
J.W. Reijnders and R.A. Duine, {\itshape Pinning of Vortices in a Bose-Einstein
  Condensate by an Optical Lattice}, Phys. Rev. Lett. 93 (2004), p. 060401.

\bibitem{reijnderspra}
---{}---{}---, {\itshape Pinning and collective modes of a vortex lattice in a
  Bose-Einstein condensate}, Phys. Rev. A 71 (2005), p. 063607.

\bibitem{pu:190401}
H. Pu, L.O. Baksmaty, S. Yi, and N.P. Bigelow, {\itshape Structural Phase
  Transitions of Vortex Matter in an Optical Lattice}, Phys. Rev. Lett. 94
  (2005), p. 190401.

\bibitem{sato:053628}
T. Sato, T. Ishiyama, and T. Nikuni, {\itshape Vortex lattice structures of a
  Bose-Einstein condensate in a rotating triangular lattice potential}, Phys.
  Rev. A 76 (2007), p. 053628.

\bibitem{polini:010401}
M. Polini, R. Fazio, A.H. MacDonald, and M.P. Tosi, {\itshape Realization of
  Fully Frustrated Josephson-Junction Arrays with Cold Atoms}, Phys. Rev. Lett.
  95 (2005), p. 010401.

\bibitem{kasamatsu}
K. Kasamatsu, {\itshape Vortex Lattices in Rotating Bose-Einstein Condensate in
  an Optical Lattice: Analogy to Uniformly Frustrated Josephson-Junction
  Arrays}, Journal of Low Temperature Physics 50 (2008), pp. 593--598.

\bibitem{PhysRevLett.51.1999}
S. Teitel and C. Jayaprakash, {\itshape Josephson-Junction Arrays in Transverse
  Magnetic Fields}, Phys. Rev. Lett. 51 (1983), pp. 1999--2002.

\bibitem{PhysRevB.31.5728}
T.C. Halsey, {\itshape Josephson-junction arrays in transverse magnetic fields:
  Ground states and critical currents}, Phys. Rev. B 31 (1985), pp. 5728--5745.

\bibitem{PhysRevB.48.3309}
J.P. Straley and G.M. Barnett, {\itshape Phase diagram for a Josephson network
  in a magnetic field}, Phys. Rev. B 48 (1993), pp. 3309--3315.

\bibitem{harper}
P.G. Harper, {\itshape Single Band Motion of Conduction Electrons in a Uniform
  Magnetic Field}, Proc. Phys. Soc. A 68 (1955), pp. 874--878.

\bibitem{hofstadter}
D.R. Hofstadter, {\itshape Energy levels and wave functions of {Bloch}
  electrons in rational and irrational magnetic fields}, Phys. Rev. B 14
  (1976), pp. 2239--2249.

\bibitem{PhysRevA.69.043609}
C. Wu, H.D. Chen, J.P. Hu, and S.C. Zhang, {\itshape Vortex configurations of
  bosons in an optical lattice}, Phys. Rev. A 69 (2004), p. 043609.

\bibitem{oktel:045133}
M.O. Oktel, M. Ni\c{t}\u{a}, and B. Tanatar, {\itshape Mean-field theory for
  Bose-Hubbard model under a magnetic field}, Phys. Rev. B 75 (2007), 
  p. 045133.

\bibitem{goldbaum:033629}
D.S. Goldbaum and E.J. Mueller, {\itshape Vortex lattices of bosons in deep
  rotating optical lattices}, Phys. Rev. A 77 (2008),  p. 033629.

\bibitem{vignolo:023616}
P. Vignolo, R. Fazio, and M.P. Tosi, {\itshape Quantum vortices in optical
  lattices}, Phys. Rev. A 76 (2007), p. 023616.

\bibitem{burkov:180406}
A.A. Burkov and E. Demler, {\itshape Vortex-Peierls States in Optical
  Lattices}, Phys. Rev. Lett. 96 (2006), p. 180406.

\bibitem{hafezi-epl}
M. Hafezi, A.S. S\o{}rensen, M.D. Lukin, and E. Demler, {\itshape
  Characterization of topological states on a lattice with Chern number},
  Europhys. Lett. 81 (2008), p. 10005.

\bibitem{bhat:060405}
R. Bhat, M.J. Holland, and L.D. Carr, {\itshape Bose-Einstein Condensates in
  Rotating Lattices}, Phys. Rev. Lett. 96 (2006),  p. 060405.

\bibitem{bhat:063606}
R. Bhat, B.M. Peden, B.T. Seaman, M. Kr\"{a}mer, L.D. Carr, and M.J. Holland,
  {\itshape Quantized vortex states of strongly interacting bosons in a
  rotating optical lattice}, Phys. Rev. A 74 (2006), p. 063606.

\bibitem{bhat:043601}
R. Bhat, M. Kr\"{a}mer, J. Cooper, and M.J. Holland, {\itshape Hall effects in
  Bose-Einstein condensates in a rotating optical lattice}, Phys. Rev. A 76
  (2007), p. 043601.

\bibitem{Halperin83}
B.I. Halperin, {\itshape Theory of the Quantized Hall Resistance}, Helv. Phys.
  Acta 56 (1983), pp. 75--102.

\bibitem{HoS96}
T.L. Ho and V.B. Shenoy, {\itshape Binary Mixtures of Bose Condensates of
  Alkali Atoms}, Phys. Rev. Lett. 77 (1996), pp. 3276--3270.

\bibitem{esry}
B.D. Esry, C.H. Greene, J.P. Burke Jr., and J.L. Bohn, {\itshape Hartree-Fock
  Theory for Double Condensates}, Phys. Rev. Lett. 78 (1997), pp. 3594--3597.

\bibitem{kempen}
E.G.M. van Kempen, S.J.J.M.F. Kokkelmans, D.J. Heinzen, and B.J. Verhaar,
  {\itshape Interisotope Determination of Ultracold Rubidium Interactions from
  Three High-Precision Experiments}, Phys. Rev. Lett. 88 (2002), p. 093201.

\bibitem{harber}
D.M. Harber, H.J. Lewandowski, J.M. McGuirk, and E.A. Cornell, {\itshape Effect
  of cold collisions on spin coherence and resonance shifts in a magnetically
  trapped ultracold gas}, Phys. Rev. A 66 (2002), p. 053616.

\bibitem{StengerISMCK98}
J. Stenger, S. Inouye, D.M. Stamper-Kurn, H.J. Miesner, A.P. Chikkatur, and W.
  Ketterle, {\itshape Spin domains in ground-state Bose-Einstein condensates},
  Nature 396 (1998), pp. 345--348.

\bibitem{kasamatsu:043611}
K. Kasamatsu, M. Tsubota, and M. Ueda, {\itshape Spin textures in rotating
  two-component Bose-Einstein condensates}, Phys. Rev. A 71 (2005), p.
  043611.

\bibitem{PhysRevA.58.4836}
P. Ao and S.T. Chui, {\itshape Binary Bose-Einstein condensate mixtures in
  weakly and strongly segregated phases}, Phys. Rev. A 58 (1998), pp.
  4836--4840.

\bibitem{PhysRevLett.78.586}
C.J. Myatt, E.A. Burt, R.W. Ghrist, E.A. Cornell, and C.E. Wieman, {\itshape
  Production of Two Overlapping Bose-Einstein Condensates by Sympathetic
  Cooling}, Phys. Rev. Lett. 78 (1997), pp. 586--589.

\bibitem{PhysRevLett.82.2228}
H.J. Miesner, D.M. Stamper-Kurn, J. Stenger, S. Inouye, A.P. Chikkatur, and W.
  Ketterle, {\itshape Observation of Metastable States in Spinor Bose-Einstein
  Condensates}, Phys. Rev. Lett. 82 (1999), pp. 2228--2231.

\bibitem{modugno}
G. Modugno, M. Modugno, F. Riboli, G. Roati, and M. Inguscio, {\itshape Two
  Atomic Species Superfluid}, Phys. Rev. Lett. 89 (2002), p. 190404.

\bibitem{thalhammer:210402}
G. Thalhammer, G. Barontini, L.D. Sarlo, J. Catani, F. Minardi, and M.
  Inguscio, {\itshape Double Species Bose-Einstein Condensate with Tunable
  Interspecies Interactions}, Phys. Rev. Lett. 100 (2008), p. 210402.

\bibitem{papp:040402}
S.B. Papp, J.M. Pino, and C.E. Wieman, {\itshape Tunable Miscibility in a
  Dual-Species Bose-Einstein Condensate}, Phys. Rev. Lett. 101 (2008), p. 040402.

\bibitem{MatthewsAHHWC99}
M.R. Matthews, B.P. Anderson, P.C. Haljan, D.S. Hall, C.E. Wieman, and E.A.
  Cornell, {\itshape Vortices in a Bose-Einstein condensate}, Phys. Rev. Lett.
  83 (1999), pp. 2498--2501.

\bibitem{muellertexture}
E.J. Mueller, {\itshape Spin textures in slowly rotating Bose-Einstein
  condensates}, Phys. Rev. A 69 (2004), p. 033606.

\bibitem{chuivortex}
S.T. Chui, V.N. Ryzhov, and E.E. Tareyeva, {\itshape Vortex states in a binary
  mixture of Bose-Einstein condensates}, Phys. Rev. A 63 (2001), p. 023605.

\bibitem{jezek}
D.M. Jezek, P. Capuzzi, and H.M. Cataldo, {\itshape Structure of vortices in
  two-component Bose-Einstein condensates}, Phys. Rev. A 64 (2001), p. 023605.

\bibitem{leonhardt}
U. Leonhardt and G. Volovik, {\itshape How to create an Alice string
  (half-quantum vortex) in a vector Bose-Einstein condensate}, JETP Lett. 72
  (2000), p.~46.

\bibitem{bargi:130403}
S. Bargi, J. Christensson, G.M. Kavoulakis, and S.M. Reimann, {\itshape
  Mixtures of Bose Gases under Rotation}, Phys. Rev. Lett. 98 (2007),  p. 130403.

\bibitem{reijnderss1}
J.W. Reijnders, F.J.M. van Lankvelt, K. Schoutens, and N. Read, {\itshape
  Rotating spin-1 bosons in the lowest Landau level}, Phys. Rev. A 69 (2004),
  p. 023612.

\bibitem{muellerho}
E.J. Mueller and T.L. Ho, {\itshape Two-Component Bose-Einstein Condensates
  with a Large Number of Vortices}, Phys. Rev. Lett. 88 (2002), p. 180403.

\bibitem{kasamatsuprl}
K. Kasamatsu, M. Tsubota, and M. Ueda, {\itshape Vortex Phase Diagram in
  Rotating Two-Component Bose-Einstein Condensates}, Phys. Rev. Lett. 91
  (2003), p. 150406.

\bibitem{woo:031604}
S.J. Woo, S. Choi, L.O. Baksmaty, and N.P. Bigelow, {\itshape Dynamics of
  vortex matter in rotating two-species Bose-Einstein condensates}, Phys. Rev.
  A 75 (2007), p. 031604.

\bibitem{kecceli:023611}
M. Ke\c{c}eli and M.O. Oktel, {\itshape Tkachenko modes and structural phase
  transitions of the vortex lattice of a two-component Bose-Einstein
  condensate}, Phys. Rev. A 73 (2006), p. 023611.

\bibitem{schweikhardspin}
V. Schweikhard, I. Coddington, P. Engels, S. Tung, and E.A. Cornell, {\itshape
  Vortex-Lattice Dynamics in Rotating Spinor Bose-Einstein Condensates}, Phys.
  Rev. Lett. 93 (2004), p. 210403.

\bibitem{nass}
E. Ardonne and K. Schoutens, {\itshape New Class of Non-Abelian Spin-Singlet
  Quantum Hall States}, Phys. Rev. Lett. 82 (1999), pp. 5096--5099.

\bibitem{barnettrefael}
R. Barnett, G. Refael, M.A. Porter, and H.P. B\"{u}chler, {\itshape Vortex
  lattice locking in rotating two-component Bose-Einstein condensates}, New
  Journal of Physics 10 (2008), p. 043030 (10pp).

\bibitem{woo:120403}
S.J. Woo, Q.H. Park, and N.P. Bigelow, {\itshape Phases of Atom-Molecule Vortex
  Matter}, Phys. Rev. Lett. 100 (2008), p. 120403.

\bibitem{ohmimachida}
T. Ohmi and K. Machida, {\itshape Bose-Einstein condensation with internal
  degrees of freedom in alkali atom gases}, J. Phys. Soc. Jpn. 67 (2008), pp.
  1822--1825.

\bibitem{Ho98}
T.L. Ho, {\itshape Spinor Bose condensates in optical traps}, Phys. Rev. Lett.
  81 (1998), pp. 742--745.

\bibitem{zhou}
F. Zhou, {\itshape Spin Correlation and Discrete Symmetry in Spinor
  Bose-Einstein Condensates}, Phys. Rev. Lett. 87 (2001), p. 080401.

\bibitem{PhysRevLett.87.010404}
M.D. Barrett, J.A. Sauer, and M.S. Chapman, {\itshape All-Optical Formation of
  an Atomic Bose-Einstein Condensate}, Phys. Rev. Lett. 87 (2001), p. 010404.

\bibitem{Stamper-KurnACIMSK98}
D.M. Stamper-Kurn, M.R. Andrews, A.P. Chikkatur, S. Inouye, H.J. Miesner, J.
  Stenger, and W. Ketterle, {\itshape Optical confinement of a Bose-Einstein
  condensate}, Phys. Rev. Lett. 80 (1998), pp. 2027--2030.

\bibitem{PhysRevLett.83.4677}
S.K. Yip, {\itshape Internal Vortex Structure of a Trapped Spinor Bose-Einstein
  Condensate}, Phys. Rev. Lett. 83 (1999), pp. 4677--4681.

\bibitem{PhysRevLett.89.030401}
T. Mizushima, K. Machida, and T. Kita, {\itshape Mermin-Ho Vortex in
  Ferromagnetic Spinor Bose-Einstein Condensates}, Phys. Rev. Lett. 89 (2002),
  p. 030401.

\bibitem{PhysRevA.66.053604}
J.P. Martikainen, A. Collin, and K.A. Suominen, {\itshape Coreless vortex
  ground state of the rotating spinor condensate}, Phys. Rev. A 66 (2002), p.
  053604.

\bibitem{PhysRevA.66.061601}
T. Kita, T. Mizushima, and K. Machida, {\itshape Spinor Bose-Einstein
  condensates with many vortices}, Phys. Rev. A 66 (2002), p. 061601.

\bibitem{homuellers1}
T.L. Ho and E.J. Mueller, {\itshape Rotating Spin-1 Bose Clusters}, Phys. Rev.
  Lett. 89 (2002), p. 050401.

\bibitem{ParedesZC02}
B. Paredes, P. Zoller, and J.I. Cirac, {\itshape Fermionizing a small gas of
  ultracold bosons}, Phys. Rev. A 66 (2002), p. 033609.

\bibitem{ReijndersvSR02}
J.W. Reijnders, F.J.M. van Lankvelt, K. Schoutens, and N. Read, {\itshape
  Quantum Hall states and boson triplet condensate for rotating spin-1 bosons},
  Phys. Rev. Lett. 89 (2002), p. 120401.

\bibitem{B.DeMarco09101999}
B. DeMarco and D.S. Jin, {\itshape {Onset of Fermi Degeneracy in a Trapped
  Atomic Gas}}, Science 285 (1999), pp. 1703--1706.

\bibitem{PhysRevLett.90.053201}
C.A. Regal, C. Ticknor, J.L. Bohn, and D.S. Jin, {\itshape Tuning $p$-Wave
  Interactions in an Ultracold Fermi Gas of Atoms}, Phys. Rev. Lett. 90 (2003),
  p. 053201.

\bibitem{hociobanu}
T.L. Ho and C.V. Ciobanu, {\itshape Rapidly Rotating Fermi Gases}, Phys. Rev.
  Lett. 85 (2000), pp. 4648--4651.

\bibitem{baranov:070404}
M.A. Baranov, K. Osterloh, and M. Lewenstein, {\itshape Fractional Quantum Hall
  States in Ultracold Rapidly Rotating Dipolar Fermi Gases}, Phys. Rev. Lett.
  94 (2005),  p. 070404.

\bibitem{osterloh:160403}
K. Osterloh, N. Barber\'{a}n, and M. Lewenstein, {\itshape Strongly Correlated
  States of Ultracold Rotating Dipolar Fermi Gases}, Phys. Rev. Lett. 99
  (2007), p. 160403.

\bibitem{gunter:230401}
K. G\"{u}nter, T. St\"{o}ferle, H. Moritz, M. K\"{o}hl, and T. Esslinger,
  {\itshape $p$-Wave Interactions in Low-Dimensional Fermionic Gases}, Phys. Rev.
  Lett. 95 (2005), p. 230401.

\bibitem{gaebler:200403}
J.P. Gaebler, J.T. Stewart, J.L. Bohn, and D.S. Jin, {\itshape $p$-Wave Feshbach
  Molecules}, Phys. Rev. Lett. 98 (2007), p. 200403.

\bibitem{PhysRevA.70.030702}
J. Zhang, E.G.M. van Kempen, T. Bourdel, L. Khaykovich, J. Cubizolles, F.
  Chevy, M. Teichmann, L. Tarruell, S.J.J.M.F. Kokkelmans, and C. Salomon,
  {\itshape $p$-wave Feshbach resonances of ultracold Li$^6$}, Phys. Rev. A 70
  (2004), p. 030702.

\bibitem{schunck:045601}
C.H. Schunck, M.W. Zwierlein, C.A. Stan, S.M.F. Raupach, W. Ketterle, A.
  Simoni, E. Tiesinga, C.J. Williams, and P.S. Julienne, {\itshape Feshbach
  resonances in fermionic [sup 6]Li}, Phys. Rev. A 71 (2005), p.
  045601.

\bibitem{inada:100401}
Y. Inada, M. Horikoshi, S. Nakajima, M. Kuwata-Gonokami, M. Ueda, and T.
  Mukaiyama, {\itshape Collisional Properties of $p$-Wave Feshbach Molecules},
  Phys. Rev. Lett. 101 (2008), p. 100401.

\bibitem{gurariereview}
V. Gurarie and L. Radzihovsky, {\itshape Resonantly paired fermionic
  superfluids}, Ann.~Phys. 322 (2007), p.~2.

\bibitem{regnaultpwavefermions}
N. Regnault and T. Jolicoeur, {\itshape Quantum Hall fractions in ultracold
  fermionic vapors}, Phys. Rev. B 70 (2004), p. 241307.

\bibitem{haldanehierarchy}
F.D.M. Haldane, {\itshape Fractional Quantization of the {H}all Effect: A
  Hierarchy of Incompressible Quantum Fluid States}, Phys. Rev. Lett. 51
  (1983), pp. 605--608.

\bibitem{levinsen:210402}
J. Levinsen, N.R. Cooper, and V. Gurarie, {\itshape Strongly Resonant $p$-Wave
  Superfluids}, Phys. Rev. Lett. 99 (2007), p. 210402.

\bibitem{varenna-ketterle}
W. Ketterle and M. Zwierlein, {\itshape Making,probing and understanding
  ultracold Fermi gases, in Ultracold Fermi Gases}, M.~Inguscio, W.~Ketterle
  and C.~Salomon,  eds.,    IOS Press, Amsterdam, 2008.

\bibitem{varenna-grimm}
R. Grimm, {\itshape Ultracold Fermi gases in the BEC-BCS crossover: a review
  from the Innsbruck perspective}, M.~Inguscio, W.~Ketterle and C.~Salomon,
  eds.,    IOS Press, Amsterdam, 2008.

\bibitem{Zwierlein01272006}
M.W. Zwierlein, A. Schirotzek, C.H. Schunck, and W. Ketterle, {\itshape
  {Fermionic Superfluidity with Imbalanced Spin Populations}}, Science 311
  (2006), pp. 492--496.

\bibitem{Partridge01272006}
G.B. Partridge, W. Li, R.I. Kamar, Y.A. Liao, and R.G. Hulet, {\itshape
  {Pairing and Phase Separation in a Polarized Fermi Gas}}, Science 311 (2006),
  pp. 503--505.

\bibitem{Zwierlein}
M.W. Zwierlein, J.R. Abo-Shaeer, A. Schirotzek, C.H. Schunck, and W. Ketterle,
  {\itshape Vortices and superfluidity in a strongly interacting Fermi gas},
  Nature 435 (2005), p. 1047.

\bibitem{Gorkov59}
P. Gorkov, {\itshape Microscopic derivation of the Ginzburg-Landau equations in
  the theory of superconductivity}, Zh.~Eksp.~Teor.~Phys. 36 (1959), p. 1918
  [JETP {\bf 9}, 1364 (1959)].

\bibitem{Gorkov60}
---{}---{}---, {\itshape The critical supercooling field in superconductivity
  theory}, Zh.~Eksp.~Teor.~Phys. 37 (1959), p. 833 [JETP {\bf 10}, 593 (1960)].

\bibitem{werthamer}
N.R. Werthamer, E. Helfand, and P.C. Hohenberg, {\itshape Temperature and
  Purity Dependence of the Superconducting Critical Field, $Hc2$. III. Electron
  Spin and Spin-Orbit Effects}, Phys. Rev. 147 (1966), pp. 295--302.

\bibitem{gurarie}
M.Y. Veillette, D.E. Sheehy, L. Radzihovsky, and V. Gurarie, {\itshape
  Superfluid Transition in a Rotating Fermi Gas with Resonant Interactions},
  Phys.~Rev.~Lett. 97 (2006), p. 250401.

\bibitem{andreevgl}
A.V. Andreev, V. Gurarie, and L. Radzihovsky, {\itshape Nonequilibrium Dynamics
  and Thermodynamics of a Degenerate Fermi Gas Across a Feshbach Resonance},
  Phys. Rev. Lett. 93 (2004), p. 130402.

\bibitem{ZhaiHo}
H. Zhai and T.L. Ho, {\itshape Critical Rotational Frequency for Superfluid
  Fermionic Gases across a Feshbach Resonance}, Phys.~Rev.~Lett. 97 (2006),
   p. 180414.

\bibitem{MollerC07}
G. M\"{o}ller and N.R. Cooper, {\itshape Density Waves and Supersolidity in
  Rapidly Rotating Atomic Fermi Gases}, Phys. Rev. Lett. 99 (2007), p.
  190409.

\bibitem{RasoltT92}
M. Rasolt and Z. Te\u{s}anovi\'{c}, {\itshape Theoretical aspects of
  superconductivity in very high magnetic fields}, Rev.~Mod.~Phys. 64 (1992),
  pp. 709--754.

\bibitem{haldanerez}
F.D.M. Haldane and E.H. Rezayi (2004),  {KITP conference, and private
  communication}.

\bibitem{yang:030404}
K. Yang and H. Zhai, {\itshape Quantum Hall Transition near a Fermion Feshbach
  Resonance in a Rotating Trap}, Phys. Rev. Lett. 100 (2008), p.
  030404.

\bibitem{readcooper}
N. Read and N.R. Cooper, {\itshape Free expansion of lowest-Landau-level states
  of trapped atoms: A wave-function microscope}, Phys. Rev. A 68 (2003),
  p. 035601.

\bibitem{shm2}
J. Sinova, C.B. Hanna, and A.H. MacDonald, {\itshape Measuring the condensate
  fraction of rapidly rotating trapped boson systems: Off-diagonal order from
  the density profile}, Phys. Rev. Lett. 90 (2003), p. 120401.

\bibitem{altman:013603}
E. Altman, E. Demler, and M.D. Lukin, {\itshape Probing many-body states of
  ultracold atoms via noise correlations}, Phys. Rev. A 70 (2004), p.
  013603.

\bibitem{baksmaty:063615}
L.O. Baksmaty, S.J. Woo, M. Banks, S. Choi, and N.P. Bigelow, {\itshape Chiral
  edge states of vortex matter}, Phys. Rev. A 72 (2005), p. 063615.

\bibitem{Paredes}
B. Paredes, P. Fedichev, J.I. Cirac, and P. Zoller, {\itshape $1/2$-Anyons in
  Small Atomic Bose-Einstein Condensates}, Phys. Rev. Lett. 87 (2001), p.
  010402.

\bibitem{papp:135301}
S.B. Papp, J.M. Pino, R.J. Wild, S. Ronen, C.E. Wieman, D.S. Jin, and E.A.
  Cornell, {\itshape Bragg Spectroscopy of a Strongly Interacting $^{85}$Rb
  Bose-Einstein Condensate}, Phys. Rev. Lett. 101 (2008), p. 135301.

\bibitem{petrov:090404}
D.S. Petrov, C. Salomon, and G.V. Shlyapnikov, {\itshape Weakly Bound Dimers of
  Fermionic Atoms}, Phys. Rev. Lett. 93 (2004), p. 090404.

\bibitem{antezza:053609}
M. Antezza, M. Cozzini, and S. Stringari, {\itshape Breathing modes of a fast
  rotating Fermi gas}, Phys. Rev. A 75 (2007), p. 053609.

\bibitem{schmiedmayer}
R. Folman, P. Kr{\" u}ger, J. Schmiedmayer, J. Denschlag, and C. Henkel,
  {\itshape Microscopic Atom Optics: From Wires to an Atom Chip},  48 (2002),
  p. 263.

\bibitem{popp:053612}
M. Popp, B. Paredes, and J.I. Cirac, {\itshape Adiabatic path to fractional
  quantum Hall states of a few bosonic atoms}, Phys. Rev. A 70 (2004), p. 053612.

\bibitem{baur-2008}
S.K. Baur, K.R.A. Hazzard, and E.J. Mueller, {\itshape Stirring trapped atoms into
  fractional quantum Hall puddles},  arXiv:0806.1517.

\bibitem{juzeliunas}
G. Juzeli\={u}nas, J. Ruseckas, P. \"{O}hberg, and M. Fleischhauer, {\itshape
  Light-induced effective magnetic fields for ultracold atoms in planar
  geometries}, Phys. Rev. A 73 (2006), p. 025602.

\bibitem{lin-2008}
Y.J. Lin, R.L. Compton, A.R. Perry, W.D. Phillips, J.V. Porto, and I.B.
  Spielman, {\itshape A Bose-Einstein Condensate in a Uniform Light-induced Vector
  Potential},  arXiv:0809.2976.

\bibitem{fetterwalecka}
 A. L. Fetter and J. D. Walecka, {\itshape Quantum Theory of Many-Particle Systems}, Dover Publications, 2003.

\bibitem{dammorf}
N. d\char39{}Ambrumenil and R. Morf, {\itshape Hierarchical classification of
  fractional quantum Hall states}, Phys. Rev. B 40 (1989), pp. 6108--6119.

\bibitem{haldanemtm}
F.D.M. Haldane, {\itshape Many-particle translational symmetries of
  two-dimensional electrons at rational landau-level filling}, Phys. Rev. Lett.
  55 (1985), pp. 2095--2098.

\bibitem{fanoortolani}
G. Fano, F. Ortolani, and E. Colombo, {\itshape Configuration-interaction
  Calculations on the Fractional Quantum {H}all Effect}, Phys. Rev. B 34
  (1986), pp. 2670--2680.

\bibitem{yoshiokaHL83}
D. Yoshioka, B.I. Halperin, and P.A. Lee, {\itshape Ground State of
  Two-Dimensional Electrons in Strong Magnetic Fields and 1/3 Quantised Hall
  Effect}, Phys. Rev. Lett. 50 (1983), pp. 1219--1222.

\end{thebibliography}

\end{document}